\documentclass[a4paper,fleqn,usenatbib]{mnras}

\usepackage{newtxtext,newtxmath}

\usepackage[T1]{fontenc}
\usepackage{ae,aecompl}

\usepackage{threeparttable}
\usepackage{verbatim} 
\usepackage[title]{appendix}

\usepackage{url}
\usepackage{times}
\usepackage{multirow}
\usepackage{lscape}

\usepackage{graphicx}
\usepackage{epstopdf}
\usepackage[labelfont=bf,labelsep=space]{caption}

\usepackage{enumitem}
\usepackage[justification=centering]{caption}
\usepackage{subfig, float}

\usepackage{natbib}
\bibpunct{(}{)}{;}{a}{}{,}

\usepackage{array}
\newcolumntype{x}[1]{>{\centering\arraybackslash\hspace{0pt}}m{#1}}

\makeatletter
\def\url@leostyle{%
  \@ifundefined{selectfont}{\def\UrlFont{\sf}}{\def\UrlFont{\small\ttfamily}}}
\makeatother
\usepackage{newtxtext,newtxmath}

\DeclareRobustCommand{\ion}[2]{%
\relax\ifmmode
\ifx\testbx\f@series
{\mathbf{#1\,\mathsc{#2}}}\else
{\mathrm{#1\,\mathsc{#2}}}\fi
\else\textup{#1\,{\mdseries\textsc{#2}}}%
\fi}

\title[WR contribution to GN-z11]{JADES NIRSpec Spectroscopy of GN-z11: Evidence for Wolf-Rayet contribution to stellar populations at 430 Myr after Big Bang?}

\author[Gunawardhana \it{et.\,al}]{M.\,L.\,P.\,Gunawardhana$^{1, 2, 3}$\thanks{E-mail: madusha.gunawardhana@sydney.edu.au}, J.\,Brinchmann$^{4, 5}$, S.\,Croom$^{1}$, A.\,J.\,Bunker$^{6}$, J.\,Bryant$^1$, \newauthor{S.\,Oh$^{7}$}
\\
$^{1}${Sydney Institute for Astronomy (SIfA), School of Physics, The University of Sydney, Sydney, NSW 2006, Australia}\\
$^{2}${Siding Spring Observatory, Research School of Astronomy and Astrophysics, Australian National University, Canberra, ACT 2611, Australia}\\
$^{3}${ARC Centre of Excellence for All Sky Astrophysics in 3 Dimensions (ASTRO 3D), Australia}\\
$^{4}$Instituto de Astrof\'isica e Ci\^{e}ncias do Espa\c{c}o, Universidade do Porto, CAUP, Rua das Estrelas, PT4150-762 Porto, Portugal\\
$^{5}$Departamento de F\'sica e Astronomia, Faculdade de Ci\^{e}ncias, Universidade do Porto, Rua do Campo Alegre 687, PT4169-007 Porto, Portugal\\
$^6$Department of Physics, University of Oxford, Denys Wilkinson Building, Keble Road, Oxford OX1 3RH, UK\\
$^7$Department of Astronomy and Yonsei University Observatory, Yonsei University, Seoul, 03722, Republic of Korea
}

\begin{document}
\color{black}

\pagerange{\pageref{firstpage}--\pageref{lastpage}} \pubyear{2002}

\maketitle

\label{firstpage}

\begin{abstract}
We investigate the unusual emission line luminosity ratios observed in the JADES NIRSpec spectroscopy of GN-z11, which reveal exceptionally strong emission lines and a significant detection of the rarely observed \ion{N}{iii}]~$\lambda1748-1753$\AA, multiplet. These features suggest an elevated N/O abundance, challenging existing models of stellar populations and nebular emission. To assess whether Wolf-Rayet (WR) stars can account for the observed line ratios, we construct a suite of stellar and nebular models incorporating high-resolution stellar spectral libraries, enabling a more accurate treatment of WR evolution and its influence on the ionising radiation field.
We find that the inclusion of WR stars is essential for reproducing the observed position of GN-z11 in the \ion{C}{iii}]/\ion{He}{ii} versus \ion{C}{iii}]/\ion{C}{iv} diagnostic plane, resolving discrepancies from previous studies. The model-derived metallicity (0.07$\lesssim$Z/Z$_{\odot}\lesssim$0.15), ionisation parameter ($\log\,U$$\approx$-2) and stellar ages are consistent with the literature estimates. However, our models under-predict the \ion{N}{iii}/\ion{O}{iii}] ratio, suggesting that WR stars alone cannot fully explain the nitrogen enrichment. This suggests that additional mechanisms, such as rapid chemical enrichment in a young, metal-poor environment, may be necessary to explain the nitrogen excess.
While our models successfully reproduce most observed line ratios, further refinements to the models are needed to fully characterise the stellar populations and the enrichment processes of high-redshift galaxies like GN-z11.
\end{abstract}

\begin{keywords}
galaxies: starburst -- galaxies: high-redshift -- galaxies: evolution -- galaxies: abundances -- galaxies: ISM -- galaxies: stellar content	
\end{keywords}

\section{Introduction}\label{sec:intro}

Since its first light, the James Webb Space Telescope (JWST) has revolutionised our understanding of the early Universe, probing deeper than ever before and challenging existing models of galaxy formation and evolution at cosmic dawn.
One of the remarkable discoveries facilitated by JWST is GN-z11, the most luminous galaxy candidate at $z>10$ in the GOODS-North field \citep{Bouwens2010, Oesch2016}. At just $\sim430$ million years after the Big Bang, GN-z11 offers a unique, yet puzzling, glimpse into the astrophysical processes driving the emergence of the first generations of galaxies. Intriguingly, its spectrum from JWST-NIRSpec \citep{Bunker2023} reveals extreme ultraviolet and optical emission lines, including the rarely observed  \ion{N}{iii}]$\lambda\lambda$1748-1753\AA, along with \ion{C}{iii}], \ion{C}{iv}] and \ion{He}{ii}] emission features.

The analysis of \cite{Bunker2023} highlights that GN-z11 occupies unique regions of emission line diagnostic diagrams, challenging existing photoionisation models for both active galactic nuclei (AGN) and starburst galaxies \citep[e.g.][]{Gutkin2016, Feltre2016}. Furthermore, the galaxy’s spectrum exhibits an unusually high nitrogen-to-oxygen (N/O) abundance ratio, raising critical questions about the chemical enrichment processes occurring in galaxies at such an early epoch.

The origin of GN-z11’s extreme line luminosities and high nitrogen-to-oxygen (N/O) abundance remains a topic of active debate. While various physical mechanisms have been proposed to explain the unusual properties of GN-z11, its true nature remains elusive. 

For instance, \cite{Cameron2023} explore several scenarios to explain the unusually high nitrogen enrichment observed in GN-z11. Their study concludes that while conventional stellar evolution models struggle to account for the elevated N/O ratio, alternative mechanisms--such as Wolf-Rayet (WR) stars, runaway stellar collisions in dense clusters facilitating the formation of very massive stars, and tidal disruption events--may provide more viable explanations. Further evidence supporting the presence of very massive stars in GN-z11 is presented by \cite{Vink2024}, who propose that stars with initial masses between 100 and 1000 M$_{\odot}$ are key contributors to nitrogen enrichment in star-forming galaxies. Due to their proximity to the Eddington limit, these massive stars undergo substantial mass loss via powerful stellar winds, enriching the surrounding interstellar medium with nitrogen. While \cite{Vink2024} assigns less weight to the WR scenario in GN-z11, a recent study by \cite{Watanabe2024} incorporating chemical evolution models with yields from various supernova types, like core-collapse supernovae, Type Ia supernovae, hypernovae, and pair-instability supernovae, suggests that these mechanisms alone do not fully account for the high N/O ratio. Instead, nitrogen enrichment may be attributed to winds from rotating Wolf-Rayet stars that undergo direct collapse.

On the other hand, \cite{Maiolino2024} highlight the detection of [\ion{Ne}{iv}]$\lambda2453$\AA, and \ion{C}{ii}$^*\lambda1335$ transitions, both of which are commonly associated with AGN activity, along with the detection of semi-forbidden nebular emission lines, suggestive of high-density gas, to argue for the presence of an AGN in GN-z11. In contrast, \cite{Senchyna2024} compare GN-z11’s spectrum to local ultraviolet datasets and note similarities with Mrk 996, where a high concentration of Wolf-Rayet stars and their CNO-processed ejecta produce a UV spectrum that closely resembles that of GN-z11. Based on detailed nebular modelling, \cite{Senchyna2024} suggest the peculiar nitrogenic features prominent in GN-z11 may be a unique signature of intense and densely clustered star formation, potentially linked to the evolutionary precursors of present-day globular clusters.

Further support for the AGN hypothesis comes from \cite{Scholtz2024}. They report extended Ly$\alpha$ emission located SW of GN-z11’s continuum centre, and an extended \ion{C}{iii}]~$\lambda1909$\AA\,emission in the same direction \citep{Maiolino2024}, which they interpreted as evidence of an AGN-driven ionisation cone. 

Conversely, several studies provide evidence supporting the starburst nature of GN-z11. For instance, \cite{Alvarez2024} utilise deep MIRI/JWST medium-resolution spectroscopy covering the rest-frame optical spectrum of GN-z11 to model its H$\alpha$ and \ion{O}{iii}$\lambda5008$\AA\,emission features. They find the line profiles to be well-represented by a narrow Gaussian component, with no indication of a dominant broad H$\alpha$ emission line typically associated with the broad-line region of an AGN, though they do not entirely rule out the possibility of a weak-line AGN contribution. Taking into account the high star formation rate and stellar mass surface densities of GN-z11, \cite{Alvarez2024} propose that the galaxy is undergoing a highly efficient starburst phase. 

Similar conclusions are drawn in the model-based studies of \cite{Kobayashi2024}, \cite{Nagele2023} and \cite{Rizzuti2024}. For instance, \cite{Kobayashi2024} propose an intermittent star formation scenario, where a quiescent phase lasting 100 Myr separates two starbursts. They find that the emergence of Wolf-Rayet stars immediately following the second starburst can account for the elevated N/O abundance observed in GN-z11. Similarly, \cite{Nagele2023} use simulations to suggest that metal-enriched supermassive stars, evolving shortly after the zero-age main sequence, could produce supersolar nitrogen levels consistent with the observations of GN-z11. Finally, \cite{Rizzuti2024} employ chemical evolutionary models incorporating various star formation histories to demonstrate that galaxies with extreme star formation rates and differential galactic winds—where the products of core-collapse supernovae are preferentially expelled—can achieve super-solar N/O abundances.

Building on the work of \cite{Bunker2023}, our study in the current paper leverages JWST observations of GN-z11 to investigate the physical properties of its stellar populations, ionising sources, and chemical enrichment. 
Using the updated stellar population synthesis code, \textsc{Starburst99}, which incorporates the latest Wolf-Rayet (WR) classifications, accurate massive stellar evolution, updated isochrones, and high-resolution stellar and WR libraries that more uniformly sample the effective temperature, surface gravity, and stellar rotation parameter space of high-mass stars, alongside photoionisation modelling (\textsc{Cloudy}), we investigate the potential contributions of massive WR stars to the extreme ionising radiation in GN-z11. We also explore mechanisms that could explain the elevated N/O abundance, aiming to bridge the gaps in our understanding of GN-z11’s unique emission properties, as well as building upon previous efforts in modelling starburst regions using self-consistent approaches to accurately reproduce both stellar and nebular properties \citep[e.g.][]{Gomes2017, Gunawardhana2020}.

The paper is structured in two main parts. In the first part, we introduce GN-z11 in \S\,\ref{sec:gn-z11}, followed by a detailed description of the construction of a comprehensive stellar and nebular model library (\S\,\ref{sec:ssp}–\S\,\ref{sec:nebular_models}). This includes modelling starbursts across stellar metallicities that permit Wolf-Rayet star formation, utilising an updated \textsc{Starburst99} code (\S,\ref{sec:ssp}) coupled with the photoionisation code \textsc{Cloudy} (\S,\ref{sec:nebular_models}) to characterise both stellar populations and nebular properties in highly star-forming regions. The second part of the paper applies these models to investigate the stellar and nebular properties of GN-z11 within the framework of various ultraviolet diagnostic diagrams (\S\,\ref{subsec:BunkerFig_interp}).

The assumed cosmological parameters are H$_0$ = 70 km\,s$^{-1}$\,Mpc$^{-1}$, $\Omega_M$ = 0.3, and $\Omega_{\Lambda}$= 0.7. We assume a \citet{Kroupa2001} stellar initial mass function (IMF) with high-mass cut-off of 120 M$_{\odot}$ throughout.

\section{JADES NIRSpec Spectroscopy of GN-z11}\label{sec:gn-z11}
The NIRSpec observations of GN-z11, the most luminous candidate z > 10 Lyman break galaxy in the GOODS-North field, were taken as a part of the JADES collaboration between the NIRSpec and NIRCam instrument science teams. This study uses the GN-z11 dataset described in detail in \cite{Bunker2023}. 

Briefly,  the GN-z11 was observed with high priority using NIRSpec in its micro-shutter array mode \citep{Ferruit2022}, comprising of four arrays of 365$\times$171 independently operable shutters, with 98"$\times$91" sky coverage. Our study uses the co-added data, with a total integration time of 3.45hrs in each medium-resolution ($R\approx1000$) G140M/F070LP (0.70--1.27\,$\mu$m), G235M/F170LP (1.66--3.07\,$\mu$m), and G395M/F290LP (2.87–5.10\,$\mu$m) grating, and 6.9hrs in the low-resolution ($R\approx100$) PRISM/CLEAR mode (0.7--5.3\,$\mu$m). The observations were processed with the reduction pipelines developed by the NIRSpec instrument science and GTO teams \citep{Cameron2023}. Based on the processed data, \cite{Bunker2023} report a redshift of $10.6034\pm0.0013$ for GN-z11.

For the analysis presented in this paper, we use the medium-resolution GN-z11 spectra presented in Figure B.1 of \citep{Bunker2023}.

In the subsequent sections, we present and discuss the development of the suite of high resolution and self-consistent stellar and nebular models, and their application to interpret the NIRSpec observations of GN-z11.

\section{Stellar Population Synthesis Modelling}\label{sec:ssp}

The JWST has significantly expanded our ability to probe much higher redshifts, allowing the observations of young, massive stellar populations that emit a substantial portion of their flux in the far-ultraviolet. The increasing number of distant galaxy observations has intensified the need to improve stellar population synthesis (SPS) models, facilitating more reliable interpretations of these galaxies' physical properties  \citep{Byrne2023, Pacifici2023}. However, a major challenge in studying high-redshift stellar populations using evolutionary population synthesis is the limited availability of comprehensive, high-resolution models with extended ultraviolet wavelength coverage. In this section, we detail our approach to incorporating theoretical stellar spectral libraries, aiming to model both the stellar and nebular features of young, massive stellar populations in a self-consistent manner. This approach leverages the spectral resolution and wavelength range provided by JWST-NIRSpec spectroscopy.

To generate high-resolution stellar population templates, we use the \textsc{Starburst99} code of \cite{Leitherer99}, which has been modified to model more accurately the key evolutionary phases of massive stars. The subsequent sections detail the modifications to the three core components of stellar population synthesis (SPS) models. In \S\,\ref{subsec:stellar_isochrones}, we discuss the implementation of the stellar isochrones used in this study, which map a star’s position in the bolometric luminosity--effective temperature plane (or equivalently, in the surface gravity--effective temperature plane) based on its mass, initial chemical composition, and age. The incorporation of the high resolution stellar libraries, including Wolf-Rayet stars, tracing different evolutionary stages of stars \citep{Vazdekis2012}, is detailed in \S\,\ref{subsec:stellar_libraries} and \S\,\ref{subsec:wr_stars}. Finally, in \S\ref{sec:nebular_models}, we describe the self-consistent construction of nebular models using the \textsc{Cloudy} photoionisation code of \citep{Ferland2013}, integrating the updated Starburst99 SPS models as input.

\subsection{The stellar evolutionary models}\label{subsec:stellar_isochrones}
For the analysis presented in this paper, we utilise two widely used single-stellar evolutionary models; the \textsc{Geneva} models \citep{Schaller1992, Charbonnel1993, Meynet1994} and the \textsc{Parsec} models \citep{Bressan2012, Chen2014, Chen2015}. Recent models that have incorporated both rotational effects in massive stars \citep[e.g.\,Geneva and MIST models;][]{Meynet1994, Paxton2013} and binary evolution \citep[e.g.\,BPASS;][]{Eldridge2009, Eldridge2017, Gotberg2019} have also been published. Compared with single-evolutionary models, the models including stellar rotation and multiple stellar evolution tend to extend the effects of different evolutionary phases of massive stars.    

While binary evolution effects offer a more complete picture of massive stellar evolution, the use of single-stellar evolutionary models in this study is both appropriate and advantageous. These models provide a well-controlled framework for isolating the impact of different evolutionary phases on spectral features. Their efficiency in exploring parameter space and fine-tuning model parameters makes them particularly well-suited for this investigation. Although binary evolution can extend the duration of certain stellar phases, allowing them to be captured even with coarse temporal sampling, by sufficiently fine sampling single-stellar evolutionary models, we can effectively ensure that these phases are accurate represented. 

\subsubsection{\textsc{Geneva} stellar evolutionary tracks}
The `\textsc{Geneva} high mass loss' (HML) isochrones \citep{Meynet1994} are generally preferred over any isochrones with standard mass loss rates (e.g.\,\textsc{Geneva} and \textsc{Padova} standard tracks) for modelling starburst galaxies, as they more accurately reproduce observations of the Wolf-Rayet (WR) phase, particularly for low-luminosity WR stars  \citep[e.g.][]{Brinchmann2008, Levesque2010, Byler2017}. These models adopt mass-loss rates approximately twice those of the `standard' grid \citep{deJager1988}, providing a more reasonable approximation of mass-loss for massive stars evolving into the WR phase. Mass-loss rates for WR stars of the Nitrogen subclass with Hydrogen-free (hereafter WNE or early-type WN), WR stars of the Carbon subclass (hereafter WC), and WR stars of the Oxygen subclass (hereafter WO) remain unchanged, except for the WR stars of the Nitrogen subclass with a specified Hydrogen mass fraction, hereafter late-type WN or WNL stars \citep{Meynet1994}.

In this analysis, we utilise the existing implementation of \textsc{Geneva} high-mass loss isochrones within \textsc{Starburst99}, covering the full range in available stellar metallicities (i.e.\,$Z$=$0.001, 0.004, 0.008, 0.02, 0.04$), assuming the solar metallicity to be $Z_{\odot}$=$0.02$. The upper initial stellar mass limit is set at 120 M$_{\odot}$, with lower mass cut-offs at 25, 20, 15, 15, 12 M$_{\odot}$, respectively, from low-to-high stellar metallicity. The HML models as implemented in \textsc{Starburst99} are, therefore, combined with the `standard' models to extend the stellar mass range down to 0.1 M$_{\odot}$, sampling around 22 different stellar mass values over 51 time-intervals.

In the next section, we detail the implementation of the \textsc{Parsec} isochrones in \textsc{Starburst99}. A key difference between the \textsc{Geneva} and \textsc{Parsec} models lies in their treatment of mass-loss rates, which significantly influence the evolution of massive stars. We also examine how these variations in mass-loss rates affect the transition of massive stars into the WR phase across different metallicities in \S\,\ref{subsubsec:link_WR_to_isochrones}. 

\subsubsection{\textsc{Parsec} stellar evolutionary tracks}

In addition to the \textsc{Geneva} models, we also utilise the  \textsc{Parsec}v1.S isochrones \footnote{The PARSECv1.2S library is downloaded from the CMD3.1 web interphase (\url{http://stev.oapd.inaf.it/cgi-bin/cmd})}  \citep{Bressan2012, Chen2014, Chen2015} for this analysis to investigate how the model assumptions change between the two sets of isochrones. The \textsc{Parsec} tracks are calculated for a scaled-solar composition, with the initial helium content linked to the initial metallicities by the relation  $Y_i = 1.78\times Z_i+0.2485$, assuming a solar metallicity of $Z_{\odot}$=$0.0152$. 
 
The \textsc{Parsec} release spans a broad range of metallicities, from $0.0001$ to $0.04$, with initial masses up to 350 M$_{\odot}$, and sample finely in stellar mass. It includes improvements in the treatment of boundary conditions for low-mass stars \citep[$\sim 0.6$ M$_{\odot}$]{Chen2014}, as well as updates to envelope overshooting and up-to-date mass-loss rates for massive stars \citep[i.e.\,$14\lesssim$ M/M$_{\odot}\lesssim$ 350]{Tang2014, Chen2015}.

We generate isochrones using evolutionary tracks from the \textsc{Parsec}v1.S library for integration into \textsc{Starburst99}. These tracks take advantage of fine sampling across time, metallicity, and stellar mass, encompassing 52 distinct stellar masses in the range $0.1\lesssim$ M/M$_{\odot}\lesssim$ 350. Each evolutionary track is sampled non-uniformly at 1\,000 timesteps to capture all stellar evolutionary phases, and cover the evolutionary stage from very near the zero-age main sequence \citep{Chen2015} to the end stages for massive stars, or up to 10 Gyr for low-mass stars.

To accurately model the critical WR phase of massive stars, we apply corrections to the effective temperatures of all \textsc{Parsec}v1.S evolutionary models of massive stars. The derivation and justification of these temperature adjustments are discussed in \S\,\ref{subsec:wr_stars}. 

\subsection{The stellar spectral library}\label{subsec:stellar_libraries}
There are several approaches to producing stellar libraries for SPS codes, each with its own strengths and limitations. For generating SPS models in star-forming regions, one critical requirement is comprehensive coverage of the effective temperature (T$_{\rm{eff}}$), surface gravity ($\log\,g$) and metallicity ([Fe/H]) parameter space. Additionally, extended wavelength coverage is essential, as different stellar phases peak at different wavelengths.

Empirical stellar libraries, while advantageous for being based on observed stellar spectra, are largely limited to stars visible within the Galaxy. As a result, they mainly encompass bright, local stars and lack sufficient coverage of the parameter space for rare, high-mass stars. This limitation makes theoretical stellar libraries more suitable for modelling young, massive stellar populations. These theoretical spectra are generated using radiative transfer processes to simulate flux from model stellar atmospheres \citep{Munari2005, Kurucz1992}, enabling coverage across the entire Hertzsprung-Russell (HR) diagram, exploring all possible atmospheric parameters.

One of the main drawbacks of theoretical libraries is their dependence on tabulated opacities and atomic/molecular absorption strengths, which can have significant uncertainties. Despite these challenges, accurately modelling young, massive stellar populations requires a stellar spectral library that covers the full parameter space of T$_{\rm{eff}}$, $\log\,g$, and [Fe/H], along with extended wavelength coverage. In this context, theoretical libraries remain the best option for capturing the evolution of high-mass stars across the desired parameter space.

We integrate the synthetic ultraviolet-blue (UVBlue) library of \cite{Rodriguez2005} into \textsc{Starburst99} to meet the high spectral resolution requirements in the ultraviolet regime. The UVBlue library, calculated at a resolution of 50\,000 using the \textsc{Atlas9} and \textsc{Synthe} codes of \cite{Kurucz1993}, consists of 1\,770 spectra covering a wavelength range of 850 -- 4700\AA. The grid spans 3000-50\,000 K in effective temperature, $0.0-5.0$ in log surface gravity at steps of $+0.5$ dex, and includes seven metallicities  ([M/H] = $-2.0, -1.5, -1.0, -0.5, +0.0, +0.3$ and $+0.5$ dex). The synthetic spectra assume solar-scaled abundances from \cite{Grevesse1998}  and employ the atomic and diatomic molecular line lists of \cite{Kurucz1992}.

The diatomic molecular lines included in the computation are C$_2$, CN, CO, H$_2$, SiO, CH, NH, OH, MgH, and SiH, excluding TiO lines for cooler stars and atomic lines with theoretical transitions (i.e.\,'predicted' lines). While uncertainties in the wavelengths and intensities of these predicted lines can introduce spurious absorption features and contaminate high-resolution spectra \citep{Munari2005, Coelho2007}, flux calibration remains essential for any high-resolution spectral library used to derive broadband colours, as the absence of predicted lines can otherwise lead to inaccuracies in colour predictions.
 As the present study focuses on exploring the evolutionary phases of massive stars, and as we do not intend to use the current library for broadband colour determinations, the uncertainties arising from the lack of predicted lines are expected to be relatively low.

\subsection{The Wolf-Rayet spectral library}\label{subsec:wr_stars}
We combine \textsc{starburst99}'s low-resolution CMFGEN library \citep{Hillier1998} with the higher-resolution Potsdam grids\footnote{The Potsdam grids (\url{http://www.astro.physik.uni-potsdam.de/~wrh/PoWR/powrgrid1.php}) cover wavelengths $>950$\AA.} of model atmospheres for Wolf-Rayet (WR) stars (Galactic, LMC, SMC, and sub-SMC), developed by \cite{Hamann2004, Sander2012, Todt2015}. The Potsdam Wolf-Rayet (PoWR) library provides extensive grids of expanding, non-local thermodynamic equilibrium, iron-group line-blanketed atmospheres for WR subtypes: WN stars, characterised by strong helium and nitrogen lines, and WC stars, defined by strong helium and carbon lines. The inclusion of iron-group line blanketing is crucial for accurately reproducing the observed spectra of WR stars, particularly WC subtypes, where numerous iron line transitions (Fe\,\textsc{iv}, Fe\,\textsc{v}, Fe\,\textsc{vi}) form a pseudo-continuum in the ultraviolet that dominates the spectral energy distribution \citep{Grafener2002}.

The PoWR grids are parameterised by luminosity (L$^*$), effective temperature (T$_{\rm{eff}}$) and wind density. The WR stars exhibit wind densities that are typically an order of magnitude higher than those of O-type stars, which are thought to be the result of multiple-photon scattering events due to their high luminosity-to-mass ratios \citep{Crowther2007, Leitherer2014}. The strong, broad emission features characteristic of WR stellar spectra originate in their powerful stellar winds, forming far from the stellar surface.  This increase in the optical thickness in mass-loss ($\dot{\rm{M}}$) winds significantly impacts WR star modelling. For instance, at a given metallicity, the minimum stellar mass required for a massive star to enter the WR phase decreases with higher $\dot{\rm{M}}$, while the WR phase duration increases. Additionally, both the time spent in each WR subtype and the surface composition are highly sensitive to mass-loss rates \citep{Meynet1994}.

Moreover, the optical thickness and strength of WR winds influence the hardness of their ionising radiation, redistributing extreme UV photons toward redder wavelengths. As a result, WR stars appear cooler and larger in size than they would in the absence of mass loss. Therefore, the shape of the ionising flux distribution, particularly the transparency below $228$\AA\,contributing to He$^0$ ionising photons, of WR stars is a strong function of both the wind density and T$_{\rm{eff}}$ \citep{Langer1989, Crowther2007}.

Given the non-negligible role of wind density in determining the features of WR stars, the PoWR library of WN and WC grids are defined as a function of wind density, parametrised in terms of a "transformed radius" \citep[][R$_{\rm{T}}$]{Schmutz1989}, and T$_{\rm{eff}}$. To integrate the PoWR library into \textsc{Starburst99}, we adopt the approach of \cite{Leitherer2014}, employing the parameterisation established by \cite{Schmutz1989}  as,
\begin{equation}
R_{\rm T} = R_* \Big( \frac{v_{\infty}}{2500} \Big/ \frac{\sqrt{D}\dot{M}}{1\times10^{-4}}\Big)^{2/3},
\label{eq:rt_calculation}
\end{equation}
where $R_*$ is the effective stellar radius in the unit of m, $v_{\infty}$ is the terminal wind velocity in the unit of km\,s$^{-1}$, and $\dot{\rm{M}}$ is the mass loss rate in M$_{\odot}$ yr$^{-1}$. Based on the observational evidence of line profile variabilities \citep{Moffat1988} and the over-prediction of the amplitude of the scattering wings in theoretical line profiles \citep{Hillier1991},  the winds of WR stars are thought to be clumped \citep{Grafener2017, Grassitelli2018}. Therefore, the wind clumping factor, ${D}$, acts to downward revise $\dot{\rm{M}}$, although the exact amount is unclear \citep{Smith2002}, for the present study, we adopt the typical values of $D$ in the literature \citep[e.g.][]{Dessart2000, Crowther2002, Smith2002, Crowther2007, Doran2013, Hainich2014, Leitherer2014} in the range $4-10$ for WC stars to $\sim4$ for WN subtypes \citep{Crowther2007}. 

Under the \cite{Schmutz1989} definition of R$_{\rm{T}}$, for a given T$_{\rm{eff}}$ and metallicity, the model spectra with the same $R_{\rm{T}}$ have approximately the same emission line equivalent widths regardless other specific wind parameters \citep{Todt2015}, meaning that for a fixed luminosity, the model depends only on R$_{\rm{T}}$ and T$_{\rm{eff}}$.  

\subsubsection{Linking WR atmospheres with stellar evolutionary tracks}\label{subsubsec:link_WR_to_isochrones}
The challenge of integrating WR atmospheres with stellar evolutionary models is a complex and long-standing problem in stellar population synthesis. This problem exists because the optical thickness of WR stellar winds causes the observed radiation to emerge at larger radii. Consequently, the observed temperatures of WR stars cannot be directly linked to the hydrostatic core temperatures (T$_{\rm{hyd}}$) provided by evolutionary models without an inward extrapolation \citep{Schmutz1989, Smith2002}. 

To this end, \cite{Maeder1990, Meynet1994} and \cite{Vazquez2007} propose a potential solution, suggesting the use of a velocity law to estimate an effective radius corresponding to an optical depth ($\tau$) of $\frac{2}{3}$. This approach allows the hydrostatic core temperature to be scaled to the temperature at this optical depth  (T$_{\rm{2/3}}$). However, as noted by \cite{Schmutz1992} and \cite{Smith2002}, the T$_{\rm{2/3}}$ derived from WR atmospheres typically remain around 30\,000 K across different WR subtypes, and still lower than T$_{\rm{hyd}}$ predicted by evolutionary models. 

To address this discrepancy, we adopt a hybrid approach, as detailed below.

Following \cite{Maeder1990, Meynet1994, Meynet2005}, and \cite{Vazquez2007}, we assume the effective radius at a $\tau\approx 2/3$ is related to the classical photospheric radius (R$_{\rm{hyd}}$) via,
\begin{equation}
R_{\text{2/3}} = R_{\rm{hyd}} + \frac{3\kappa\lvert{\dot M}\rvert}{8 \pi v_{\infty}},
\label{eq:reff_corr}
\end{equation}
where $\kappa$ is the opacity, and $\lvert{\dot M}\rvert$ is the mass-loss rate in M$_{\odot}$ yr$^{-1}$. Then using the Stefan-Bolzmann law, and assuming $L\approx L_{\rm{2/3}}$, T$_{\rm{2/3}}$ is calculated.  

The challenge, then, is how to determine an accurate temperature for WR stars, given that T$_{\rm{hyd}}$ is generally too-high, while T$_{\rm{2/3}}$  is typically too low. To address this, we adopt the weighted mean temperature approach proposed by \cite{Smith2002}, which is based on analysing the distribution of WR temperatures from \textsc{Starburst99} at solar metallicity. In this formalism, the WR temperature (T$_{\rm{WR}}$) is defined as, 
\begin{equation}
T_{\rm{WR}} = 0.6 \times T_{\rm{hyd}} + 0.4 \times T_{2/3}.
\label{eq:weighted_Twr}
\end{equation}
 
The formalism in Eq.\,\ref{eq:weighted_Twr} is used in correcting the temperatures during the WR phase only \citep{Vazquez2007, Leitherer2014}.

\subsubsection{The Wolf-Rayet stellar classification}
\begin{figure*}
\begin{center}
\includegraphics[trim={0.2cm 2.4cm 2.0cm 3.8cm},clip, width=0.99\textwidth]{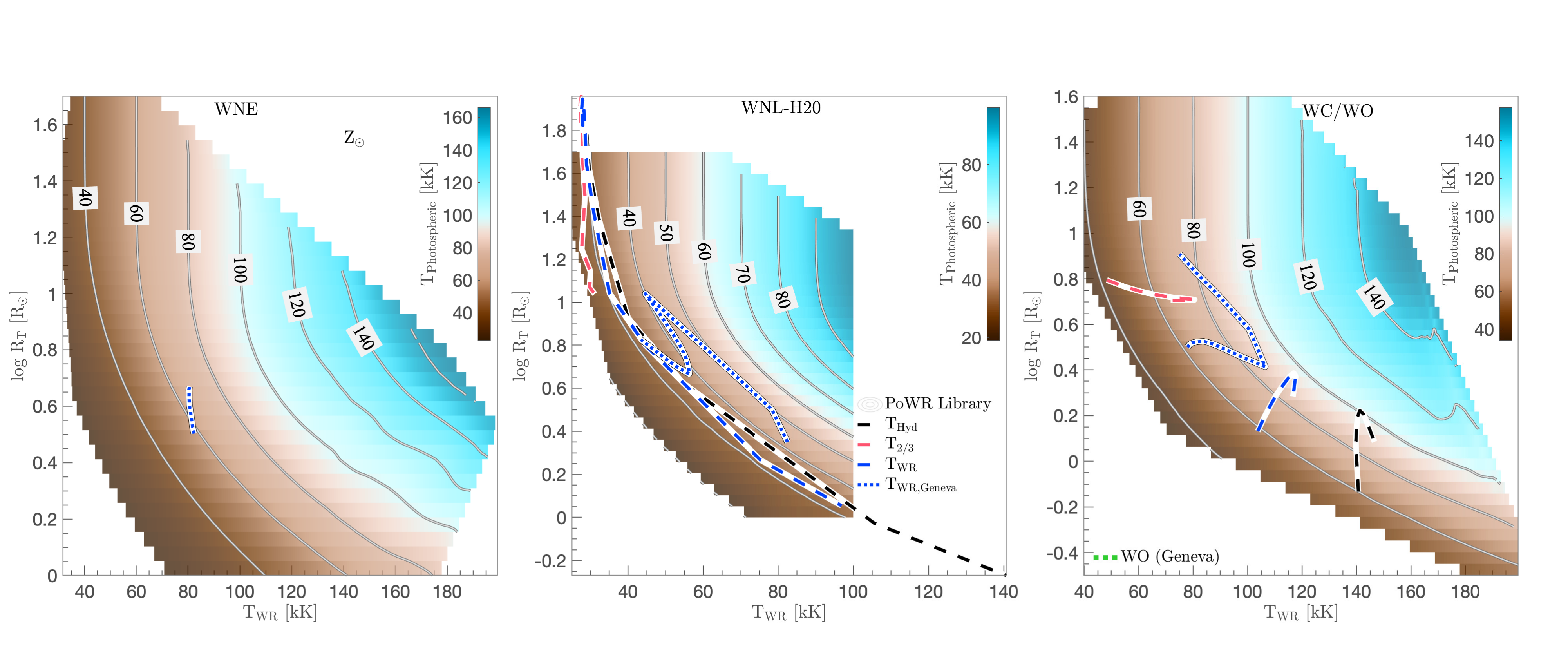}
\includegraphics[trim={.1cm 2.4cm 2.0cm 3.8cm},clip, width=0.99\textwidth]{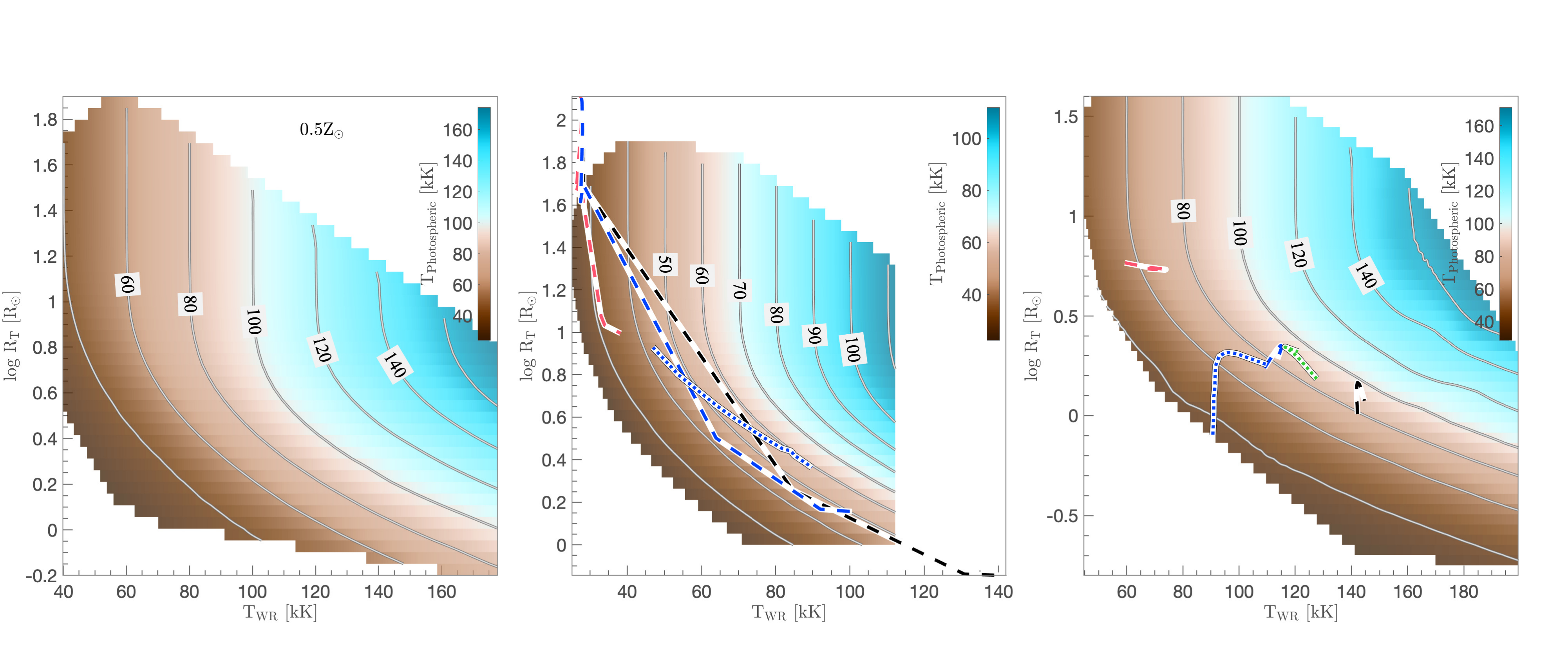}
\includegraphics[trim={.2cm 1.4cm 2.0cm 3.8cm},clip, width=0.99\textwidth]{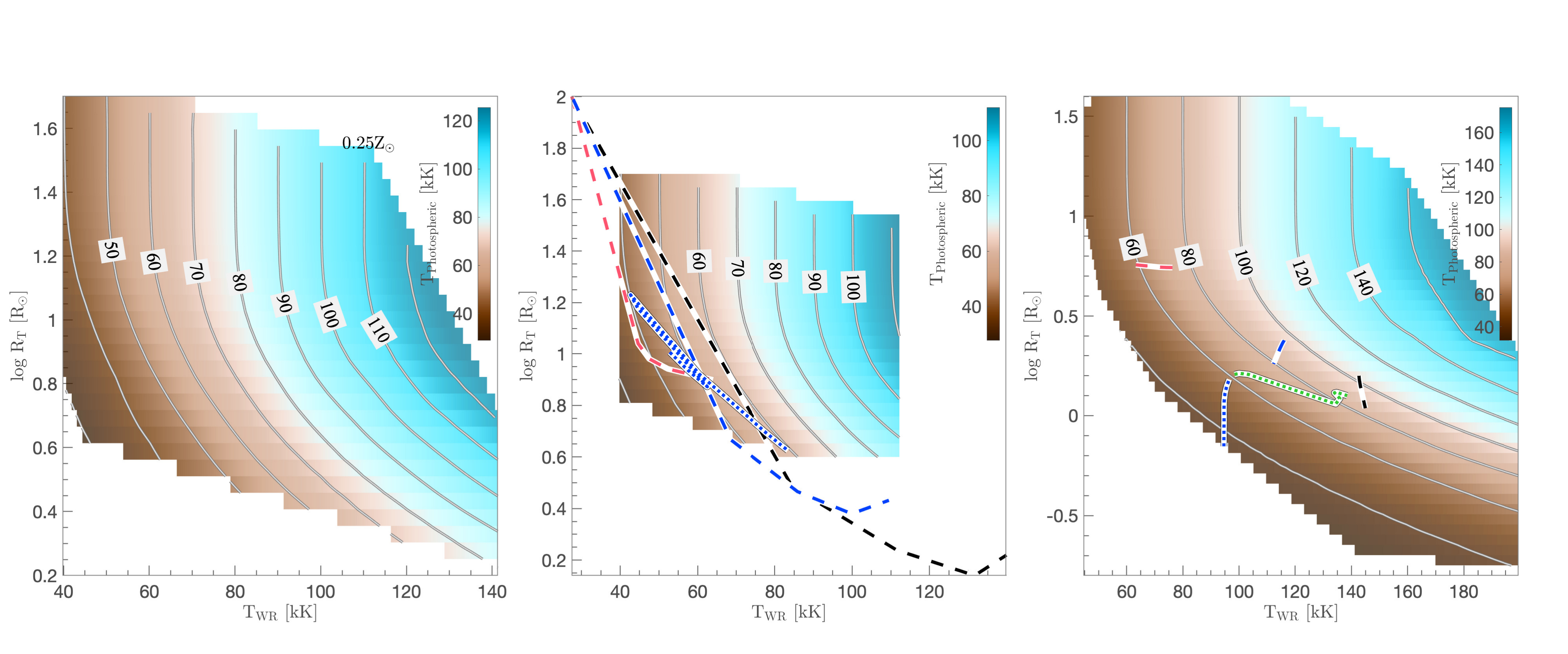}
\caption{On the selection of WR spectra as a function of T$_{\rm{WR}}$ [10$^3$ K] and R$_{\rm{T}}$ [R$_{\odot}$] for \textsc{Parsec} and \textsc{Geneva} stellar isochrones from PoWR library. The panels (left-to-right, top-to-bottom) show PoWR grids for WNE, WNL-H20, and WC spectra at solar, LMC-like, SMC-like, and sub-SMC-like ($\sim$0.7 Z${\odot}$) metallicities. The contours and color gradients represent photospheric temperatures as a function of T$_{\rm{WR}}$ and R$_{\rm{T}}$. Overlaid on each grid are evolutionary tracks for a 120 M$_{\odot}$ star from \textsc{Parsec} (dashed) and \textsc{Geneva} (dotted) isochrones as it enters its WR phase. For the \textsc{Parsec} models, we show the selection of WR spectra with respect to T$_{\rm{hyd}}$ (black-dashed), T$_{2/3}$ (red-dashed) and T$_{\rm{WR}}$ (blue-dashed), and their respective R$_{\rm{T}}$s discussion in \S\,\ref{subsec:wr_stars}. For the \textsc{Geneva} models, we only show the selection for T$_{\rm{WR}}$. Note that the panels displaying only the PoWR model grids indicate metallicities where \textsc{Parsec} or \textsc{Geneva} models do not predict the formation of WR stars. The green dotted lines indicate the WO evolutionary phase, which appears only in the \textsc{Geneva} models for 120 M$_{\odot}$ stars. For this phase to occur in \textsc{Parsec} models require higher stellar masses.}
\label{fig:linking_sst_with_powr}
\end{center}
\end{figure*}
\begin{figure*}
\ContinuedFloat
\begin{center}
\includegraphics[trim={.1cm 1.4cm 2.0cm 3.8cm},clip, width=0.99\textwidth]{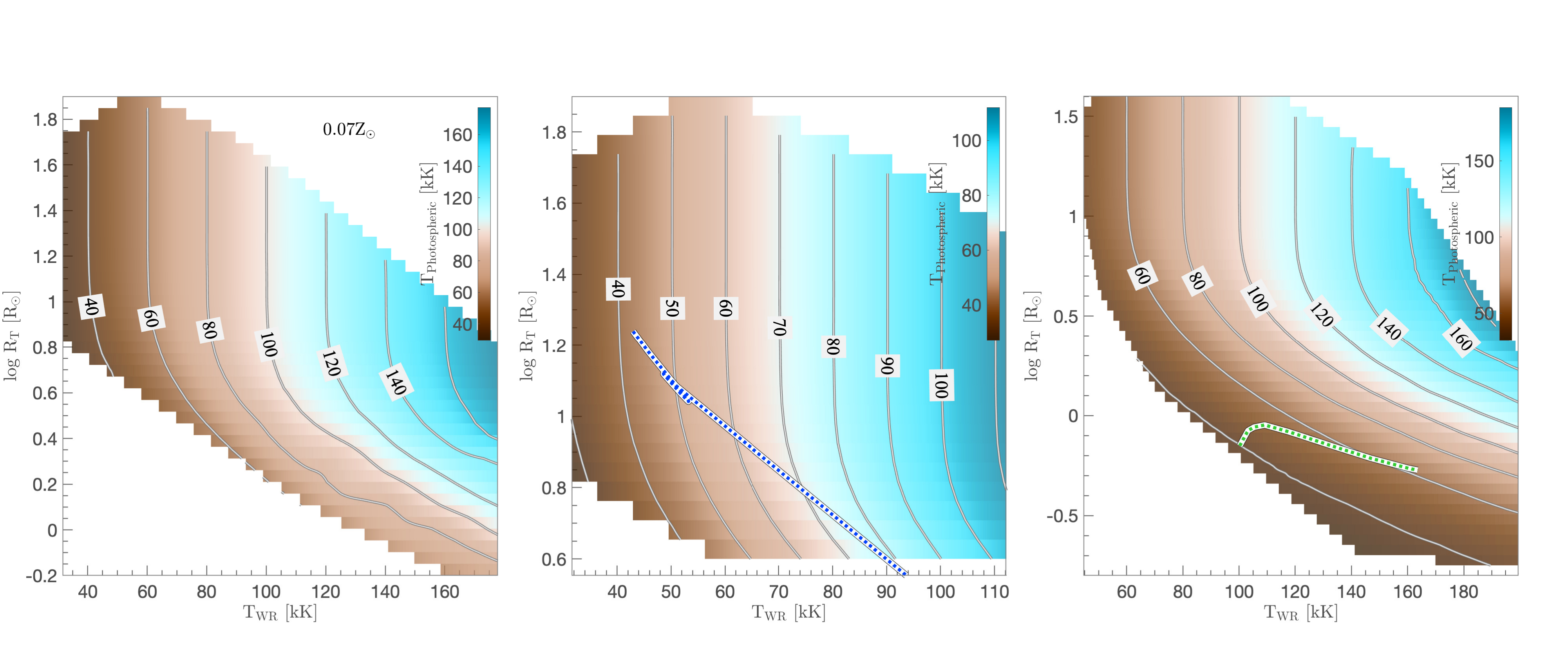}
\caption{(Continued) The selection of WR spectra as a function of T$_{\rm{WR}}$ [10$^3$ K] and R$_{\rm{T}}$ [R$_{\odot}$] for \textsc{Parsec} and \textsc{Geneva} stellar isochrones from PoWR library for the sub-SMC-like ($\sim$0.7 Z${\odot}$) stellar metallicity.}
\end{center}
\end{figure*}

To become a WR star, a massive star must meet specific criteria for initial mass, T$_{\rm{eff}}$, and surface abundances of hydrogen (H), carbon (C), nitrogen (N), and oxygen (O). The minimum mass threshold, M$_{\rm{min}}$, is metallicity-dependent, increasing as metallicity decreases, while T$_{\rm{eff}}$ and surface abundances are dictated by the isochrones. According to single-star evolutionary theories, a star with an initial mass of M$_{\rm{min}}$ of 20 M$_{\odot}$ at solar metallicity can become a late-type WN star if its surface hydrogen fraction drops below 0.4 and T$_{\rm{eff}}$ exceeds 25\,000 K. At the same metallicity, a higher M$_{\rm{min}}$ is required for the star to evolve into an early-type WN or a WC star \citep{Georgy2012, Georgy2015}.

While the observational criteria for identifying WR stars are well established \citep{Crowther2007}, classifying WR subtypes along a stellar evolutionary track of a given star is more challenging. There are several WR classification schemes exist in the literature, \citep[e.g.][]{Leitherer99, Georgy2012, Chen2015}. In this study, we adopt a scheme similar to  \cite{Georgy2012} and \cite{Leitherer99} - a star is considered capable of entering the late-type WN phase if its zero-age main sequence mass exceeds M$_{\rm{min}}$ for the given metallicity, its T$_{\rm{eff}}$ exceeds 25\,000 K, and its surface H-abundance falls below 0.4. A star with negligible surface H-mass fraction and a C-to-N abundance ratio less than 1 while maintaining the T$_{\rm{eff}}$ requirement is classified as an early-type WN star. If the C-abundance $\geq$ N-abundance, the star is classified as a WC or WO subtype. We further differentiate between WC and WO subtypes based on the surface abundance ratio of C+O to helium.

In Figure 1, we illustrate the evolution of WR spectra for a 120 M$_{\odot}$ star as it enters the WR phase, sampled in steps of 0.1 Myr. Each panel, from left-to-right, and top-to-bottom, shows the complete PoWR library grids for WNE, WNL-H20\footnote{For this study, we select the PoWR grid for the WR stars of the Nitrogen subclass with $20\%$ Hydrogen mass fraction, i.e.\,WNL-H20.}, and WC subtypes at solar, LMC, SMC, and sub-SMC metallicities, respectively. The colour coding and the contours represent the stellar photospheric temperatures at specified T$_{\rm{WR}}$ and R$_{\rm{T}}$.  
  
As discussed earlier, we apply a correction to the stellar effective temperatures of massive stars in the \textsc{Parsec} stellar isochrones during their WR phases. In Figure \ref{fig:linking_sst_with_powr}, the dashed lines illustrate how WR spectral selection varies depending on whether T$_{\rm{hyd}}$ (black-dashed), T$_{2/3}$ (red-dashed) or T$_{\rm{WR}}$ (blue-dashed) is used. 
The need for this temperature correction is especially evident in the selection of WNL stars, where T$_{\rm{hyd}}$ is overestimated for WNL WR-subtype at solar, LMC, and SMC metallicities (i.e.\,T$_{\rm{hyd}}$ extends to beyond the PoWR grids towards higher T$_{\rm{WR}}$) while T$_{2/3}$ tends to be underestimated. The weighted mean temperature correction generally aligns WR temperatures to be within the respective PoWR grids. The correction, however, remains inadequate for the SMC grid, where all selections extend beyond the PoWR grids at the lower temperature end. Note that the temperature corrections would be tapering off at the lower temperatures, as the primary goal is to address overly high temperatures.

Figure \ref{fig:linking_sst_with_powr} also shows the WNE (left column of panels), WC and WO (right column of panels) selections guided by the respective isochrones. Since the PoWR library lacks a separate WR spectral grid for WO subtypes, we substitute WC spectra in the population synthesis analysis. Notably, no WR evolution is observed at sub-SMC metallicities with \textsc{Parsec} isochrones, as they do not allow a 120 M$_{\odot}$ star to meet the conditions necessary to evolve into the WR phase.

In the same figure, we also present WR spectral selections based on \textsc{Geneva} high-mass loss isochrones (dotted lines). The \textsc{Geneva} isochrones already incorporate a temperature correction for WR stars, so we rely on their provided corrections to estimate the weighted mean temperatures shown in Figure \ref{fig:linking_sst_with_powr}. One key difference between \textsc{Parsec} and \textsc{Geneva} isochrones is that \textsc{Geneva} allows massive stars to achieve WR conditions at metallicities lower than that permitted by the \textsc{Parsec} isochrones (see the bottom panels of Figure\,\ref{fig:linking_sst_with_powr}). This discrepancy is mainly due to the higher mass-loss rates in the \textsc{Geneva} models. These higher mass-loss rates at given metallicity enable massive stars to enter the WR phase at earlier stages of evolution compared to \textsc{Parsec}. Consequently, at a given age, the higher mass-losses allow more low-mass stars (with $M>M_{\rm{min}}$) to enter the WR phase with \textsc{Geneva} than with \textsc{Parsec}. Overall, achieving WR star formation at sub-SMC and lower metallicities typically requires high mass-loss rates or binary interactions.

Finally, since the PoWR grids begin at a wavelength of $950$\AA, we supplement the spectra at shorter wavelengths by integrating them with the low-resolution CMFGEN library available in \textsc{Starburst99} \citep{Hillier1998}, selecting models based on the closest matching T$_{\rm{eff}}$.

\section{Photoionisation Modelling}\label{sec:nebular_models}
Predicting the ionising spectrum is crucial for modelling young, massive stellar populations in star-forming regions. These spectra are typically dominated by a combination of nebular continuum, strong emission lines, and broad emission features characteristic of WR stars during intense starbursts. Accurate modelling of these strong nebular emissions is essential when comparing SPS models to observations of young stellar populations. In contrast, nebular emissions can generally be disregarded when modelling older stellar populations.

In this analysis, we generate photoionisation models using the version 23.00 of the \textsc{Cloudy} code \citep{Chatzikos2023} in a self-consistent manner, using the latest \textsc{Starburst99} templates as input. For the gas-phase chemical composition, we adopt the 30-element abundance prescription from  \cite{Gutkin2016}, which primarily assumes solar-scaled abundances compiled by \cite{Bressan2012}, except for a few elements.
In this framework, the solar (photospheric) metallicity is Z$_{\odot}=0.01524$, consistent with the  \textsc{Parsecv1.S} isochrones. \cite{Gutkin2016} also fine-tuned the solar oxygen and nitrogen abundances from \cite{Bressan2012} to better match the observed properties of Sloan Digital Sky Survey galaxies in several optical diagnostic diagrams. For non-solar metallicities, the abundances of primary nucleosynthesis elements are assumed to scale linearly with the ISM metallicity.

Nitrogen, being both a primary\footnote{Nitrogen is synthesised as a primary element in CNO cycles during the Hydrogen-burning of stars in the $4\lesssim$ M/M$_{\odot} \lesssim 8$.} and secondary\footnote{Nitrogen is also synthesised as secondary nucleosynthesis element from CO products of previous generations of stellar populations \citep{Gutkin2016, Dopita2000}} nucleosynthesis element, requires special treatment. As a secondary nucleosynthesis element, its abundance is expected to scale with stellar metallicity \citep{Gutkin2016}. To relate the primary+secondary N-abundance to O, we use the relationship proposed by \cite{Groves2004} and adopted by \cite{Gutkin2016}. Additionally, we assume a solar C/O ratio of (C/O)$_{\odot}=0.44$.

To account for the depletion of refractory metals onto dust grains, we also adopt the ISM depletion factors from \cite{Gutkin2016}. The metals depleted from the gas phase contribute to grain formation, which significantly affects the scattering and absorption of incident radiation, radiation pressure, collisional cooling, and photoelectric heating of the gas. Depleting key cooling agents from the gas phase reduces gas cooling efficiency, thus increasing the electron temperature and, thereby, enhancing cooling through more energetic optical transitions \citep{Dopita2002, Groves2004, Gutkin2016}. The extent of metal depletion is characterised by the gas-to-metal mass fraction, $\xi_d$. Following \cite{Gutkin2016}, we adopt $\xi_d = 0.36$, indicating that $36\%$ of heavy elements by mass are locked in dust grains.

While $\xi_d$ is inherently metallicity-dependent, in this study, we assume a constant $\xi_d$ across all metallicities. Since we focus only on metallicities capable of producing WR stars, the uncertainties introduced by this assumption are relatively minor, and we discuss its implications for model predictions and comparisons in \S,\ref{subsec:photomodels_predictions}.

For constructing the \textsc{Cloudy} models of star-forming regions, we assume a simple spherical geometry. While not entirely realistic, the studies of star-forming regions dominated by young, massive stars and their birth clouds find a spherical approximation to be sufficient \citep[e.g.][]{Efstathiou2000, Siebenmorgen2007}. Additionally, we use emission line luminosity ratios throughout this study to minimise the impact of the underlying assumption of a spherical geometry.

\subsection{Grid of \textsc{Cloudy} photoionisation models}
\begin{figure*}
\begin{center}
\includegraphics[width=0.325\textwidth, trim={0.8cm 0.1cm 1.8cm 1.2cm},clip]{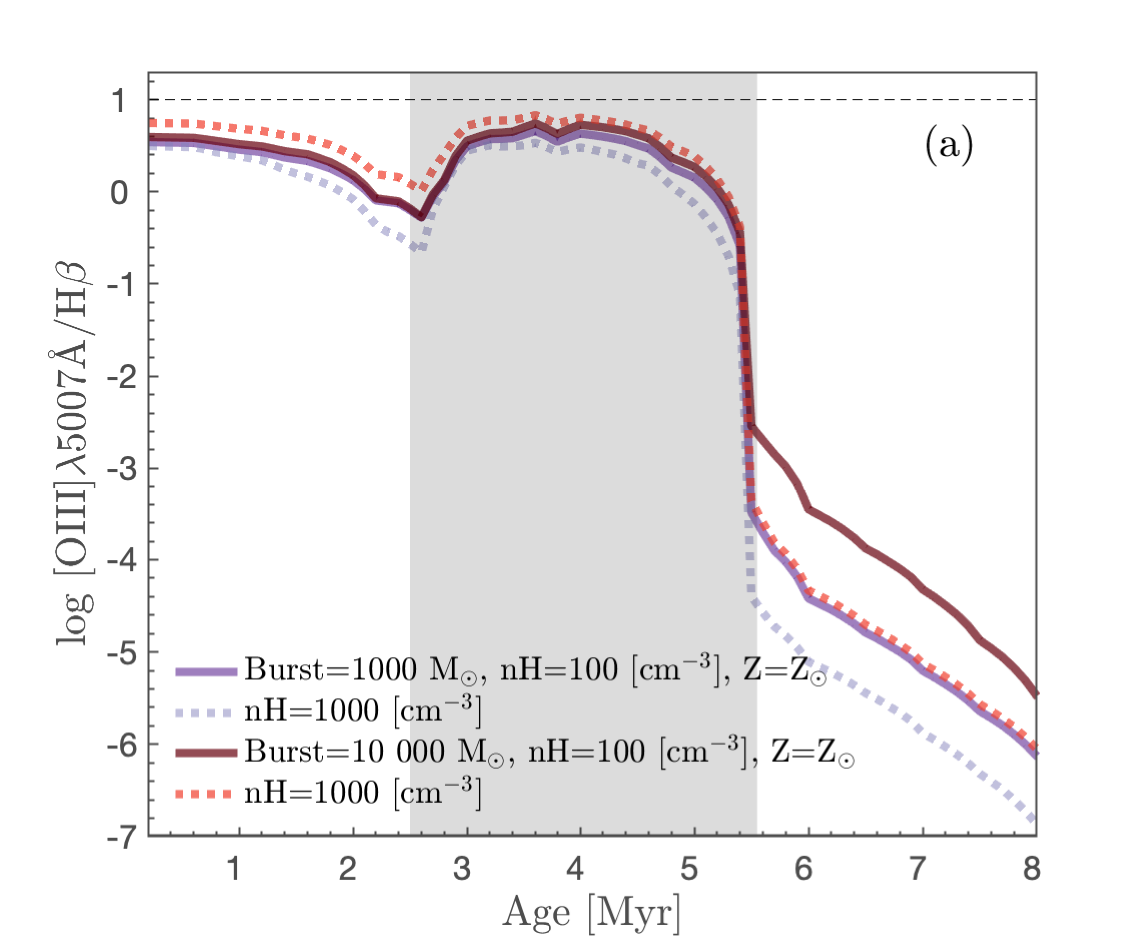}
\includegraphics[width=0.325\textwidth, trim={0.8cm 0.1cm 1.8cm 1.2cm},clip]{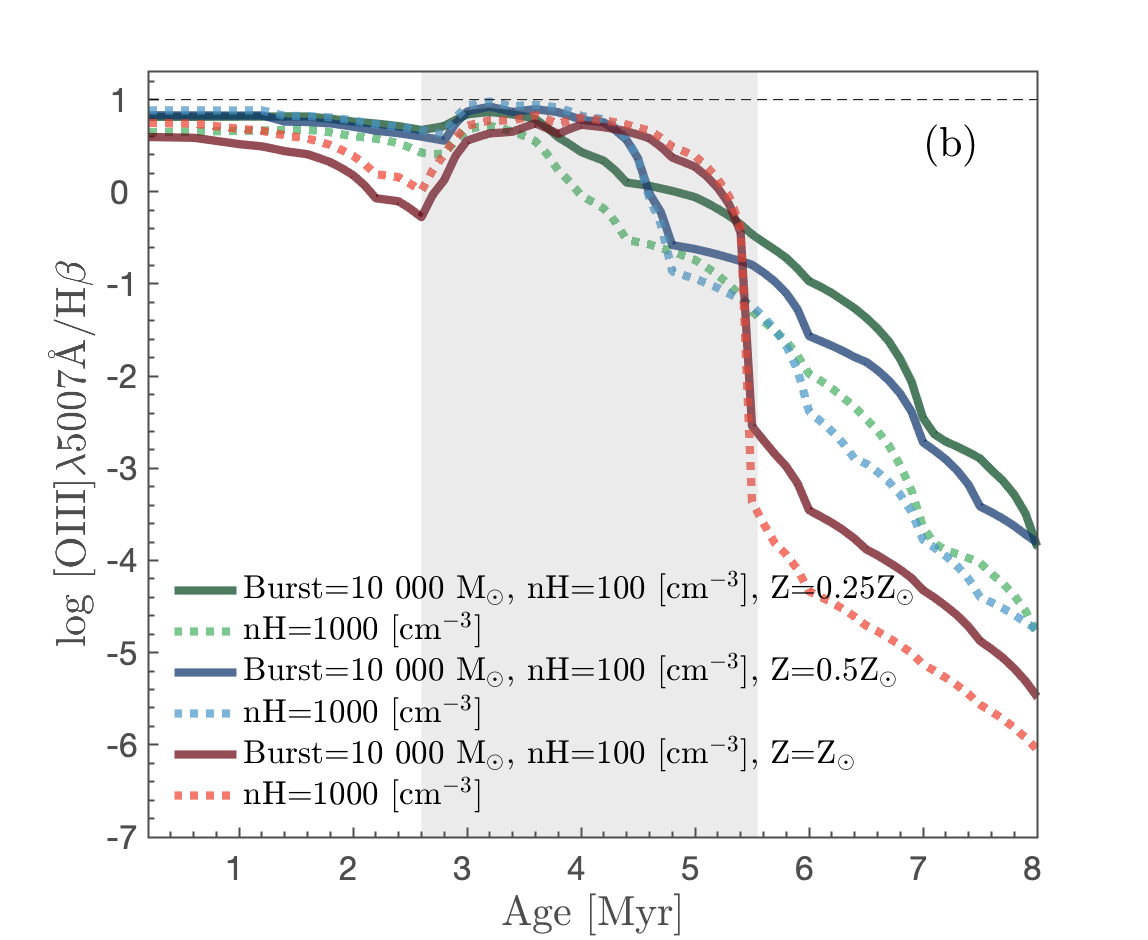}
\includegraphics[width=0.335\textwidth, trim={0.0cm 0.cm 0.cm 0.cm},clip]{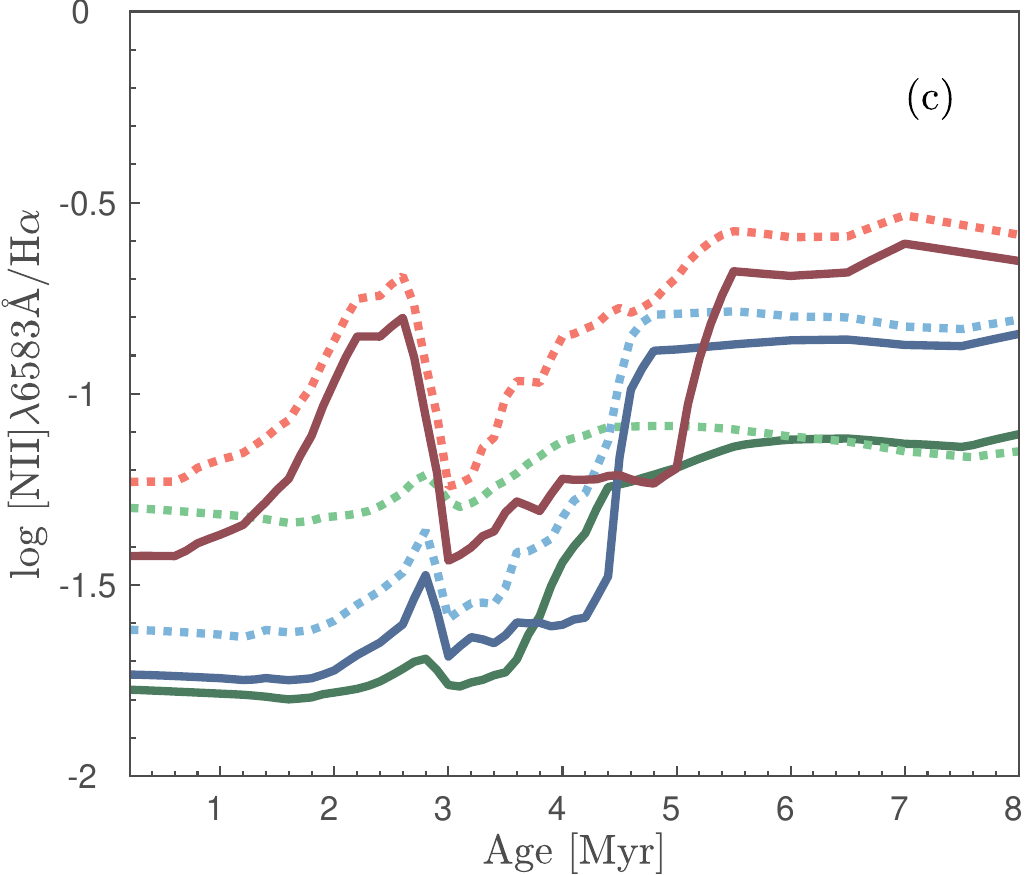}
\includegraphics[width=0.352\textwidth, trim={0.cm 0.cm 0.cm 0.cm},clip]{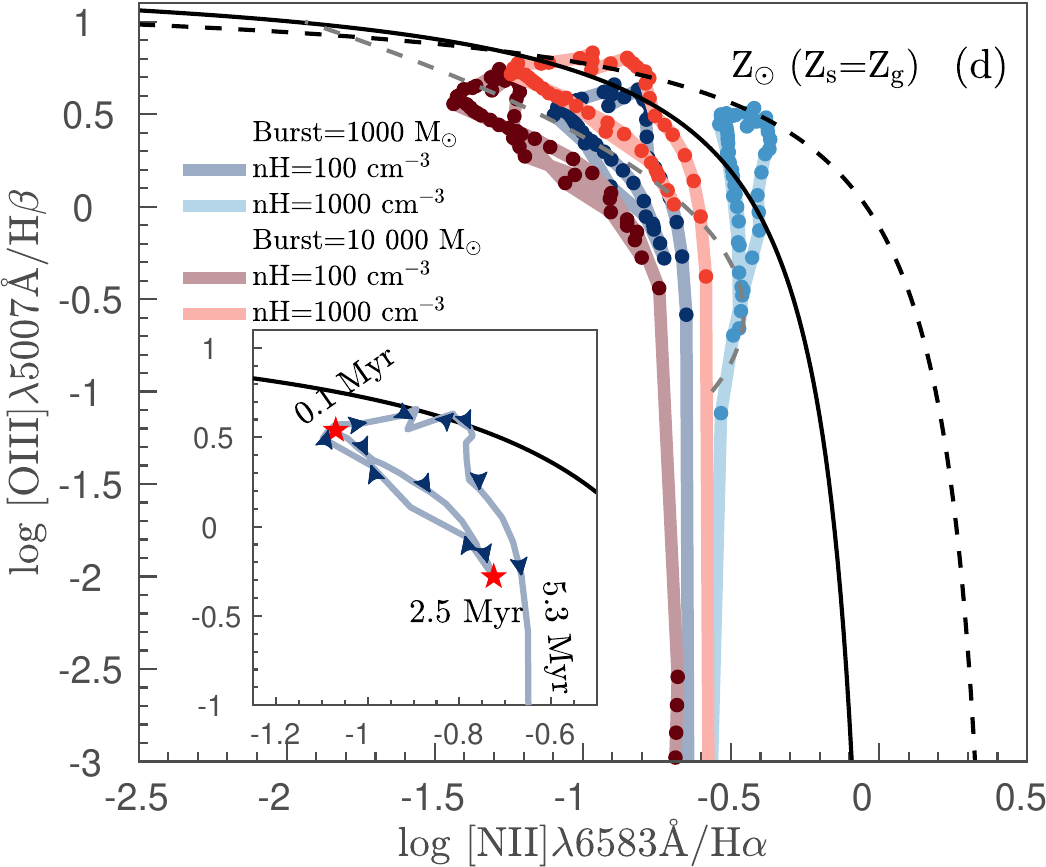}
\includegraphics[width=0.353\textwidth, trim={0.5cm 0.2cm 2.9cm 2.3cm},clip]{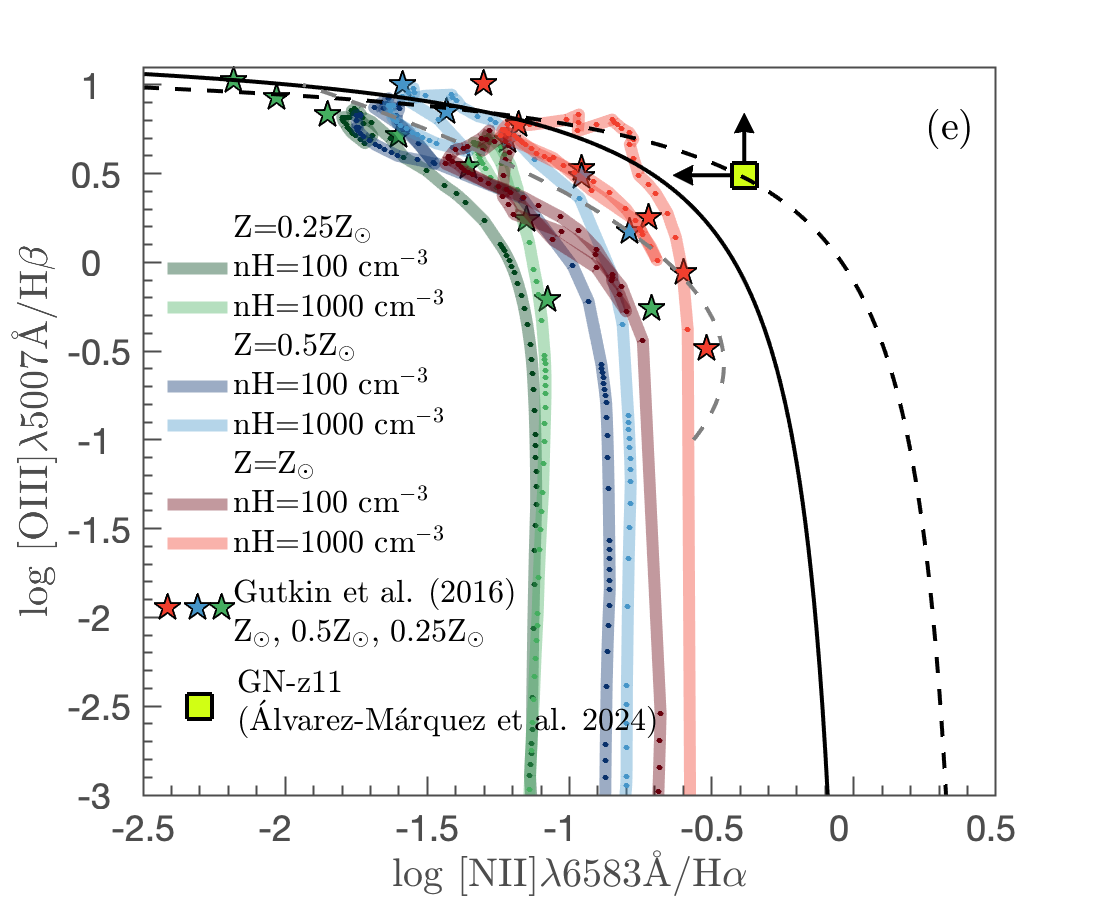} 
\caption{The behaviour of the predicted nebular lines used in BPT diagnostics \citep[][]{Baldwin1981}. (a) The evolution of [\ion{O}{iii}]~$\rm{\lambda}$5007\AA/H$\beta$ line luminosity ratio as a function of $n_H$ (solid and dotted lines) at solar metallicity (Z$_{\rm{s}}$=Z$_{\rm{g}}$) for bursts of star formation of strengths 1000 and 10\,000 $M_{\odot}$ for a model HII region of a fixed radius. (b) and (c) show the evolution of [\ion{O}{iii}]~$\rm{\lambda}$5007\,\AA/H$\beta$ and [\ion{N}{ii}]~$\rm{\lambda}$6584\,\AA/H$\alpha$ as a function of $n_H$ for SMC, LMC, and solar metallicities for a fixed starburst of 10\,000 M${\odot}$. The shaded regions in (a) and (b) show the timespan over which the WR phase is significant. The beginning and the duration of the WR phase of massive stars are largely functions of stellar metallicity.(d) demonstrates the predicted behaviour in the BPT diagram for starbursts of strengths 1000 and 10\,000 M$_{\odot}$ for a solar stellar and gas metallicity. For each burst, we plot two $n_H$ values (100 and 1000 cm$^{-3}$), shown as darker and lighter shades, respectively. The black dashed and solid lines show the \citet{Kewley2001} and \citet{Kauffmann2003} demarcations, respectively, and the dashed grey line denotes the fit to the SDSS data from \citet{Brinchmann2008}. In the inset, we illustrate the general evolutionary behaviour of the tracks for solar metallicity. The two red stars denote the first age point (0.1 Myr) of the track and the age at which the WR stars first appear (2.5 Myr), with the arrows tracing the evolution of the line luminosities in the BPT plane. Finally, (e) shows the evolution in the BPT plane for SMC-like, LMC-like, and solar-like metallicities for a fixed burst strength of 10\,000 M$_{\odot}$ and for two different $n_H$ values. For comparison, we overplot the \citet{Gutkin2016} models for the same metallicities as stars. Also, overplotted in lime green is the JWST-MIRI spectroscopy-based optical line ratios for GN-z11 from \citet{Alvarez2024}, with arrows denoting the upper limits.}
\label{fig:diagnostics_vs_obs}
\end{center}
\end{figure*}
This section outlines the development of a comprehensive suite of photoionisation models designed to investigate the ultraviolet emission line properties of star-forming galaxies.

In \textsc{Cloudy} modelling, the ionisation parameter, $U$, determines the intensity of the ionising continuum, and depends on the number of ionising photons ($Q_H$), which characterises the strength of a starburst, and Hydrogen density ($n_H$). $U$ is defined as,
\begin{equation}
U = \frac{Q_H}{4\pi R^2\times n_H\times c}, 
\label{eq:ionisation_relation}
\end{equation}
where $c$ is the speed of light and $R$ is the radius of the ionised region, assumed to be fixed at 3pc. While we fix the radius for our model HII regions, the radius can be varied posteriori, permitting starbursts of any magnitude to be generated provided the $U$ range probed is within the range of input to \textsc{Cloudy}.

Table\,\ref{tab:model_libraries} summarises the sampling of key model parameters considered in this study. For a single stellar generation, most ionising photons are emitted within the first 10 Myr, during which the effects of WR stars are also most pronounced. Therefore, we sample younger ages ($<10$ Myr) at a finer time resolution of 0.1 Myr, while the gas density $n_H$ is sampled at a coarser resolution to establish lower and upper bounds for the model grid. Throughout this analysis, we assume that gas-phase metallicity (Z$_{\rm{ISM}}$) matches the stellar metallicity.
\begin{table}
\caption{Grid sampling of the main parameters of the photoionisation model library}
\begin{threeparttable}[h]
\centering
\begin{tabular}{l|c|}
\hline
 Parameter						& Sampled range					\\
\hline
Ages [Myr] 						& 0.2 Myr $-$ 500 Myr\tnote{a}\, (94 different values)	         \\
Stellar metallicity [$Z_{\rm{s}}$] 		&  0.001 $-$  0.04\tnote{b}\,  (5 different values)         \\
IMF upper mass cut [M$_{\odot}$]             & 120   							 \\
Gas metallicity [$Z_{\rm{g}}$] 		        &  0.001 $-$  0.04\tnote{c}\,  (5 different values)   \\
$n_{\rm{H}}$ [cm$^{-3}$]				&  100, 500, 1000\tnote{c}  				   \\			
Burst strengths [log M$_{\odot}$]  		& 2 $-$  5\tnote{d} 	  \\
\hline
\end{tabular}
\begin{tablenotes}
     \item[a] {The ages are variably sampled.  In the models, the young ages (i.e.\,$<10$ Myr) are sampled at 0.1 Myr steps in order to fully capture the evolutionary effects of massive stars, particularly the effects of the crucial short-lived WR phase. The ages in the range $\sim10-100$ Myr sampled at steps of 10 Myrs, and the $100-500$ Myr in steps of 100 Myr.}
     \item[b] {The present day (photospheric) solar metallicity is assumed to be $Z_{\odot}=0.0152$, matching with \textsc{Parsec} isochrones and \cite{Gutkin2016} abundances. Note that the solar metallicity assumed for the \textsc{Geneva} isochrones is 0.02.}
     \item[c] {We assume $Z_{\rm{ISM}}$=$Z_{s}$ (i.e.\,gas-phase metallicities are assumed to be equal to stellar metallicity).}
     \item[d] {Any range of burst strengths can be assumed provided the corresponding evolution of $\log U$ is in the $-0.5\lesssim U \lesssim 4$ range. We assume a fixed radius for our model HII regions. For an HII region of a different radius, these strengths need to be scaled following Eq.\,\ref{eq:ionisation_relation}. }  
\end{tablenotes}
    \end{threeparttable}%
\label{tab:model_libraries}
\end{table}%

The most common diagnostics for characterising the nature of nebular excitation are the BPT diagnostics \citep{Baldwin1981}. As a preliminary check, before examining the ultraviolet emission line properties, we assess the predicted behaviour of [\ion{O}{iii}]~$\rm{\lambda}$5007\,\AA/H$\beta$ and [\ion{N}{ii}]~$\rm{\lambda}$6584\,\AA/H$\alpha$ from our new model suite in Figure\,\ref{fig:diagnostics_vs_obs}. 

Figure\,\ref{fig:diagnostics_vs_obs}a illustrates the evolution of the [\ion{O}{iii}]~$\rm{\lambda}$5007\,\AA/H$\beta$ ratio for starbursts of 1000 M$_{\odot}$ and 10\,000 M$_{\odot}$, with $n_H=100$ and 1000 [cm$^{-3}$], assuming $Z_{\rm{ISM}}=Z=Z_{\odot}$. The grey shading indicates the period during which massive stars evolve into WR stars in a single stellar generation. As stars enter the WR phase, the ionising continuum intensifies, leading to a sharp increase in the [\ion{O}{iii}]~$\rm{\lambda}$5007\,\AA/H$\beta$ ratio. Conversely, this ratio rapidly drops as massive stars exit the WR phase. The impact of $n_H$ and starburst strength is minimal at earlier ages ($<6$ Myr), while significant variations in the [\ion{O}{iii}]~$\rm{\lambda}$5007\,\AA/H$\beta$ ratio are observed at later ages ($>6$ Myr), although these ratios are too low to be detectable in H II regions.

Figure\,\ref{fig:diagnostics_vs_obs}b and c illustrate the evolution of [\ion{O}{iii}]~$\rm{\lambda}$5007\AA/H$\beta$ and [\ion{N}{ii}]~$\rm{\lambda}$6584\AA/H$\alpha$ for SMC, LMC, and solar metallicities, assuming a starburst of strength $10\,000$ M$_{\odot}$ and $n_H=100$ and $1000$ [cm$^{-3}$]. 
The shaded region, the same as in Figure\,\ref{fig:diagnostics_vs_obs}a, represents the ages where WR star effects become significant at for a $Z_{\rm{ISM}}=Z=Z_{\odot}$. As discussed in \S\,\ref{subsec:wr_stars}, both the onset and duration of the WR phase depend on mass-loss rates and metallicity. The M$_{\rm{min}}$ threshold for WR formation is lower at higher metallicities, allowing more low-mass stars to enter the WR phase than at low metallicities, thereby extending the duration of the influence of WR stars on the stellar population's evolution.

The impact of the metallicity dependence of ionising photon from WR stars on BPT diagnostics is evident in panels (b) and (c), where the peak in [\ion{O}{iii}]~$\rm{\lambda}$5007\AA/H$\beta$ (and the decline in [\ion{N}{ii}]~$\rm{\lambda}$6584\AA/H$\alpha$) occur at younger ages for higher metallicity models compared to sub-solar models, reflecting the longer WR durations at higher metallicities. In contrast, the WR effects in sub-solar models are less pronounced and taper off more gradually due to the higher M$_{\rm{min}}$ threshold, resulting in fewer massive stars entering the WR phase.
Finally, increasing $n_H$ enhances both [\ion{O}{iii}]~$\rm{\lambda}$5007\AA/H$\beta$ and [\ion{N}{ii}]~$\rm{\lambda}$6584\AA/H$\alpha$ due to a higher rate of collisional excitation.

Figure\,\ref{fig:diagnostics_vs_obs}d and e present the BPT diagnostic plots. Like Figure\,\ref{fig:diagnostics_vs_obs}a, Figure\,\ref{fig:diagnostics_vs_obs}d focuses exclusively on solar metallicity models, allowing us to distinguish the effects of varying burst strengths and $n_H$ values at a fixed radius. The blue and brown solid lines represent the evolution of BPT diagnostics for starbursts of 1000 and 10\,000 M$_{\odot}$, respectively, with the darker and lighter shades of each colour denoting $n_H=100$ and 1000 [cm$^{-3}$], respectively. The black solid and dashed lines denote the demarcations from \citet{Kewley2001} and \citet{Kauffmann2003}, separating Active Galactic Nuclei (AGNs) from star-forming galaxies and composites, and pure star-forming systems, respectively. The dashed grey line represents the best fit to the low-redshift SDSS data from \citet{Brinchmann2008}.

As the starburst strength is defined by $Q_H$ (Eq.\,\ref{eq:ionisation_relation}), an increase in $Q_H$ will, in turn, increase $U$, extending the evolutionary tracks towards higher [\ion{O}{iii}]~$\rm{\lambda}$5007\,\AA/H$\beta$ and lower [\ion{N}{ii}]~$\rm{\lambda}$6584\AA/H$\alpha$ regime in the BPT plane. Conversely, for a given $Q_H$, an increase in $n_H$ will lower $U$, thus shifting the evolutionary tracks towards higher [\ion{N}{ii}]~$\rm{\lambda}$6584\AA/H$\alpha$. 

To illustrate the effects of WR stars on the evolution of BPT diagnostics, a zoomed-in view of the solar model is shown in the inset of Figure\,\ref{fig:diagnostics_vs_obs}d. The red stars mark the start of the track and the age at which massive stars begin entering the WR phase, while the embedded arrows indicate the direction of evolution. Up to $\sim2.5$ Myr, the solar model evolves as expected, with the [\ion{O}{iii}]~$\rm{\lambda}$5007\,\AA/H$\beta$ gradually declining due to the aging of young, ionising stars. At $\sim2.5$ Myr, however, the most massive stars start transitioning into the WR phase, and within $\lesssim0.5$ Myr, the hot, exposed cores of WR stars significantly boost the influx of ionising photons, increasing $U$ and inducing a rapid jump in [\ion{O}{iii}]~$\rm{\lambda}$5007\,\AA/H$\beta$. 

As the WR phase progresses, massive stars, from the most massive (i.e.\,120 M$_{\odot}$) down to M$_{\rm{min}}$, gradually enter the WR phase. After the initial sharp increase in [\ion{O}{iii}]~$\rm{\lambda}$5007\,\AA/H$\beta$, the ratio continues to rise, albeit at a slower pace, as additional stars evolve into WR stars. Similarly,  [\ion{N}{ii}]~$\rm{\lambda}$6584\AA/H$\alpha$ ratio also shows an increase. The decline in [\ion{O}{iii}]~$\rm{\lambda}$5007\,\AA/H$\beta$ begins as massive stars start exiting the WR phase. As stars from 120 M$_{\odot}$ to M$_{\rm{min}}$ transition out of the WR phase, the [\ion{O}{iii}]~$\rm{\lambda}$5007\,\AA/H$\beta$ ratio shows an initial gradual decline, followed by a sharp drop, coinciding with the exit of all $>$M$_{\rm{min}}$ stars from the WR phase.  

Figure\,\ref{fig:diagnostics_vs_obs}e presents the BPT diagrams for SMC (green), LMC (blue), and solar (red) models, assuming a burst strength of $10\,000$ M$_{\odot}$, and $n_H=100$ and 1000 [cm$^{-3}$]. Notably, the enhanced influx of ionising photons can propel the predicted line ratios into the composite (SF+AGN) regime. In higher metallicity models, the line ratios may even extend into the AGN region. We also overplot the \cite{Gutkin2016} models, spanning a similar metallicity range, in Figure\,\ref{fig:diagnostics_vs_obs}e to illustrate that our models are consistent with the literature. 

Also overplotted in Figure\,\ref{fig:diagnostics_vs_obs}e is the GN-z11 BPT measurement based on JWST/MIRI medium-resolution spectroscopy \citep{Alvarez2024}, with arrows indicating the upper limits of the measurements. The optical line ratios place GN-z11 within the composite region, suggesting contributions from both starburst activity and an AGN. According to \cite{Alvarez2024}, this is consistent with the properties of local low-metallicity starbursts and high-$z$ luminous galaxies detected at similar redshifts to GN-z11. Notably, our starburst models exhibit a systematic shift towards the composite region as the electron density increases, particularly over the evolutionary phases where Wolf-Rayet (WR) stars dominate. This trend suggests that our models will naturally encompass the observed position of GN-z11 on the BPT diagram at higher densities.

These excursions of the models shown in Figure\,\ref{fig:diagnostics_vs_obs}d and e into the AGN/composite regime is primarily driven by the presence of the WR stars. In fact, a significant portion of the WR stellar phase of a stellar population is spent trekking towards and residing within this regime. This is particularly evident in solar models, where the WR effects are more pronounced.

\subsubsection{On the excursions in the BPT plane - starburst strengths and durations based on sound-crossing time}\label{subsubsec:soundcrossing}
To better understand the temporal nature of these excursions in the BPT plane, we analysis the likely sound-crossing times for different starbursts. 

\begin{figure}
\begin{center}
\includegraphics[width=0.48\textwidth, trim={.0cm 0.0cm 0.0cm 0.0cm},clip]{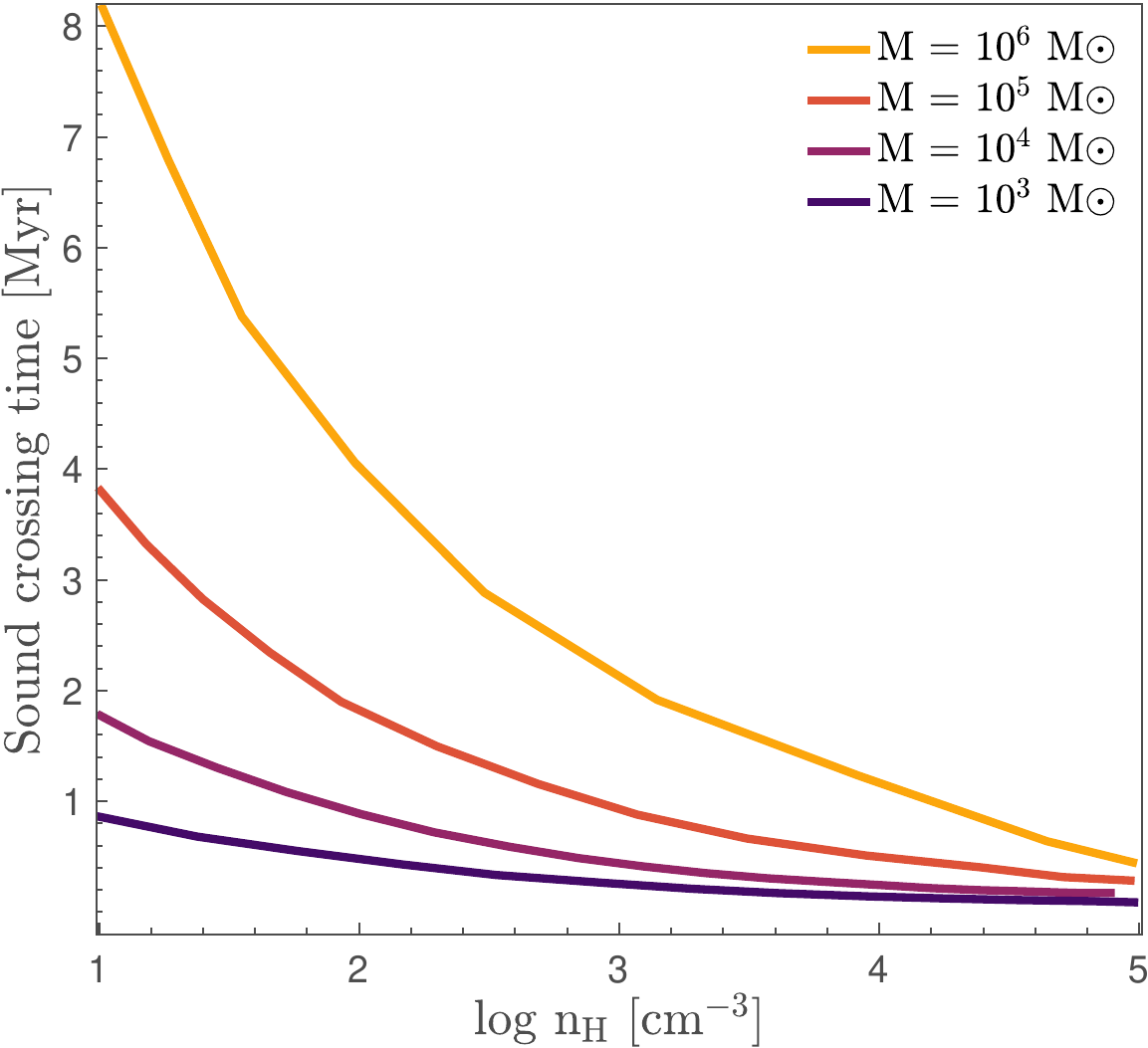}
\caption{The sound-crossing time as a function of n$_{H}$ for starbursts of different strengths.}
\label{fig:soundCrossing_time}
\end{center}
\end{figure}
Star formation within a gaseous cloud can be triggered and propagated through various mechanisms. In this analysis, we consider a simplified model in which an external or internal perturbation propagates through the cloud, triggering star formation. The characteristic velocity at which information travels through the cloud is approximately set by the sound speed in the medium.

\begin{figure}
\begin{center}
\includegraphics[width=0.48\textwidth, trim={.0cm 0.0cm 0.0cm 0.0cm},clip]{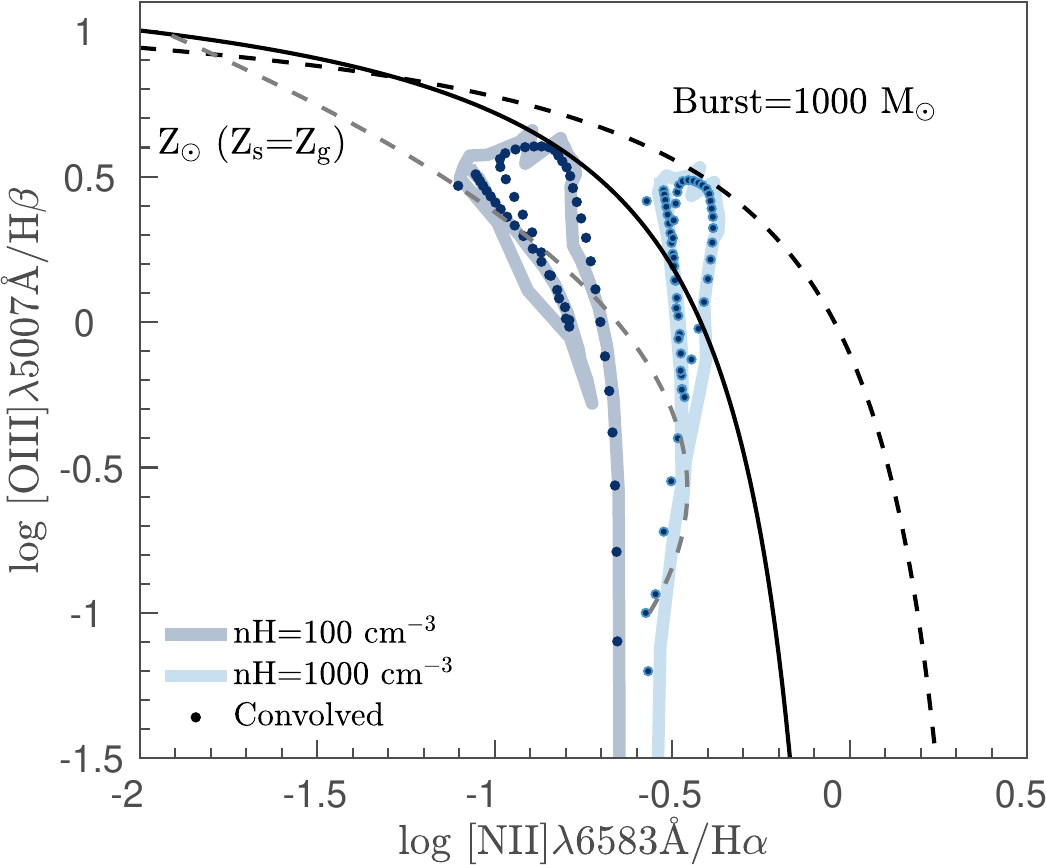}
\caption{The model predictions for a starburst of 1000 M${\odot}$ at solar metallicity, with n${\rm{H}}$ values of 100 and 1000 cm$^{-3}$ (solid lines, same as shown in Figure\,\ref{fig:diagnostics_vs_obs}), are compared with the same models convolved with a 1 Myr tophat function (filled-in circles of the same colour). The convolution with the 1 Myr tophat function demonstrates that the features in the models persist over longer timescales.}
\label{fig:tphat_convolve}
\end{center}
\end{figure}
Assuming a cloud of roughly constant density, we can estimate its characteristic size based on its total mass. Dividing this size by the sound speed provides an approximate timescale over which perturbations can traverse the cloud, triggering star formation.

Figure\,\ref{fig:soundCrossing_time} illustrates the sound crossing time as a function of n$_H$ for different burst strengths, assuming a temperature of $T = 10^4$ K and a pure hydrogen composition. This simple analysis suggests that for starbursts with an average density of n$_H = 10^3$ cm$^{-3}$, a typical duration of $\sim$1 Myr is a reasonable expectation. Several studies \citep[e.g.][]{Senchyna2024, Maiolino2024}, however, highlight that the densities in GN-z11 can be higher than 10$^3$ cm$^{-3}$. As shown in Figure,\ref{fig:soundCrossing_time}, starburst durations of $\lesssim$1 Myr can be expected in such cases.

Assuming a starburst duration of $\sim 1$ Myr, Figure\,\ref{fig:tphat_convolve} presents the convolution of the two n$_{\rm{H}}$ models for a starburst of 1000 M$_{\odot}$ (indicated as solid lines and also shown in Figure\,\ref{fig:diagnostics_vs_obs}d) with a tophat function of 1 Myr duration. Given that GN-z11 is likely experiencing an extreme starburst, a 1 Myr smoothing timescale is a realistic approximation of its starburst activity. The convolved models highlight that these excursions are not brief fluctuations but persist over sufficiently long timescales to be observable in extreme starbursting environments. 

Notably, starburst durations of $\lesssim$1 Myr, corresponding to n$_H > 10^{3}$ cm$^{-3}$ in GN-z11, as suggested by \cite{Senchyna2024, Maiolino2024}, would have even less minimal impact on the smoothing out fluctuations driven by WR populations observed in Figure,\ref{fig:diagnostics_vs_obs}d.

Furthermore, Figure\,\ref{fig:diagnostics_vs_obs}b and c demonstrate that the interval between the emergence of the first WR stars and the disappearance of the last WR stars spans several million years across the metallicity range studied, meaning that starbursting regions hosting WR stars can have emission line ratio representative of AGN and composites.

In the following sections, we focus on model predictions in the ultraviolet regime for Z$\lesssim$Z$_{\odot}$, which is the primary focus of our study.

\subsection{Ultraviolet emission line properties}\label{subsec:uv_line_props}
\begin{figure}
\begin{center}
\includegraphics[width=0.4\textwidth, trim={0.cm 0.8cm 0.05cm 0.cm},clip]{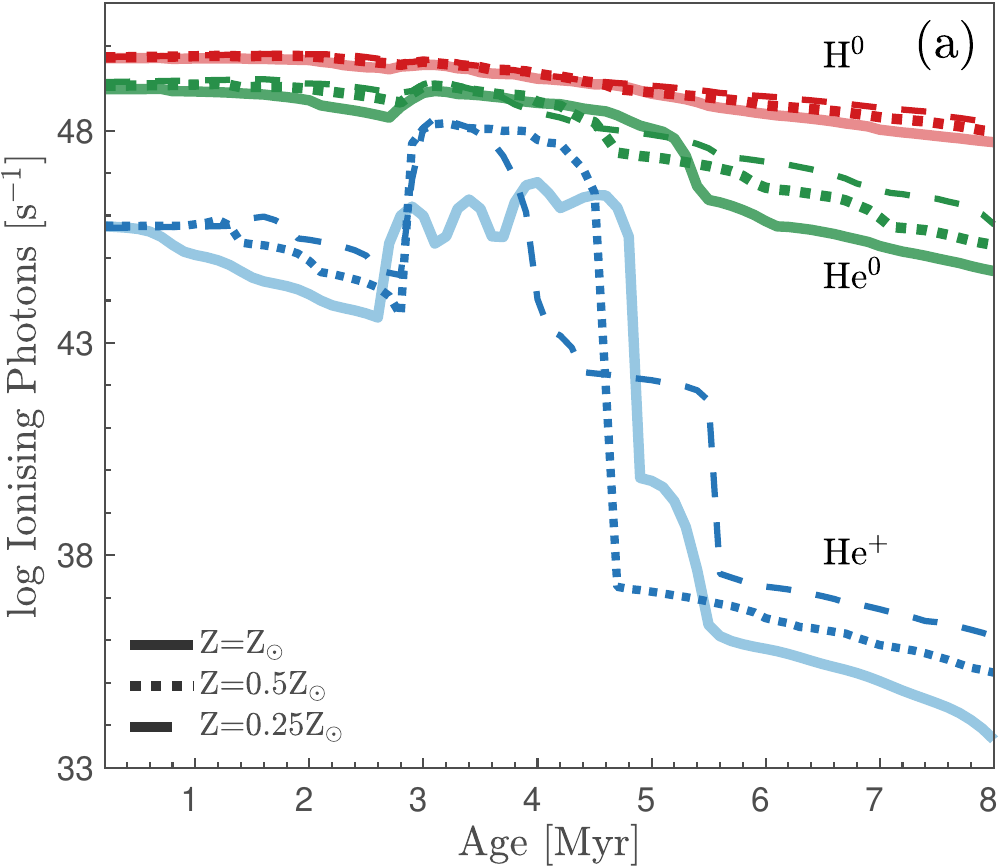}
\includegraphics[width=0.41\textwidth, trim={0.cm 0.9cm 0.cm 0.cm},clip]{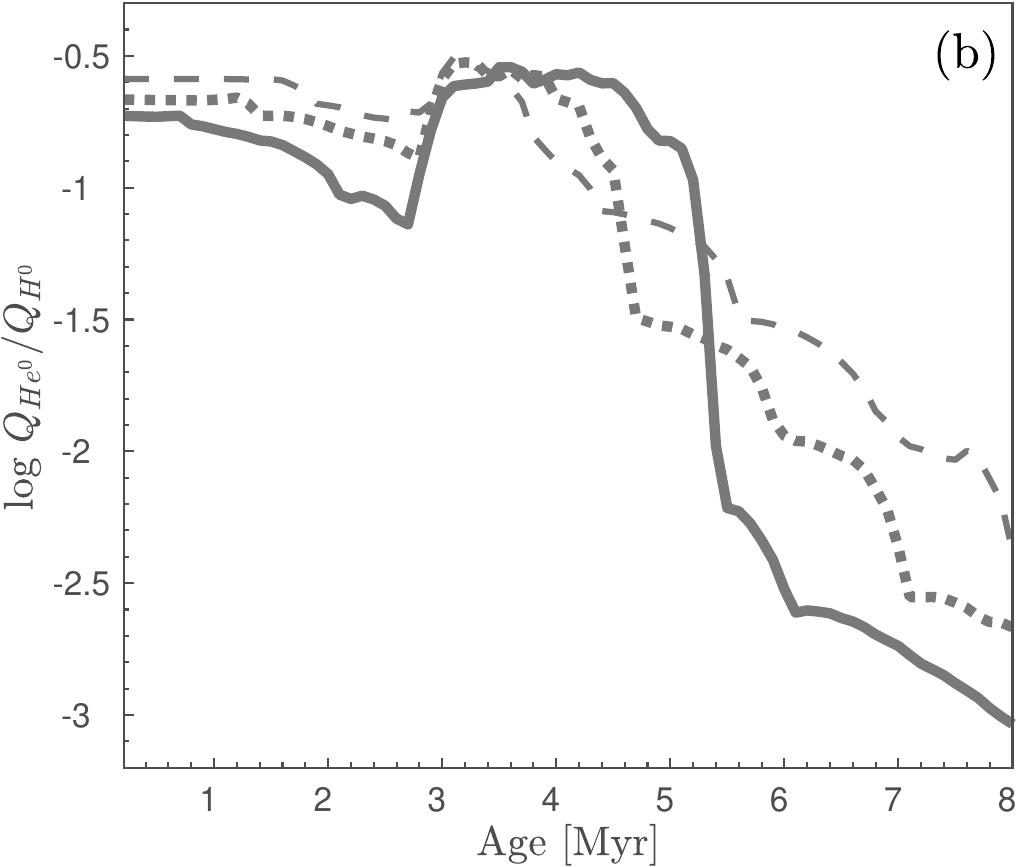}
\includegraphics[width=0.41\textwidth, trim={0.cm 0.cm 0.cm 0.cm},clip]{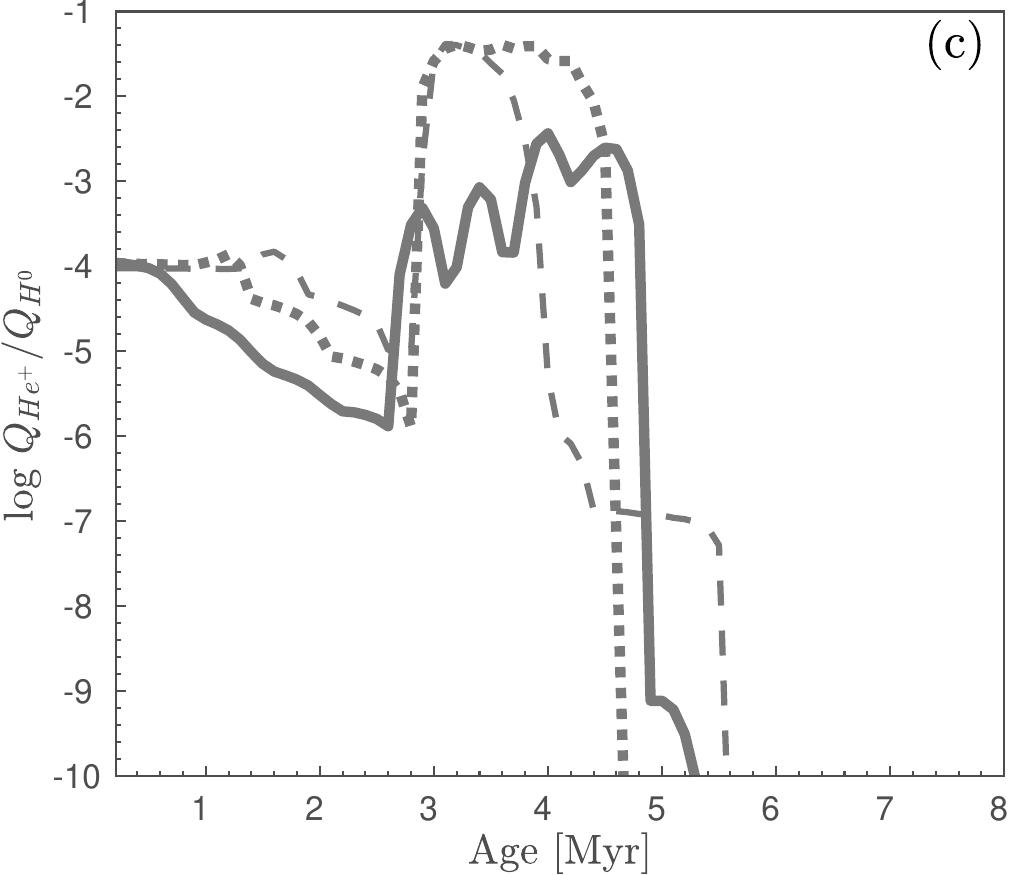}
\caption{The evolution in ionising luminosity and hardness of ionising spectra. (a) The evolution of the number of ionising photons ($Q$) for ages $\lesssim10$ Myr in the ionising continua of H$^0$ ($\lesssim912$\AA), He$^0$ ($\lesssim504$\AA), He$^+$ ($\lesssim228$\AA) at $Z_{\odot}$ (solid) and $0.5Z_{\odot}$ (dotted) for a starburst of 1000 M$_{\odot}$ at time steps of 0.1 Myr. In (b) and (c), the evolution of the hardness of ionising spectra parameterised as $\log Q_{\rm{He^0}}/Q_{\rm{H^0}}$ and $\log Q_{\rm{He^+}}/Q_{\rm{H^0}}$, respectively.}
\label{fig:evolution_of_ionisingFlux}
\end{center}
\end{figure}
\begin{figure}
\begin{center}
\includegraphics[width=0.45\textwidth, trim={0.cm 0.cm 0.05cm 0.cm},clip]{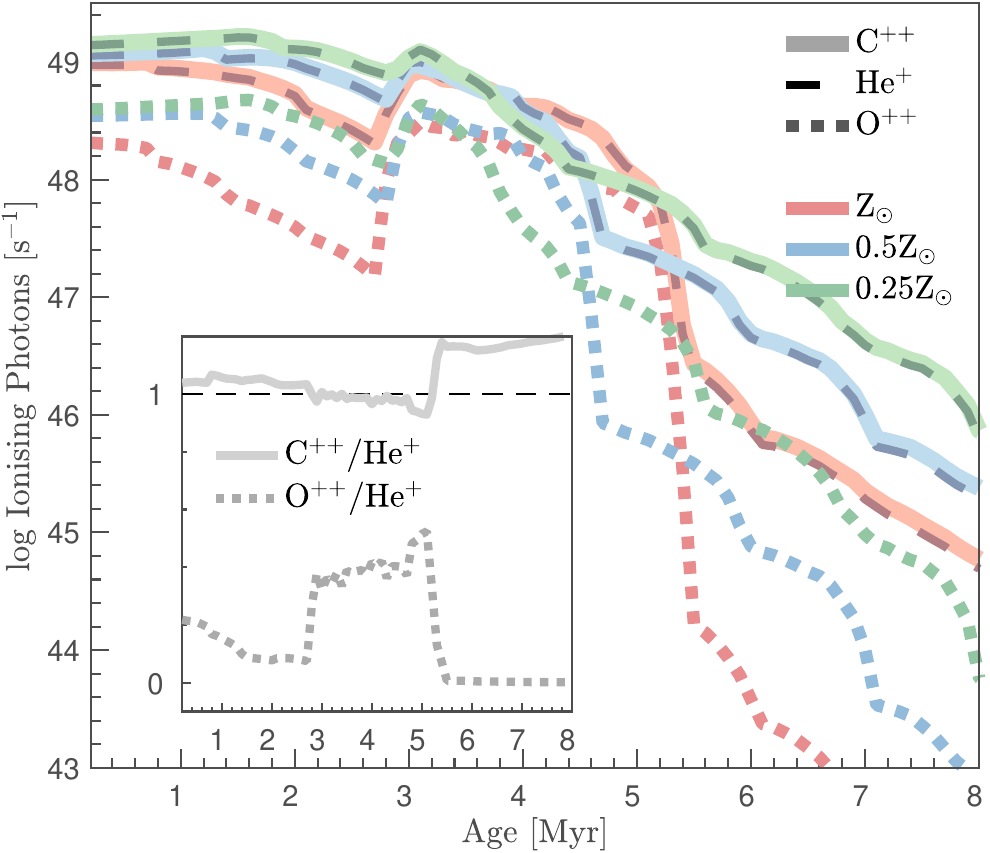}
\caption{The evolution of the number of ionising photons ($Q$) for ages $\lesssim10$ Myr in the ionising continua of C$^{++}$ (i.e.\,$258.9\lesssim\lambda$ [\AA]$\lesssim508.5$), He$^{+}$ (i.e.\,$504\lesssim\lambda$ [\AA]$\lesssim228$) and O$^{++}$ (i.e.\,$225.7\lesssim\lambda$ [\AA]$\lesssim353$) species at $Z=Z_{\odot}$ (red), $0.5Z_{\odot}$ (blue) and $0.5Z_{\odot}$ (green) for a starburst of 1000 M$_{\odot}$ at time steps of 0.1 Myr. The inset presents the evolution of the ratio of C$^{++}$ to He$^+$ (solid line) and O$^{++}$ to He$^+$ (dotted lines), emphasising, in particular, the differences in evolution between C$^{++}$ and He$^+$, which appears as overlapping in the main plot.}
\label{fig:evolution_of_ionisingFlux2}
\end{center}
\end{figure}
\begin{figure*}
\begin{center}
\includegraphics[width=0.335\textwidth, trim={0.0cm 0.cm 0.cm 0.cm},clip]{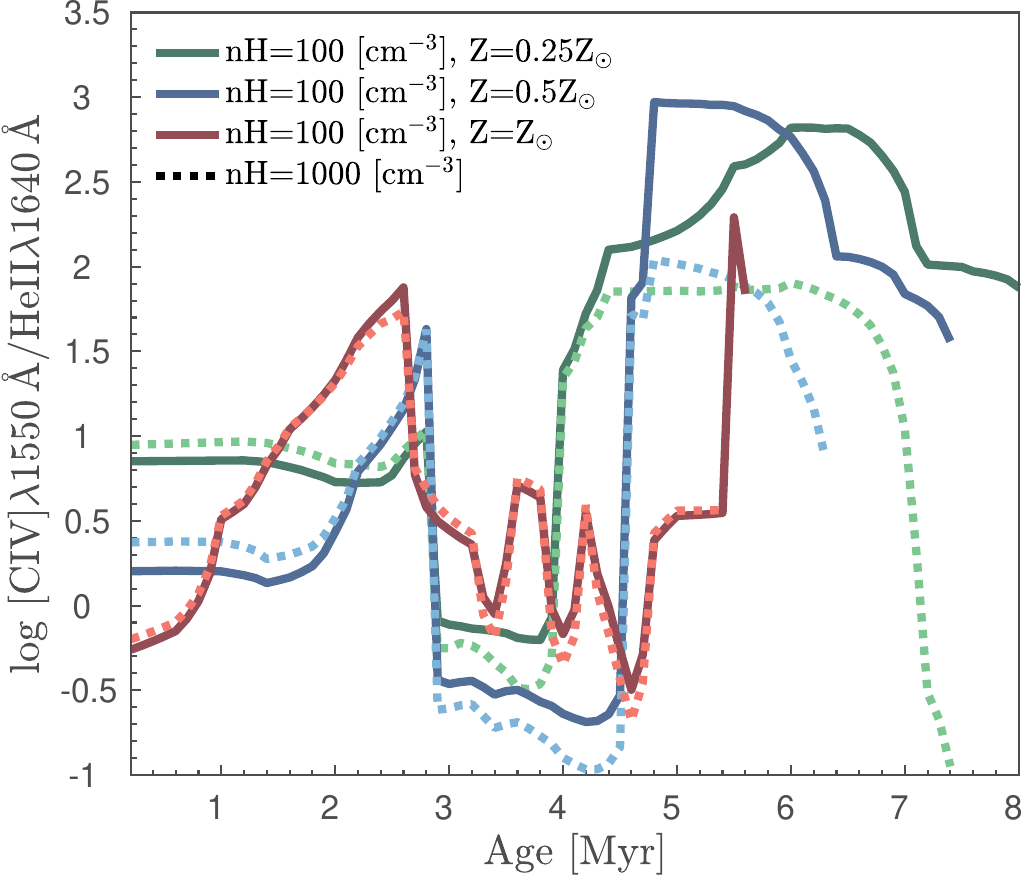}
\includegraphics[width=0.325\textwidth, trim={0.cm 0.cm 0.cm 0.cm},clip]{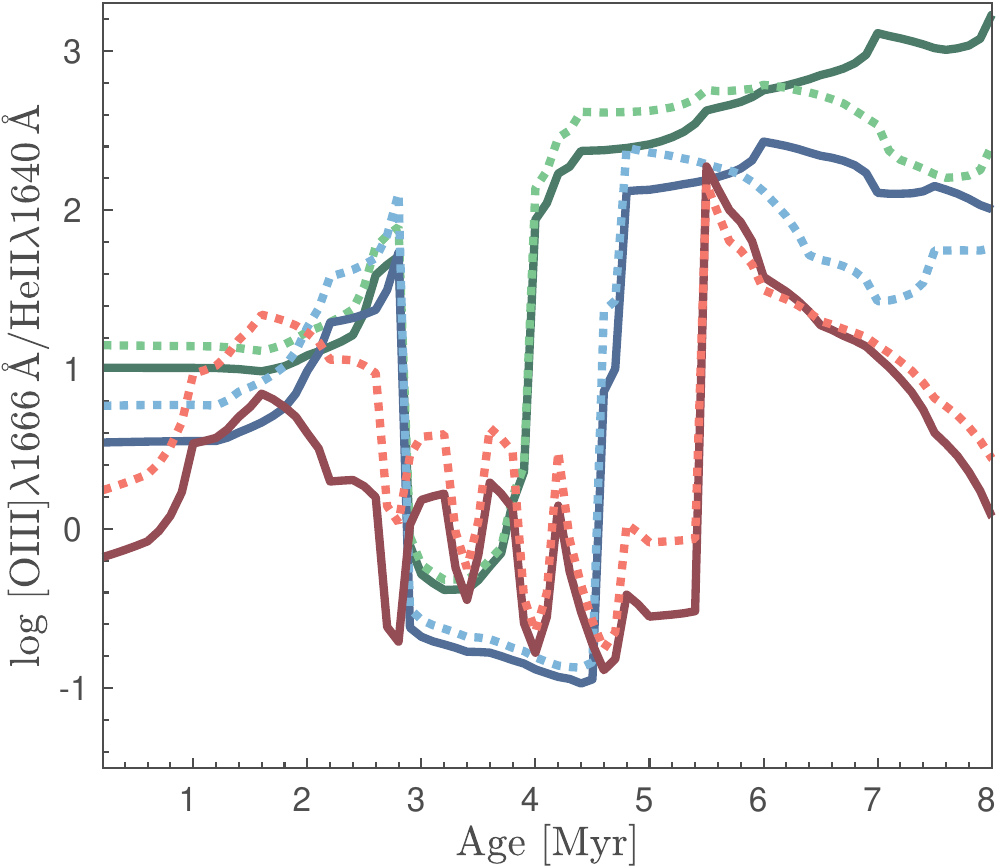}
\includegraphics[width=0.325\textwidth, trim={0.cm 0.cm 0.cm 0.cm},clip]{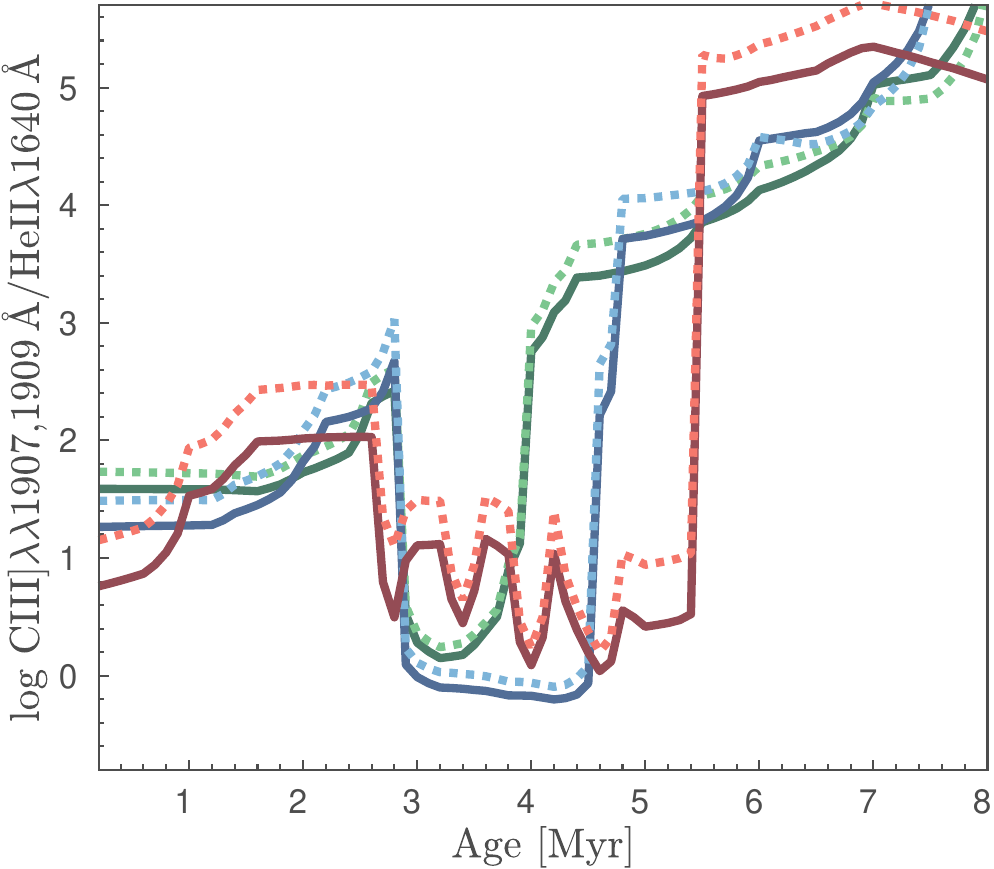}
\includegraphics[width=0.325\textwidth, trim={0.cm 0.cm 0.cm 0.cm},clip]{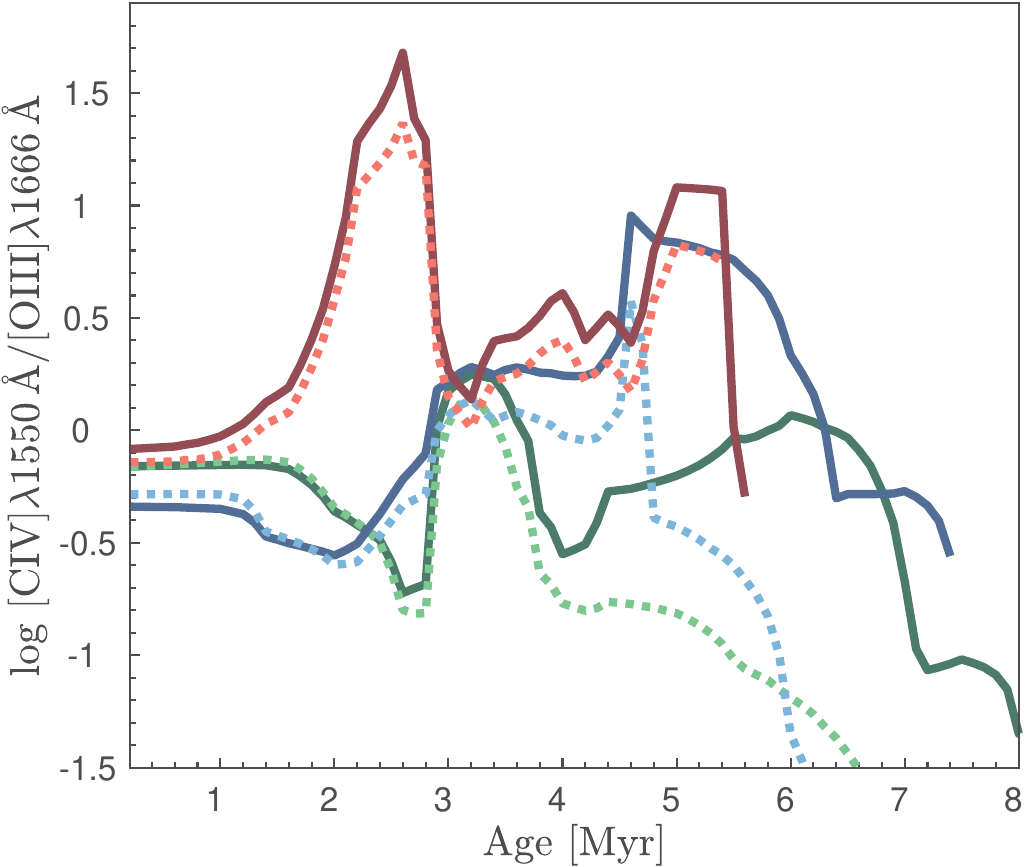}
\includegraphics[width=0.325\textwidth, trim={0.cm 0.cm 0.cm 0.cm},clip]{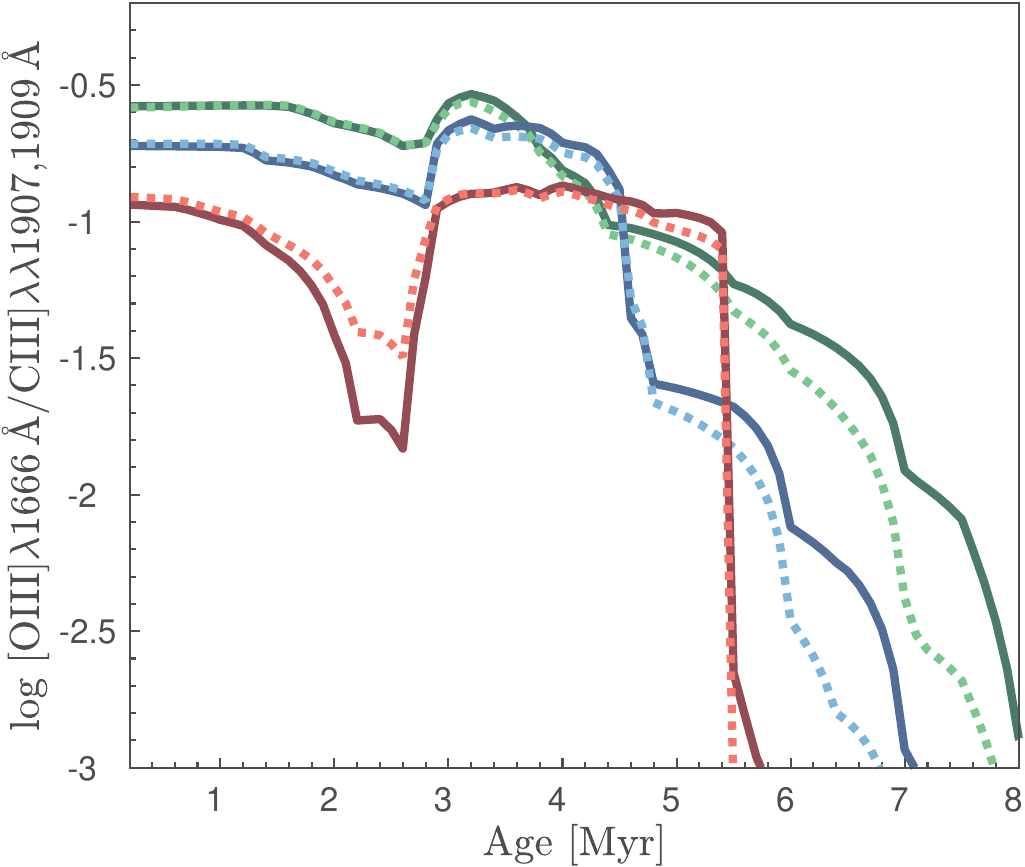}
\includegraphics[width=0.325\textwidth, trim={0.cm 0.cm 0.cm 0.cm},clip]{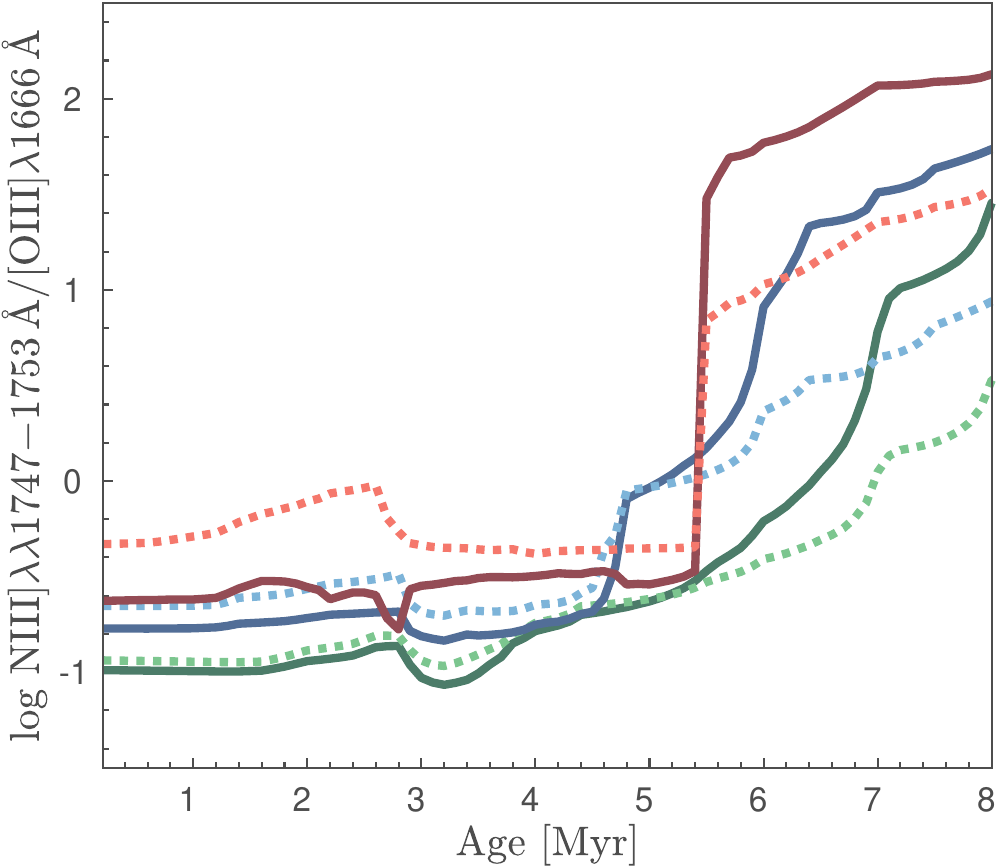}
\includegraphics[width=0.325\textwidth, trim={0.cm 0.cm 0.cm 0.cm},clip]{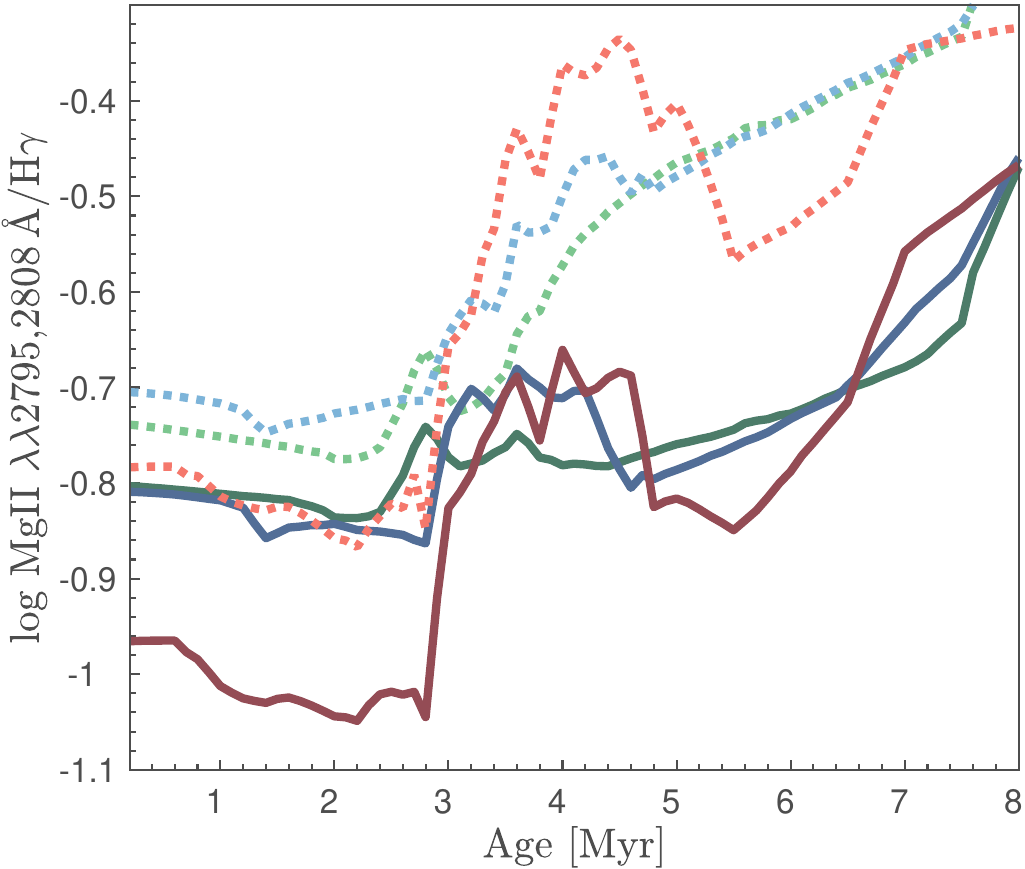}
\includegraphics[width=0.325\textwidth, trim={0.cm 0.cm 0.cm 0.cm},clip]{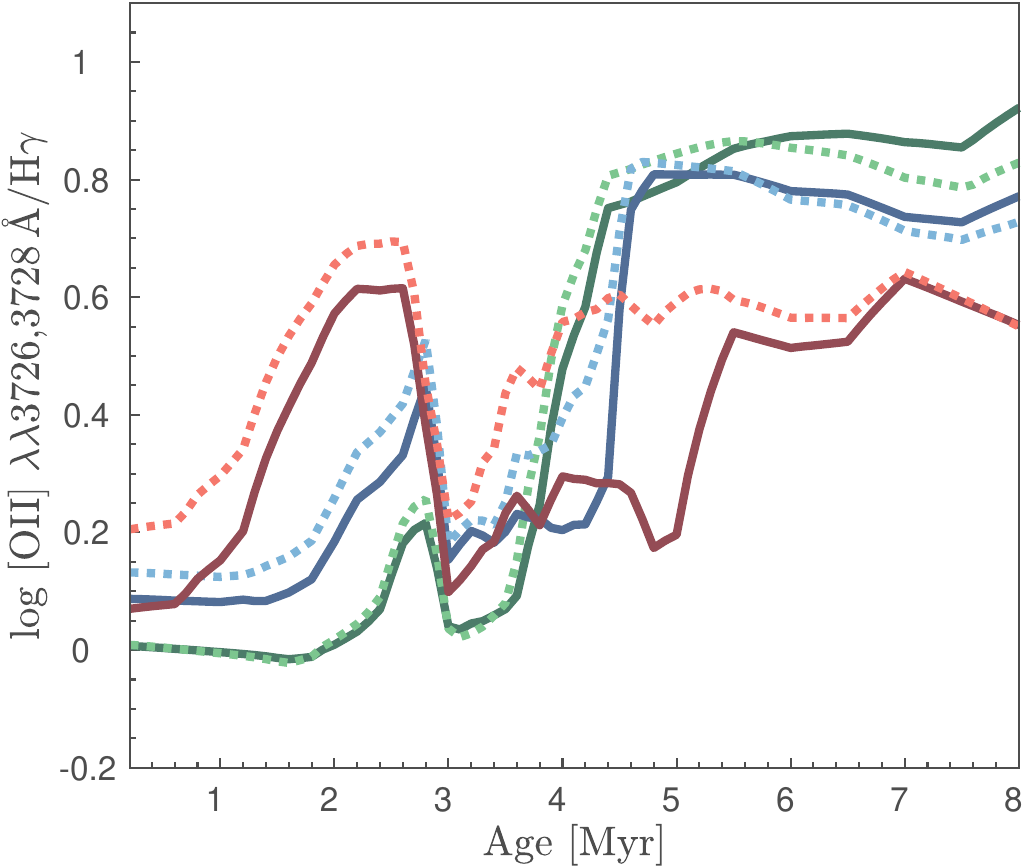}
\includegraphics[width=0.325\textwidth, trim={0.cm 0.cm 0.cm 0.cm},clip]{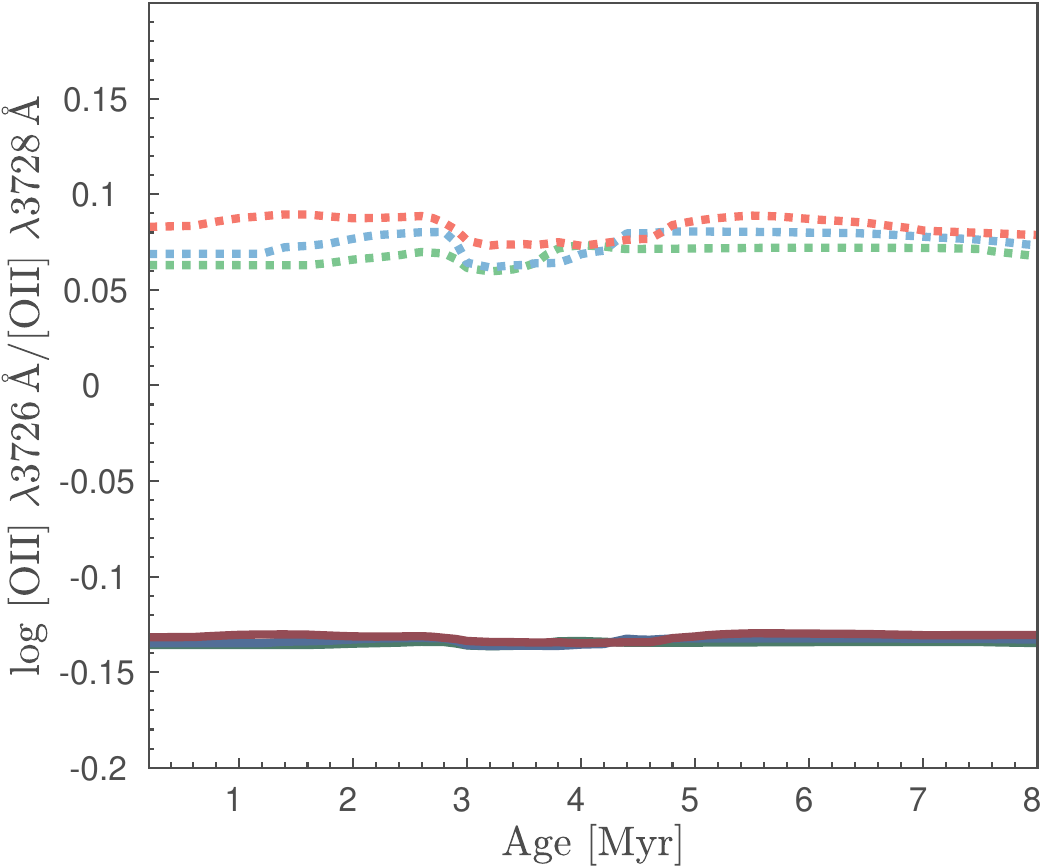}
\caption{The evolution of different line luminosity ratios as a function of n$_{\rm{H}}$ and metallicity for a burst of star formation of strength 10\,000 M$_{\odot}$ for a model HII region of a fixed radius. The evolutionary behaviour for $n_H=100$ (solid lines) and $1000$ (dotted lines) [cm$^{-3}$], and $Z=Z_{\odot}$ to $0.25Z_{\odot}$ (different colours) are considered. The gas and stellar metallicities are approximately similar.}
\label{fig:uv_diagnostics}
\end{center}
\end{figure*}

The diagnostics diagrams involving combinations of collisionally excited metal lines or line multiplets, such as \ion{N}{v}\,$\lambda1240$\AA\,(multiplet), \ion{O}{i}\,$\lambda1304$\AA\,(triplet), \ion{S}{iv}\,${\lambda}1397$\AA + \ion{O}{iv}]\,$\lambda1402$\AA, [\ion{N}{iv}]\,$\lambda1483$\AA + \ion{N}{iv}]\,$\lambda$1487\AA\, \ion{C}{iv}\,$\lambda\lambda$1548,1550\AA, \ion{He}{ii}\,$\lambda$1640\AA\,(Balmer recombination line), \ion{O}{iii}]\,$\lambda\lambda$1661,1666\AA, \ion{N}{iii}\,$\lambda$1750\AA\,(multiplet), [\ion{C}{iii}]\,$\lambda$1907\AA + \ion{C}{iii}]\,$\lambda$1909\AA, \ion{Si}{ii}\,$\lambda1814$\AA\,(multiplet), [\ion{Si}{iii}]\,$\lambda1883$\AA + \ion{Si}{iii}]\,$\lambda1892$\AA, [\ion{O}{iii}]\,$\lambda2321$\AA, [\ion{Ne}{iv}]\,$\lambda2424$\AA, [\ion{O}{ii}]\,$\lambda\lambda2470, 2471$\AA, [\ion{Ne}{iii}]\,$\lambda3343$\AA, [\ion{Ne}{v}]$\lambda3426$\AA, [\ion{O}{ii}]\,$\lambda\lambda3726, 3729$\AA, [\ion{Ne}{iii}]\,$\lambda3868$\AA, are useful for identifying the nature of ionising sources and constraining interstellar gas parameters and the shape of the ionising radiation \citep{Groves2004, Feltre2016}. Most the lines listed above are detected in GN-z11. In addition, several higher-order Hydrogen Balmer lines, such as H5--H8, are also detected with high signal-to-noise.

Among these, the collisionally excited \ion{C}{iv}]\,$\lambda$1550\AA\,and \ion{C}{iii}]\,$\lambda$1909\AA\,are some of the most commonly detected ultraviolet emission lines in spectra \citep{Stark2014, Stark2015}. Together with the \ion{He}{ii}\,$\lambda$1640\AA\,Balmer recombination line, the diagnostics based on the \ion{C}{iv}]\,$\lambda$1550\AA\,and \ion{C}{iii}]\,$\lambda$1909\AA\,offer better discrimination between AGN and shock-ionised gas than optical diagnostics, as these lines are predicted to be significantly stronger in shocks than in AGNs \citep{Villar-Martin1997, Allen1998, Groves2004}. 

In highly star-forming systems, however, the effects of massive stellar evolution must be considered when interpreting diagnostics involving \ion{C}{iv}]\,$\lambda$1550\AA\, and \ion{C}{iii}]\,$\lambda$1909\AA\,and \ion{He}{ii}\,$\lambda$1640\AA. At metallicities of $Z_{\rm{ISM}}>0.006$, for instance, the nebular \ion{He}{ii}\,$\lambda$1640\AA\,can be affected by the broad stellar emission originating from WR stars, which are hotter, have denser winds, and are Helium enriched. Additionally, as metallicity increases, the number of energetic photons capable of producing nebular \ion{He}{ii}\,$\lambda$1640\AA\,drop, while stellar \ion{He}{ii}\,$\lambda$1640\AA\,emission rises due to an increase in the number of stars entering the WR phase, as the minimum mass required for WR formation decreases \citep{Meynet1995, Chandar2004}. The interpretation of  \ion{C}{iv}]\,$\lambda$1550\AA\,also presents challenges, as at $Z_{\rm{ISM}}>0.006$, it can display P-Cygni absorption profiles due to the high mass-loss rates of O-type stars, which also depend on metallicity \citep{Walborn1984}.

The ionising photons in the Lyman continuum ($Q_{\rm{H^0}}$) are primarily produced by hot main-sequence stars, while WR stars contribute significantly to the He$^0$ and He$^+$ ionising flux. In Figure\,\ref{fig:evolution_of_ionisingFlux}a, we present the evolution of ionising photon production [s$^{-1}$] over the $0.2-8$ [Myr] period, focusing on the ionising continua for H$^0$ ($\lesssim912$\AA), He$^0$ ($\lesssim504$\AA), He$^+$ ($\lesssim228$\AA) at solar metallicity ($Z_{\odot}$ solid lines), half-solar metallicity ($0.5Z_{\odot}$ dotted lines), and quater-solar metallicity ($0.25Z_{\odot}$ dashed lines). Sub-panels (b) and (c) of Figure\,\ref{fig:evolution_of_ionisingFlux} depict the hardness of the ionising spectrum, expressed as the ratios of  He$^{0}$ to H$^{0}$ ionising luminosity and He$^{+}$ to H$^{0}$ ionising luminosity, respectively, for the same metallicities shown in the main panel.

The most notable differences appear during the WR phase, from $\sim2.5-6$ [Myr], particularly in the He$^+$ continuum ($\lesssim228$ [\AA], the upper edge of the He$^+$ ionising region). These differences are primarily driven by the changing WR-to-O star ratio, with a higher proportion of stars entering the WR phase as metallicity increases. Furthermore, as metallicity rises, the He$^{+}$ continuum becomes increasingly sensitive to contributions from the various WR subtypes. This sensitivity manifests as the zig-zag pattern in Figure\,\ref{fig:evolution_of_ionisingFlux}c, reflecting variations in the relative abundances of WR subtypes and O stars.

In Figure\,\ref{fig:evolution_of_ionisingFlux2}, we track the evolution of ionising continuum for three higher-ionisation lines commonly observed in the ultraviolet spectra of star-forming galaxies: C$^{++}$ ($258.9\lesssim\lambda$ [\AA]$\lesssim508.5$), He$^{+}$ ($504\lesssim\lambda$ [\AA]$\lesssim228$) and O$^{++}$ ($225.7\lesssim\lambda$ [\AA]$\lesssim353$) at $Z_{\odot}$, $0.5Z_{\odot}$ and $0.25Z_{\odot}$. The inset highlights the evolution of the ratios of C$^{++}$/He$^{+}$ and O$^{++}$/He$^{+}$, offering a closer examination of the evolutionary trends seen in the main panel. While the C$^{++}$ and He$^{+}$ ionising luminosities exhibit broadly similar patterns, the inset underscores the variations between the two, though they appear to be relatively subtle compared to O$^{++}$.

Model predictions for several commonly used ultraviolet diagnostics, which are also observed in the GN-z11 spectrum, are presented in Figure\,\ref{fig:uv_diagnostics}\footnote{For the remainder of this paper, we focus on models with solar and sub-solar metallicities, excluding super-solar models.}. The ratios \ion{C}{iv}]\,$\lambda$1550\AA, \ion{O}{iv}]\,$\lambda$1666\AA\,and \ion{C}{iii}]\,$\lambda$1909\AA\,with respect to \ion{He}{ii}\,$\lambda$1640\AA\,are shown in panels (a)–(c) of Figure\,\ref{fig:uv_diagnostics}.

The ionisation potentials of the species involved increase from \ion{C}{iii} (24.38 eV) to \ion{He}{ii} (24.59 eV), \ion{O}{iii} (35.12 eV), and \ion{C}{iv} (47.89 eV) \citep{Leitherer2011, Steidel2014, Feltre2016}. Consequently, the number of photons energetic enough to produce higher-ionisation species decreases as the stellar population ages. This trend is particularly evident in panel (c), where the \ion{C}{iii}/\ion{He}{ii} ratio shows an increase at ages $\lesssim2.5$ [Myr] and $\gtrsim6$ [Myr], reflecting the relatively faster decline of \ion{He}{ii} ionising photons compared to \ion{C}{iii}.

During the WR phase ($\sim$2.5-6 Myr), the reverse occurs, with the \ion{C}{iii}/\ion{He}{ii} ratio showing a depression as \ion{He}{ii} ionising photons are boosted by the hot, exposed cores of WR stars. The zig-zag pattern observed in the solar metallicity models is due to the sensitivity of the He$^+$ ionising flux to the WR subtype-to-O-star ratio, which becomes more pronounced at higher metallicities.

Both \ion{N}{iv}]~$\lambda\lambda1483, 1487$\AA\,and \ion{N}{iii}]~$\lambda1747-1753$\AA\,(multiplet $1747\leq\lambda\leq1753$\AA) are clearly detected in the GN-z11 NIRSpec/prism spectrum, implying extremely elevated Nitrogen abundance in this system \citep{Cameron2023}. In Figure\,\ref{fig:uv_diagnostics}f, we show the evolutionary behaviours of Nitrogen line luminosities with respect to Oxygen. 
Finally, in Figure\,\ref{fig:uv_diagnostics}i, the ratio of the [\ion{O}{ii}]\,$\lambda\lambda3727, 3729$\AA, a well-known gas density diagnostic, is shown.

\section{On the analysis of JADES NIRSpec Spectroscopy of GN-z11} \label{subsec:BunkerFig_interp}
\begin{figure}
\begin{center}
\includegraphics[width=0.473\textwidth, trim={.7cm 0.1cm 1.9cm 1.3cm},clip]{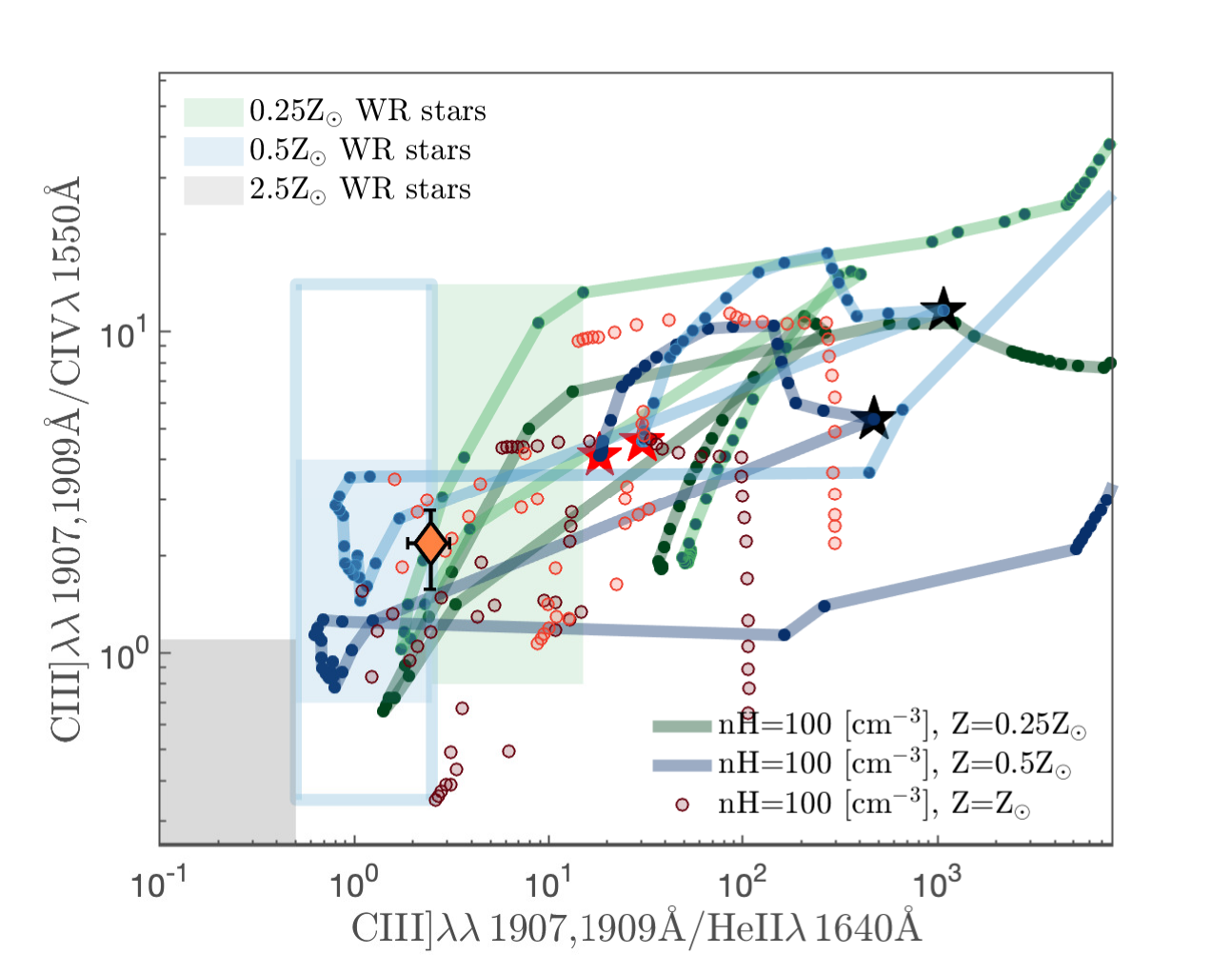}
\caption{Our \textsc{Starburst99}/\textsc{Parsec} + \textsc{Cloudy} models' equivalent of \citet{Bunker2023} Figures\,4. The behaviour of the predicted C\textsc{III}]$\,\rm{\lambda\lambda}1907,1909$\AA\, to C\textsc{IV}$\,\rm\lambda1550$\AA\, versus C\textsc{III}]$\,\rm{\lambda\lambda}1907,1909$\AA\, to He\textsc{ii}$\,\rm\lambda1640$\AA\,for a starburst of strength $10\,000$ M$_{\odot}$ for a model HII region of a fixed radius. The different colours correspond to sub-solar to solar metallicities, while the lighter shading of the same colour denotes $\rm{nH}$ values (100 and 1\,000 [cm$^{-3}$]). Out of the three sub-solar metallicities considered in this study, the lowest metallicity that still allows the formation of WR stars as guided by the \textsc{Parsec} stellar tracks is 0.25Z$_{\odot}$. Therefore, we do not show the lowest metallicity models of 0.07Z$_{\odot}$. Each track is sampled at a time resolution of 0.1 [Myr], and for clarity, we show the solar metallicity models only as data points. The shaded regions of the same colour for 0.5Z$_{\odot}$ and 0.25Z$_{\odot}$ indicate the parameter space sampled by the predicted line ratios as the massive stars enter the WR evolutionary phase. Furthermore, for the 0.5Z$_{\odot}$ models, we also show the region sampled if the starburst strength is varied from 1000 M$_{\odot}$ (upper boundary) to 100\,000 M$_{\odot}$ (lower boundary) as a blue open rectangle. The gray shaded region denote regime sampled by super-solar metallicity models (the tracks and the individual data points are not shown for clarity). The red and black stars denote the start ($\sim$0.2 Myr) of the tracks, and the first appearance of WR stars, respectively for the Z$=0.5$Z$_{\odot}$ tracks, and the GN-z11 line flux ratio from \citet{Bunker2023} is shown as orange diamond for reference.  }
\label{fig:reproduce_buncker_figs}
\end{center}
\end{figure}
\begin{figure*}
\begin{center}
\includegraphics[width=0.49\textwidth, trim={1.5cm .5cm 2.5cm 1.3cm},clip]{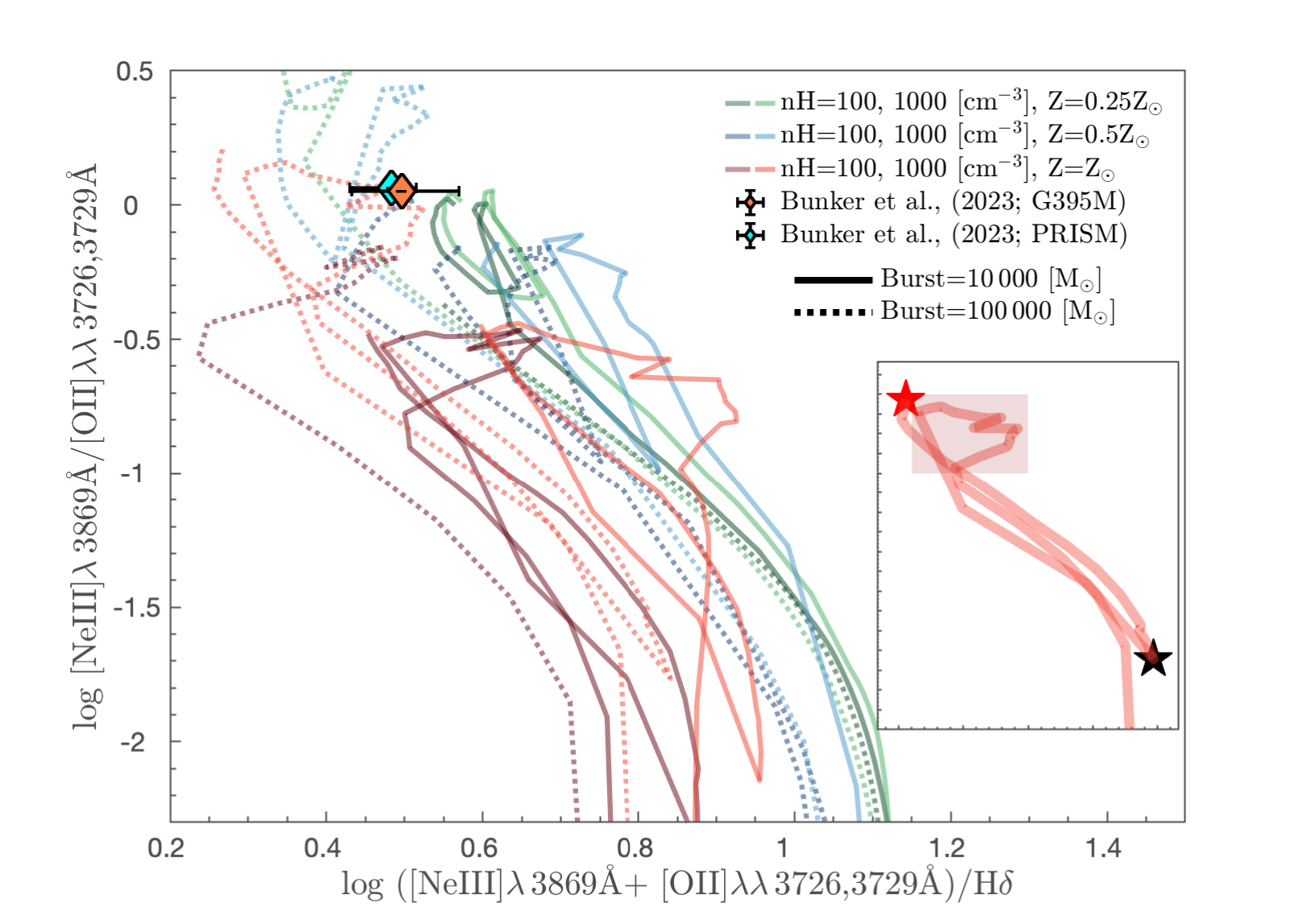}
\includegraphics[width=0.49\textwidth, trim={0.cm .0cm 0.0cm 0.cm},clip]{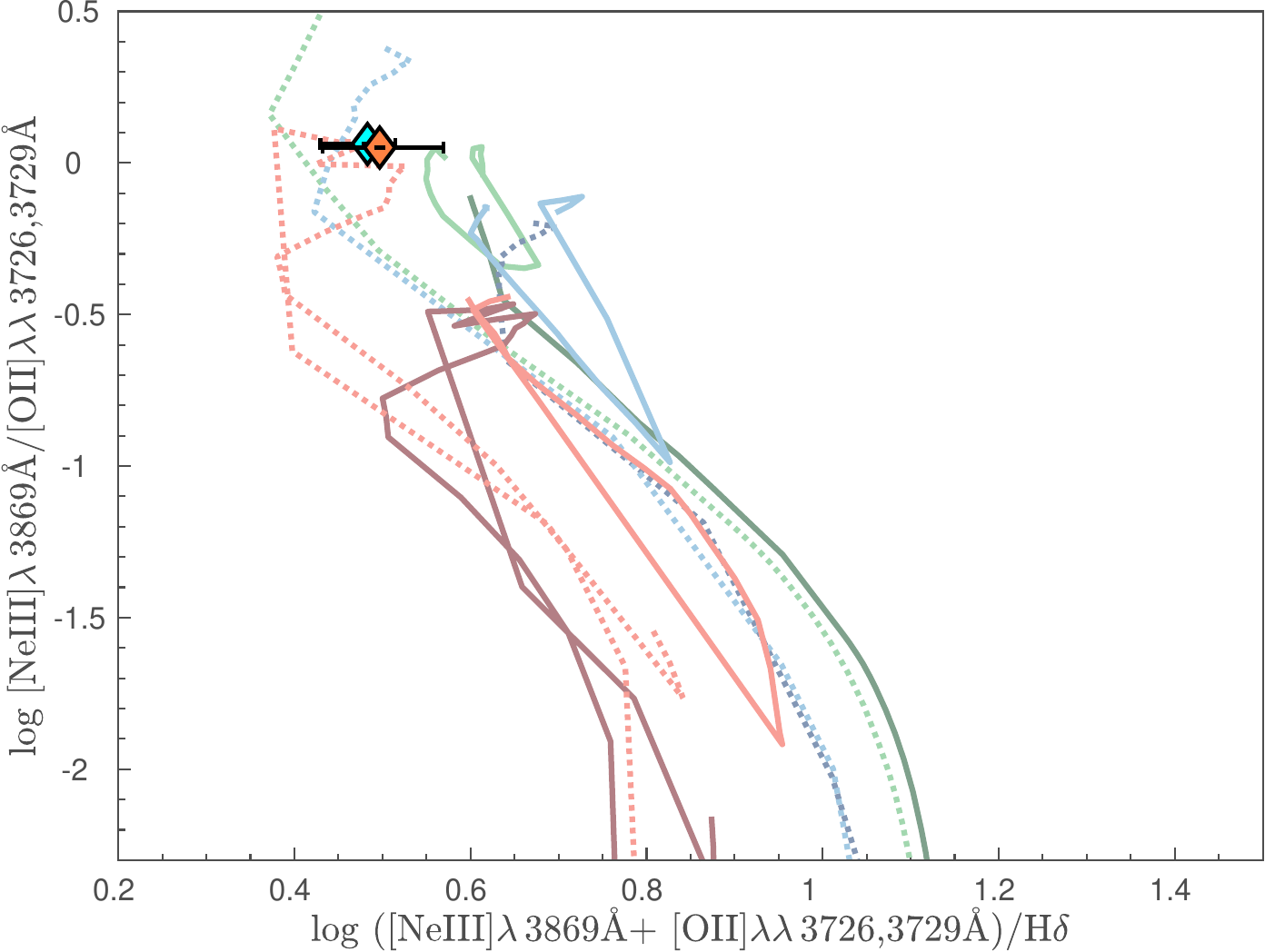}
\caption{our \textsc{Starburst99}/\textsc{Parsec} + \textsc{Cloudy} models' equivalent of \citet{Bunker2023} ] Figures\,8 \citep[see also Figure 8 of][]{Bunker2023}. Left Panel: The predicted evolution of [\ion{Ne}{iii}]~$\rm{\lambda}$3869\AA/[\ion{O}{ii}]~$\rm{\lambda\lambda}$3726,3729\AA\,versus ([\ion{Ne}{iii}]~$\rm{\lambda}$3869\AA + [\ion{O}{ii}]~$\rm{\lambda\lambda}$3726,3729\AA) / H$\delta$ for bursts of star-formation of strengths 10\,000 and 100\,000 M$_{\odot}$ (solid and dotted lines, respectively). The colour coding corresponds to different metallicities and is the same as in Figure\,\ref{fig:reproduce_buncker_figs}. The data points denote the line luminosity ratios reported in \citet{Bunker2023} based on G395M and PRISM/CLEAR observations. In the inset, we show the behaviour of the solar metallicity model, with the red and black stars indicating the start of the track and the point of the first appearance of WR stars. As in Figure\,\ref{fig:reproduce_buncker_figs}, the shaded region in the inset represents the parameter space the model points primarily occupy as massive stars starts to evolve in to WR stars. Right Panel: Same as in the left panel, except illustrating only the predictions in the $-2.0\lesssim \log U\lesssim -1.0$.}
\label{fig:reproduce_buncker_figs8}
\end{center}
\end{figure*}

In this section, we focus on the disagreement between the measured GN-z11 flux ratios and the model predictions reported in \cite{Bunker2023}. In their Figure 4, the observed flux ratio of GN-z11 is compared to the predicted \ion{C}{III}]\,$\rm{\lambda\lambda}1907,1909$\AA\,/\ion{C}{IV}\,$\rm\lambda1550$\AA\,versus \ion{C}{III}]\,$\rm{\lambda\lambda}1907,1909$\AA/\ion{He}{ii}\,$\rm\lambda1640$\AA\,diagnostics for star-forming galaxies \citep[using the models of ][]{Gutkin2016} and AGNs \citep[usin the models of ][]{Feltre2016}. These models consider metallicities in the range  $0.066\lesssim Z/Z_{\odot}\lesssim 0.131$ and hydrogen number densities $2\lesssim \log n_{\rm{H}}/\rm{cm}^{-3}\lesssim 4$, and \cite{Bunker2023} report that GN-z11 lies outside the parameter space sampled by either model set.

To explore this discrepancy, we replicate the diagnostic diagram from Figure 4 of \cite{Bunker2023}, overlaying a subset of our models based on \textsc{Parsec} stellar evolutionary tracks alongside the measured fluxes for GN-z11 (as reported in Table 1 of \cite{Bunker2023}). Figure\,\ref{fig:reproduce_buncker_figs} shows the predicted \ion{C}{III}]\,$\rm{\lambda\lambda}1907,1909$\AA\, to \ion{C}{IV}\,$\rm\lambda1550$\AA\,versus \ion{C}{III}]\,$\rm{\lambda\lambda}1907,1909$\AA\,to \ion{He}{ii}\,$\rm{\lambda}1640$\AA\,diagnostics for $Z/Z_{\odot} \gtrsim 0.25$, $2\lesssim \log n_{\rm{H}} \lesssim 4$, and a starburst of $10\,000$ M$_{\odot}$. The colour coding follows that of Figure\,\ref{fig:uv_diagnostics}, and the shaded regions correspond approximately to the epochs dominated by WR star contributions for each respective metallicity.

The open blue rectangle highlights the parameter space sampled by starbursts with strengths of $1000-100\,000$ M$_{\odot}$ at $0.5$Z$_{\odot}$, while the orange data point represents the observed GN-z11 flux ratios \citep{Bunker2023}. For clarity, we only display models corresponding to solar metallicity as red circles. Additionally, in figures derived from \textsc{Parsec} models, we restrict our focus to sub-solar metallicities ($\gtrsim$0.25Z$_{\odot}$) that still permit WR star formation. As noted earlier, of the sub-solar metallicities explored in this study,  0.25Z$_{\odot}$ is the lowest metallicity that allows WR star formation under \textsc{Parsec} stellar evolutionary tracks. In contrast, \textsc{Geneva} high-mass loss tracks enable WR star formation down to a metallicity of 0.07Z$_{\odot}$.

The H II region models shown in Figure\,\ref{fig:reproduce_buncker_figs}  extend to overlap the GN-z11 data, in contrast to the \cite{Gutkin2016} models presented in Figure 4 of \cite{Bunker2023}. This difference primarily arises from the distinct underlying assumptions between the two sets of models. The star-forming models of \cite{Gutkin2016} assume a continuous star formation rate of 1 M$_{\odot}$ yr$^{-1}$ over 100 Myr, establishing a steady-state population of H II regions within a model galaxy before deriving predictions for nebular emission.

In contrast, our models track the evolution of a single stellar population, in which the majority of ionising photons are produced during a period of $<10$ Myr. This approach effectively captures the temporal evolution of an individual HII region following a burst of star formation. By comparison, the models of \citet{Gutkin2016} simulate ensembles of HII regions representative of an entire galaxy, thereby averaging over the distinctive features that may arise from a single, actively star-forming HIIregion.

Each approach has its advantages and limitations. Our single-burst approach allows for a detailed exploration of how different evolutionary stages of massive stars influence the behaviour of predicted emission line luminosities, explicitly characterising the evolution of massive stars within an H II region. On the other hand, the steady-state approach of \cite{Gutkin2016}, which assumes ongoing star formation until 100 Myr, models a galaxy-scale population of H II regions. While GN-z11 is classified as a galaxy, our focus is on understanding the physical mechanisms--particularly the evolution of massive stars--that could produce the unusual line luminosities detected in its NIRSpec spectrum. To achieve this, tracing the impact of individual evolutionary phases of massive stars on the observed line luminosities requires studying a single starburst event.

As shown in Figure\,\ref{fig:reproduce_buncker_figs}, it is indeed possible to populate the region of the \ion{C}{iii}]\,$\rm{\lambda\lambda}1907,1909$\AA\, to \ion{C}{iv}\,$\rm\lambda1550$\AA\,versus \ion{C}{iii}]\,$\rm{\lambda\lambda}1907,1909$\AA\,to \ion{He}{ii}\,$\rm{\lambda}1640$\AA\,diagnostic plane at the location of GN-z11 through contributions from WR stars.

In Figure\,\ref{fig:reproduce_buncker_figs8}, we present our models' equivalent of Figure 8 from \cite{Bunker2023}: the [\ion{Ne}{iii}]~$\rm{\lambda}$3869\AA/[\ion{O}{ii}]~$\rm{\lambda\lambda}$3726,3729\AA\,versus ([\ion{Ne}{iii}]~$\rm{\lambda}$3869\AA + [\ion{O}{ii}]~$\rm{\lambda\lambda}$3726,3729\AA) / H$\delta$ diagnostic diagram, with the GN-z11 measurements overplotted as orange and cyan data points. Like oxygen, neon is a dominant ionising species and a principal coolant in star-forming regions \citep{Cunha2006}. It is produced during the later stages of massive stellar evolution via carbon burning and is expected to closely track oxygen abundance in H II regions \citep{Levesque2014}.

The [\ion{Ne}{iii}]~$\rm{\lambda}$3869\AA/[\ion{O}{ii}]~$\rm{\lambda\lambda}$3726,3729\AA\,ratio effectively probes the the ionising continuum shape \citep{Perez-Montero2007, Levesque2014} as the ionisation thresholds of N$^{++}$ (303\AA, ionisation potential of 40.96 eV) and O$^+$ (911\AA, ionisation potential of 13.62 eV) spans a wide ultraviolet wavelength range. This extended wavelength range provides greater sensitivity to shorter wavelengths, dominated by ionising photons from young massive stars.

Similar to earlier figures, the evolutionary behavior of [\ion{Ne}{iii}]~$\rm{\lambda}$3869\AA/[\ion{O}{ii}]~$\rm{\lambda\lambda}$3726,3729\AA\,versus ([\ion{Ne}{iii}]~$\rm{\lambda}$3869\AA + [\ion{O}{ii}]~$\rm{\lambda\lambda}$3726,3729\AA) / H$\delta$ is shown in the left panel of Figure\,\ref{fig:reproduce_buncker_figs8} for $n_H=100$ and 1000 [cm$^{-3}$], with starburst strengths of $10\,000$ (solid lines) and $100\,000$ M$_{\odot}$ (dotted lines). The inset in Figure\,\ref{fig:reproduce_buncker_figs8}b provides a close-up of the evolution for the solar metallicity n$_H=1000$ [cm$^{-3}$] model, emphasising the behaviour of the predicted line ratios during the epochs dominated by WR stars when their ionising flux contribution is most significant.

The right panel of Figure\,\ref{fig:reproduce_buncker_figs8} replots the same evolutionary tracks as the left panel but maps them to the ionisation parameter range $-2.0\lesssim \log U \lesssim -1.0$. 

Our motivation for highlighting the parameter space sampled by models over $-2.0\lesssim \log U \lesssim -1.0$ is to emphasise the $Z$-$U$ dependence in our models, particularly since \cite{Bunker2023} reports a $\log U$ of $\sim -2.0$ for GN-z11. Overall, our models within this ionisation parameter range overlap with the GN-z11 measurements. While the 100\,000 M$_{\odot}$ models show the best agreement with GN-z11, it is important to note that the ratio ([\ion{Ne}{iii}]~$\rm{\lambda}$3869\AA + [\ion{O}{ii}]~$\rm{\lambda\lambda}$3726,3729\AA) / H$\delta$ can exhibit two solutions depending on metallicity. Specifically, for $Z/Z_{\odot}\lesssim0.3$, the model tracks progressively shift toward higher ([\ion{Ne}{iii}]~$\rm{\lambda}$3869\AA + [\ion{O}{ii}]~$\rm{\lambda\lambda}$3726,3729\AA) / H$\delta$ ratios with increasing metallicity, while for $Z/Z_{\odot}\gtrsim0.3$, the trend reverses, shifting toward lower ratios.

The two solutions of ([\ion{Ne}{iii}]~$\rm{\lambda}$3869\AA + [\ion{O}{ii}]~$\rm{\lambda\lambda}$3726,3729\AA) / H$\delta$ is partially evident in Figure\,\ref{fig:reproduce_buncker_figs8}, where the 0.25$Z_{\odot}$ models are sandwiched between the solar and 0.5$Z_{\odot}$ models, though closer to the 0.5$Z_{\odot}$ models, with some overlap likely resulting from differences in WR effects at varying metallicities.

Given the dependence of ([\ion{Ne}{iii}]~$\rm{\lambda}$3869\AA + [\ion{O}{ii}]~$\rm{\lambda\lambda}$3726,3729\AA) / H$\delta$ on metallicity, it is possible that $Z/Z_{\odot}\lesssim0.25$ models within $-2.0\lesssim \log U \lesssim -1.0$, incorporating WR effects\footnote{Recall that the \textsc{Parsec} stellar evolutionary models do not allow WR star formation at $Z/Z_{\odot}\lesssim0.25$, whereas \textsc{Geneva} evolutionary models do.}, with moderate to high n$_H$ values ($\sim$1000 [cm$^{-3}$]), may provide better alignment with GN-z11.
Finally, \cite{Bunker2023} report a metallicity range of $0.07\lesssim Z/Z_{\odot}\lesssim0.15$  for  GN-z11, which is consistent with our findings.

\subsection{The stellar populations in the GN-z11}\label{subsec:ssps_in_gnz11}
\begin{figure*}
\begin{center}
\includegraphics[trim={.0cm 0.cm 0.0cm 0.cm},clip, width=\textwidth,angle=0]{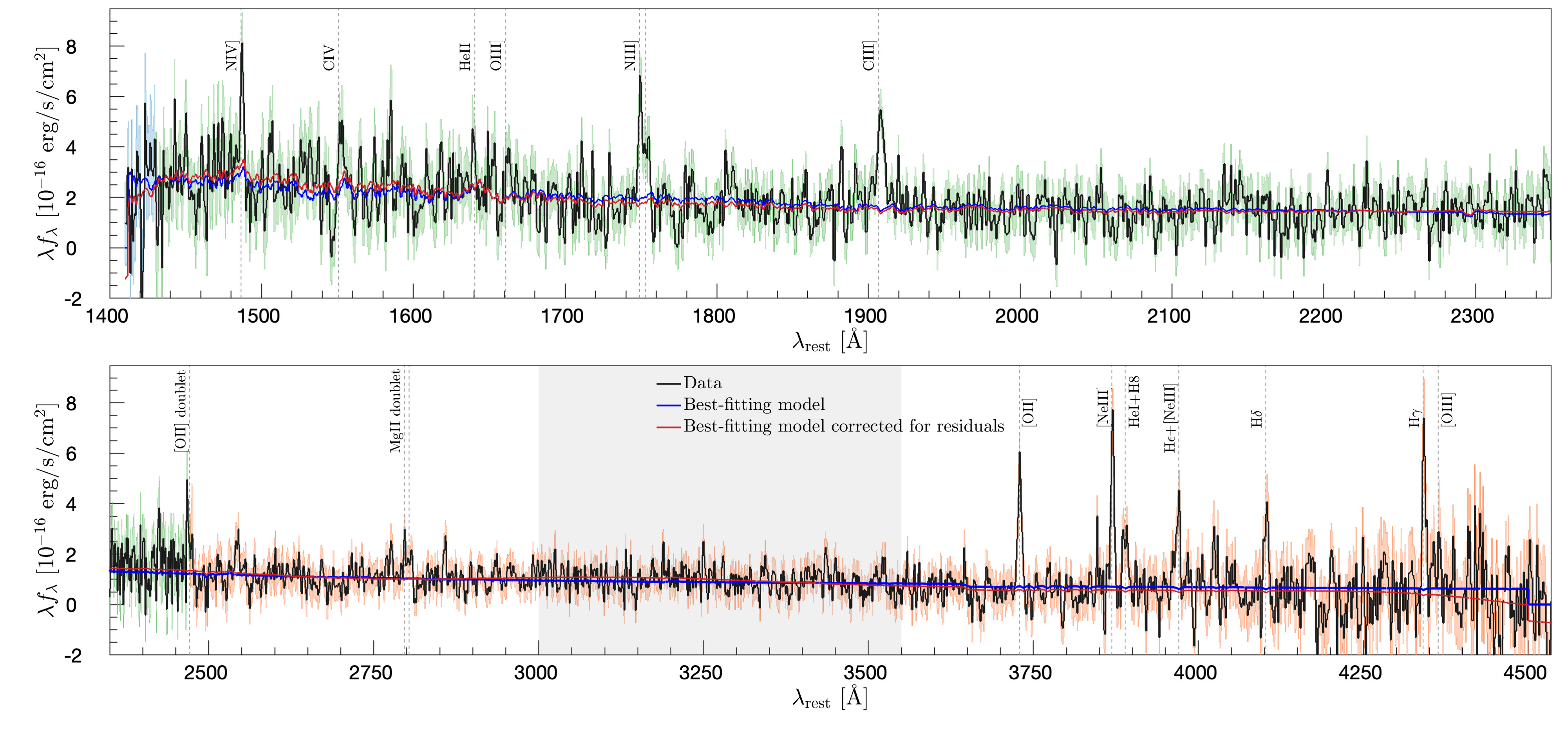}
\caption{The best-fitting continuum model at $Z_s=Z_{\odot}$ with (blue solid line) and without (red solid line) a residual correction to GN-z11 medium resolution G140M (blue), G235M (green) and G395M (red) grating data shown in black. The prominent emission lines observed in the spectrum are labelled, with the grey-shaded region denoting the detection of continuum excess in the rest-frame UV between 3000\AA\,and 3550\AA\,reported by \citet{Ji2024}.}
\label{fig:best-fitting_gnz11}
\end{center}
\end{figure*}
\begin{figure*}
\begin{center}
\includegraphics[trim={.0cm 0.cm 0.0cm 0.cm},clip, width=\textwidth,angle=0]{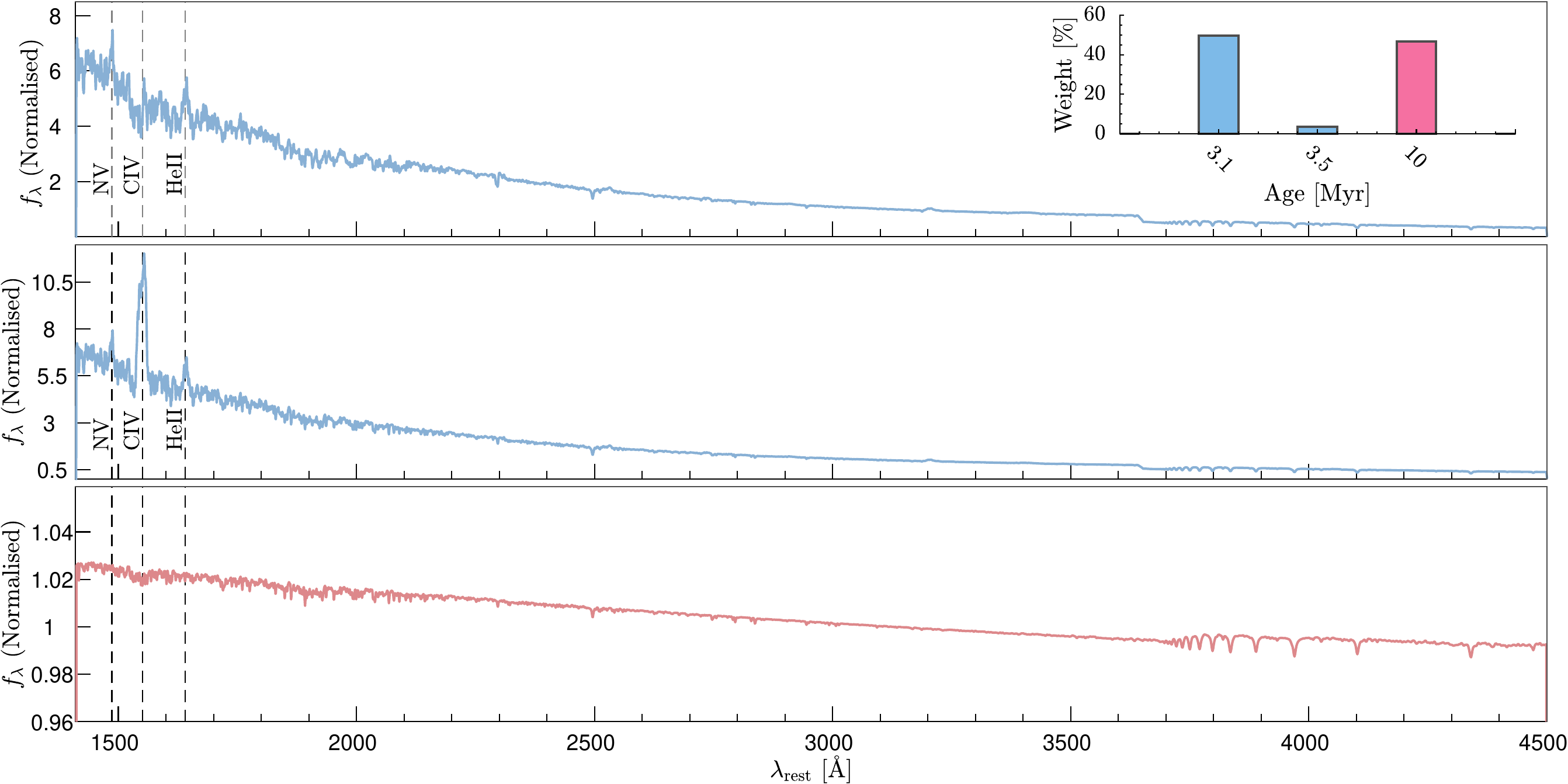}
\caption{The light-weighted and dust-obscured individual stellar population templates contributing to the best-fitting model shown in Figure\,\ref{fig:best-fitting_gnz11}. At solar metallicity, a $\sim3.1-3.5$ Myr old stellar population appear to contribute $>$50\% by weight to the best-fitting model of GN-z11 (blue), and the rest by a $\sim10$ Myr population (red). Also shown are the prominent carbon, helium and nitrogen broad emission peaks unique to WR stars. The inset shows the age distribution and the respective light weights. The templates are normalised at $\sim3000$\AA.}
\label{fig:best-fitting_gnz11_templates}
\end{center}
\end{figure*}

For modelling the spectrum of a star-forming region rich in both stellar and emission features, a spectral fitting methodology that leverages the full range of nebular and stellar information to extract the star formation history (SFH) and physical properties is typically preferred \citep{Gunawardhana2020}. However, our study focuses on understanding the impact of the WR phase on nebular emission and stellar features. Accordingly, the \textsc{Starburst99} + \textsc{Cloudy} model library we constructed is restricted to metallicities that permit WR star formation as dictated by the input stellar isochrones. Furthermore, throughout our analysis, we assume  $Z=Z_{\rm{ISM}}=Z_s$.

We fit the \textsc{Starburst99} + \textsc{Cloudy} models to the GN-z11 spectrum using model-fitting routines built on \textsc{Platefit}, a code originally designed to perform non-negative least-squares fitting with dust attenuation treated as a free parameter, aiming to find the best-fitting stellar continuum model for a given spectrum \citep{Tremonti2004, Brinchmann2004, Gunawardhana2020}. To model the continuum, we combine stellar and nebular continua for a given stellar metallicity and treat dust attenuation as a free parameter using a simple attenuation curve, $\tau(\lambda)\propto \lambda^{-1.3}$, assuming similar extinction between young and old stellar populations. This exponent is suitable for describing the midrange optical properties of dust grains between those of the Milky Way, LMC, and SMC \citep{Charlot2000}. Additionally, we find that the choice of alternative common attenuation laws in the literature \citep[e.g.,][]{Prevot1984, Calzetti2000} has a minimal impact on the fitting results.

By design, \textsc{Platefit} identifies the best-fitting stellar continuum model for a given spectrum. However, to address our aim of investigating the impact of WR stars, we extract the best-fitting linear combination of light-weighted models for each metallicity listed in  Table\,\ref{tab:model_libraries}. Since the input templates to \textsc{Platefit} represent the evolution of a single stellar generation following a burst, our approach assumes that the SFH of GN-z11 can be approximated as a sum of discrete bursts for a given metallicity. The relative weights of individual models contributing to the SFH are defined relative to their luminosity at 2000\AA.

We combine the GN-z11  NIRSpec medium-resolution grating data from G140M (blue), G235M (green), and G395M (red) to perform \textsc{Platefit} stellar and nebular continuum fitting as a function of $Z_s$. This process extracts the individual light-weighted stellar populations likely contributing to the SFH of GN-z11. We determine an approximate velocity dispersion during the continuum fitting, using trial velocity dispersions to converge on the value that minimises the chi-squared statistic. The resulting best-fitting $Z=Z_{\odot}$ model is presented in Figure\,\ref{fig:best-fitting_gnz11}, with grating data color-coded as G140M (blue), G235M (green), and G395M (red).

The figure also compares the best-fitting model with (red) and without (blue) a correction for residuals to account for any subtle variations in the continuum, noting that the correction primarily affects the bluest and reddest regions of the spectrum. Overall, we find that the best-fitting model without additional residual corrections adequately represents the GN-z11 spectrum. The best-fitting continuum model is then subtracted to extract the nebular emission luminosities presented in \S\,\ref{subsec:photomodels_predictions}.

Figure\,\ref{fig:best-fitting_gnz11_templates} shows the individual light-weighted stellar and nebular continua templates contributing to the best-fitting Z$_{\rm{s}}=$Z$_{\odot}$ model in Figure\,\ref{fig:best-fitting_gnz11}. The youngest stellar templates, aged $3.1-3.5$ Myr, are shown in blue and contribute approximately over 50\% (by light-weight) to the total stellar population of GN-z11 at solar metallicity. The remaining contribution comes from a 10 Myr stellar population. 

Prominent features in the youngest templates include broad emission peaks associated with WR stars and the Balmer Jump at 3645\AA, characteristic of young stellar populations. 

Moreover, as previously discussed, we determine the best-fitting linear combination of stellar and nebular continuum models for all stellar metallicities that allow for WR star formation. The best-fitting models for Z$_{\rm{s}}$ = 0.5Z${\odot}$ and 0.25Z$_{\odot}$ assign similar weights to populations with ages of $\lesssim$3.5 Myr, with the remaining contribution distributed at 30 Myr for Z$_{\rm{s}}$ = 0.5Z${\odot}$ and 10 Myr for Z$_{\rm{s}}$ = 0.25Z${\odot}$. These model-derived ages are broadly consistent with the stellar population age of $\sim$19 Myr estimated by \cite{Bunker2023} using \textsc{Beagle}-SED fitting.

Finally, to further investigate the role of WR stars in GN-z11, we subtract the best-matching stellar and nebular continuum models for each metallicity considered and fit Gaussian profiles to the prominent emission lines (see Figure\,\ref{fig:best-fitting_gnz11}) to measure their fluxes.

In the following subsections, we compare and analyse GN-z11 within the framework of various ultraviolet emission line diagnostic schemes constructed based on the \textsc{Starburst99}/\textsc{Parsec} + \textsc{Cloudy} library of models.

\subsection{AGN Diagnostics of GN-z11 in the Ultraviolet}
\begin{figure}
\begin{center}
\includegraphics[width=0.473\textwidth, trim={1.4cm 0.5cm 2.6cm 1.4cm},clip]{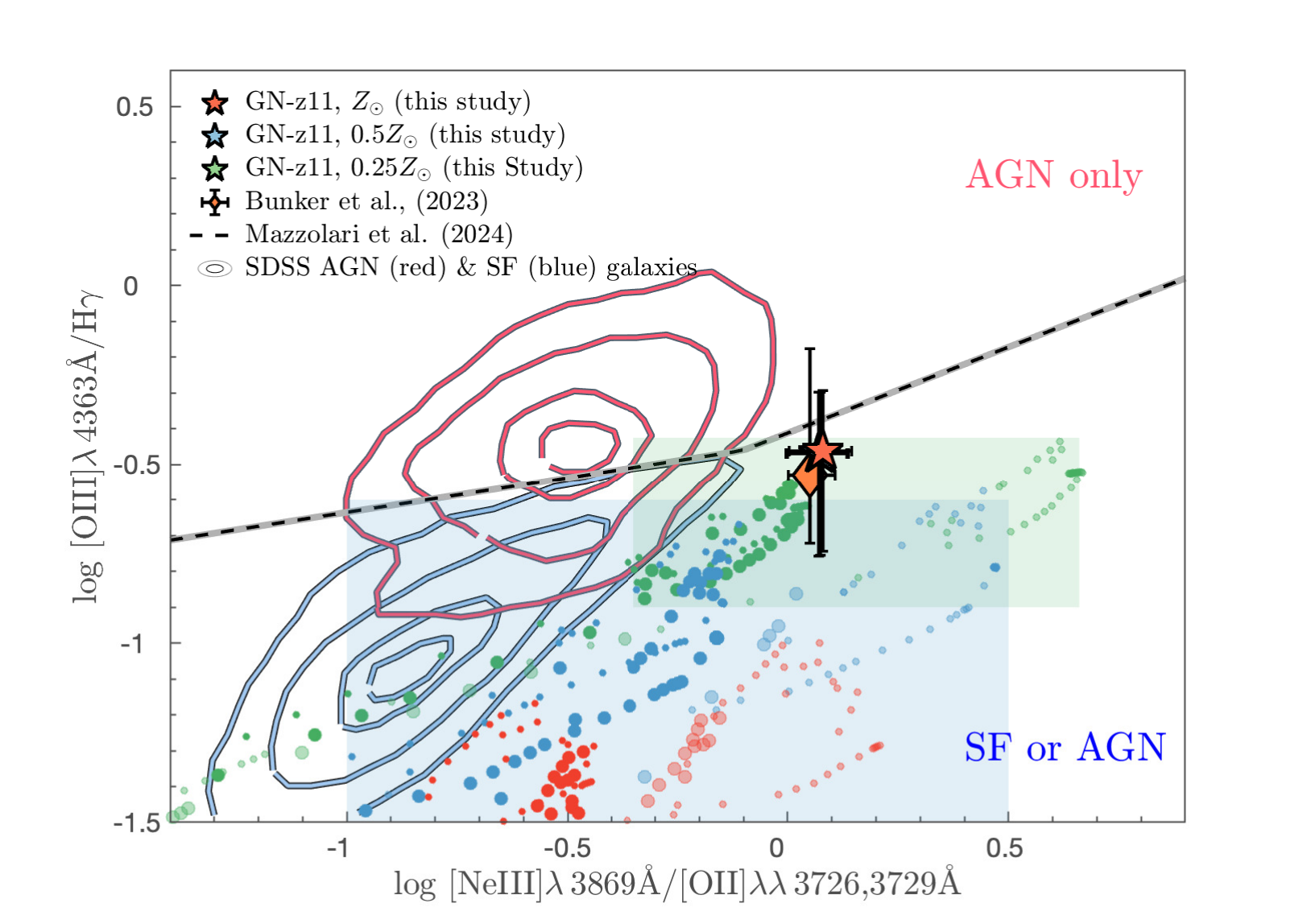}
\caption{The new AGN diagnostics based on the [\ion{O}{iii}]~$\rm{\lambda}$4363\AA\,auroral line introduced by \citet{Mazzolari2024}. The \textsc{Starburst99}/\textsc{Parsec} + \textsc{Cloudy} predictions of the [\ion{Ne}{iii}]~$\rm{\lambda}$3869\AA/[\ion{O}{ii}]~$\rm{\lambda\lambda}$3726,3729\AA versus [\ion{O}{iii}]~$\rm{\lambda}$4363\AA/H$\gamma$ line ratios used by \citet{Mazzolari2024} for discriminating AGNs. Similarly to previous figures, the models correspond to starbursts of strengths 10\,000 and 100\,000 M$_{\odot}$ (filled in circles with decreasing symbol sizes) with different colours representing SMC-like, LMC-like and Solar metallicities. The lighter and darker shadings of the same colour denote $n_H$ of 100 and 1000 [cm$^{-3}$], respectively. The shaded regions in each panel approximately denote the parameter space sampled by SMC- and LMC-like metallicity models during the stellar population ages dominated by WR stars. Also shown are the GN-z11 measurements from \citet{Bunker2023} (orange diamond) and the measurements based on this study (red, blue and green stars) for different metallicities. The contours are reproduced from Figure\,2 of \citet{Mazzolari2024}, and denote the distributions of the SDSS AGN (red) and SF (blue) galaxies for 90\%, 70\%, 30\%, and 10\% of the populations, with the dashed black line indicating their AGN demarkation. }
\label{fig:new_agn_discriminator}
\end{center}
\end{figure}

In Figure,\ref{fig:new_agn_discriminator}, we present the \textsc{Starburst99}/\textsc{Parsec} + \textsc{Cloudy} model predictions within a newly introduced ultraviolet AGN diagnostic framework of \cite{Mazzolari2024}. They propose a novel diagnostic method based on the [\ion{O}{iii}]~$\rm{\lambda}$4363\AA\,auroral line in combination with  [\ion{Ne}{iii}]~$\rm{\lambda}$3869\AA, [\ion{O}{ii}]~$\rm{\lambda\lambda}$3726,3729\AA, and H$\gamma$ to distinguish AGN activity from star formation. 
Within this framework, our models are shown for stellar populations with SMC- (green), LMC- (blue), and solar-like (red) stellar metallicities, considering starbursts of strengths 10\,000 and 100\,000 M$_{\odot}$ (filled in circles with decreasing symbol sizes) and gas densities of n$_H$ = 100 cm$^{-3}$ and 1000 cm$^{-3}$. As in previous figures, shaded regions corresponding to SMC- and LMC-like metallicities (the same colour scheme) indicate the parameter space occupied during stellar evolutionary phases dominated by WR stars. The dashed line represents the AGN demarcation threshold introduced by \cite{Mazzolari2024}, while the contours, reproduced from Figure 2 of \cite{Mazzolari2024}, illustrate the 90\%, 70\%, 30\%, and 10\% distributions of Sloan Digital Sky Survey (SDSS) AGN and star-forming (SF) galaxy populations.

The data points correspond to the measurements from \cite{Bunker2023} (diamond symbol) and emission line fluxes derived by subtracting the best-fitting stellar continuum model for each stellar metallicity considered in the present analysis (star symbols).

Notably, our starburst models exhibit a systematic offset from the SDSS distribution of star-forming galaxies, likely due to their design to represent the starburst-driven properties of HII regions, as well as the fact that most of the model points shown correspond to stellar evolutionary phases dominated by WR populations. Importantly, the model predictions overlap with the GN-z11 measurements derived both in this study and in \cite{Bunker2023}, further supporting the potential role of WR-driven enrichment in shaping the ultraviolet emission properties of GN-z11.

\subsection{Exploring Stellar Populations in GN-z11 Through Other Metrics in the Ultraviolet}\label{subsec:photomodels_predictions}
\begin{figure*}
\begin{center}
\includegraphics[width=0.479\textwidth, trim={0.5cm .3cm 1.9cm 1.0cm},clip]{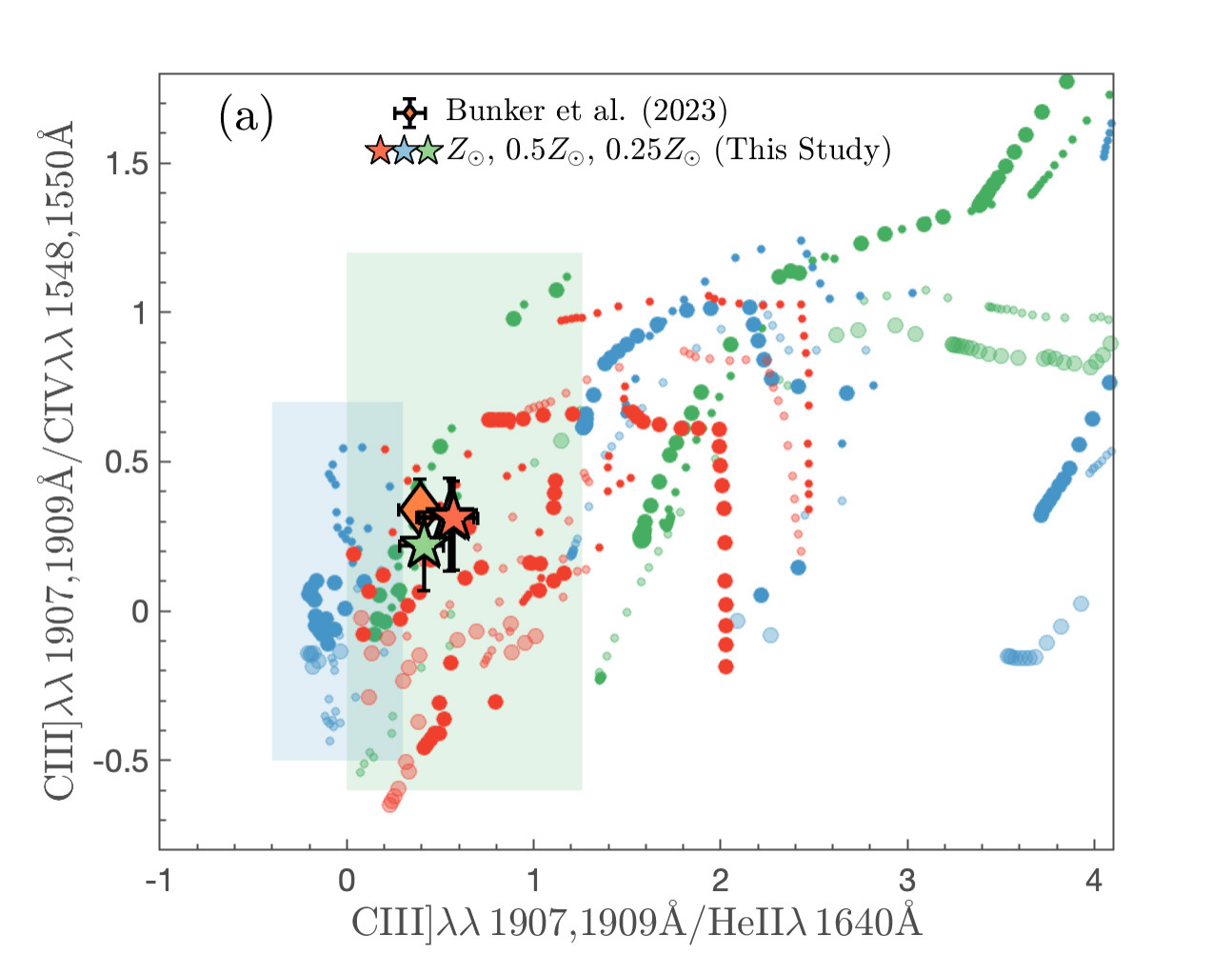}
\includegraphics[width=0.51\textwidth, trim={1.3cm .3cm 2.3cm 1.2cm},clip]{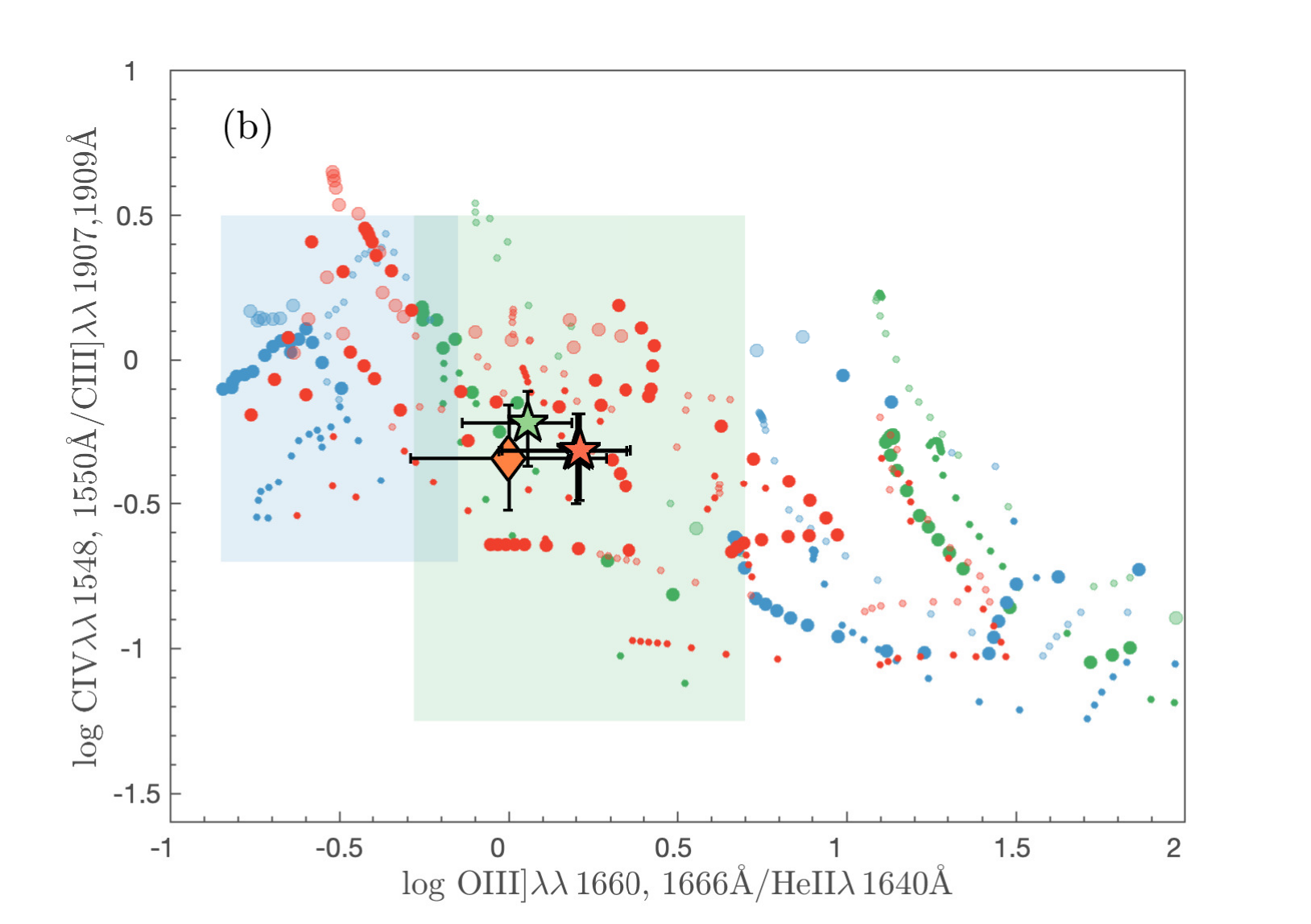}
\includegraphics[width=0.495\textwidth, trim={1.3cm .3cm 2.3cm 1.2cm},clip]{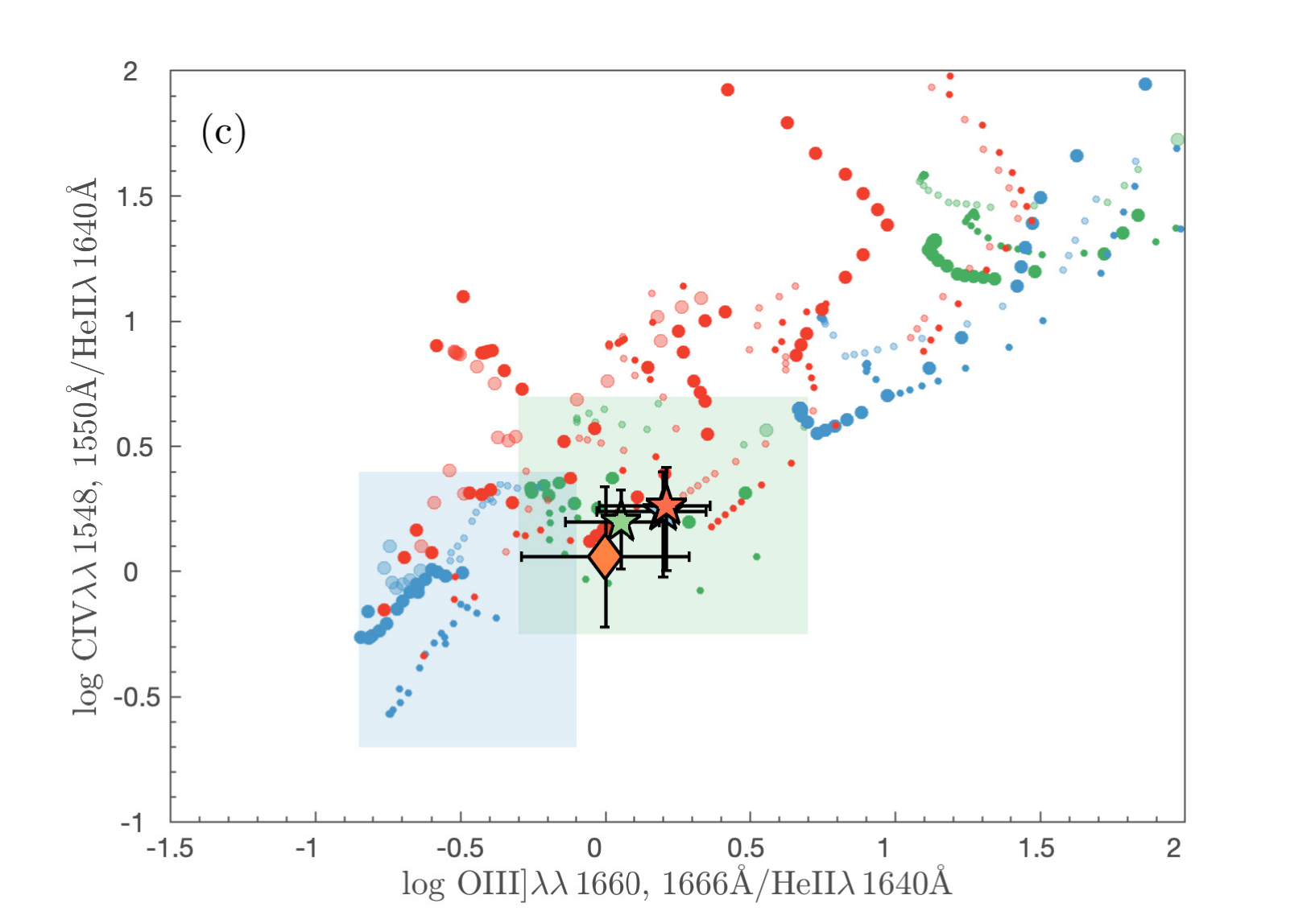}
\includegraphics[width=0.495\textwidth, trim={1.3cm .3cm 2.3cm 1.2cm},clip]{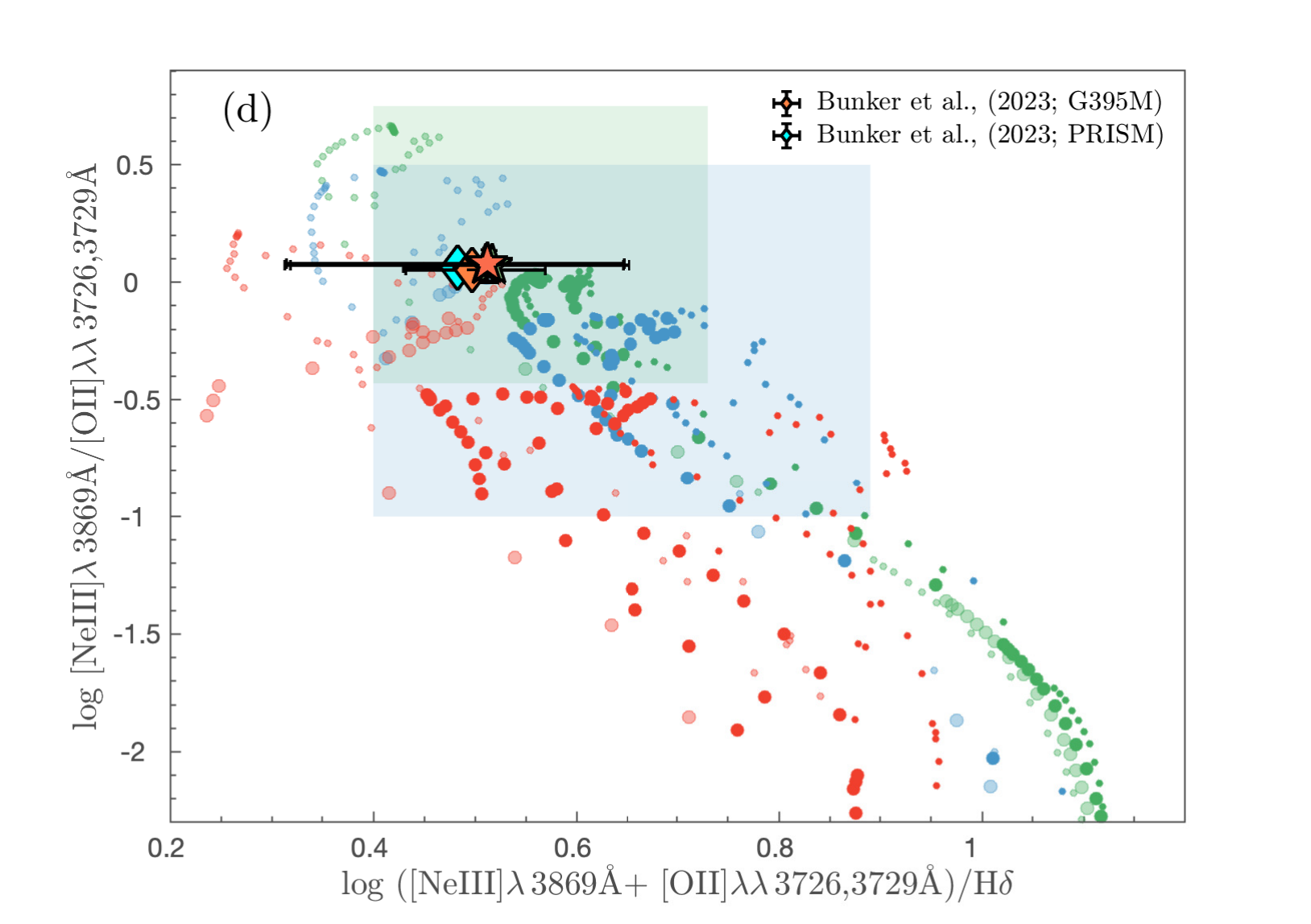}
\includegraphics[width=0.495\textwidth, trim={1.3cm .3cm 2.3cm 1.2cm},clip]{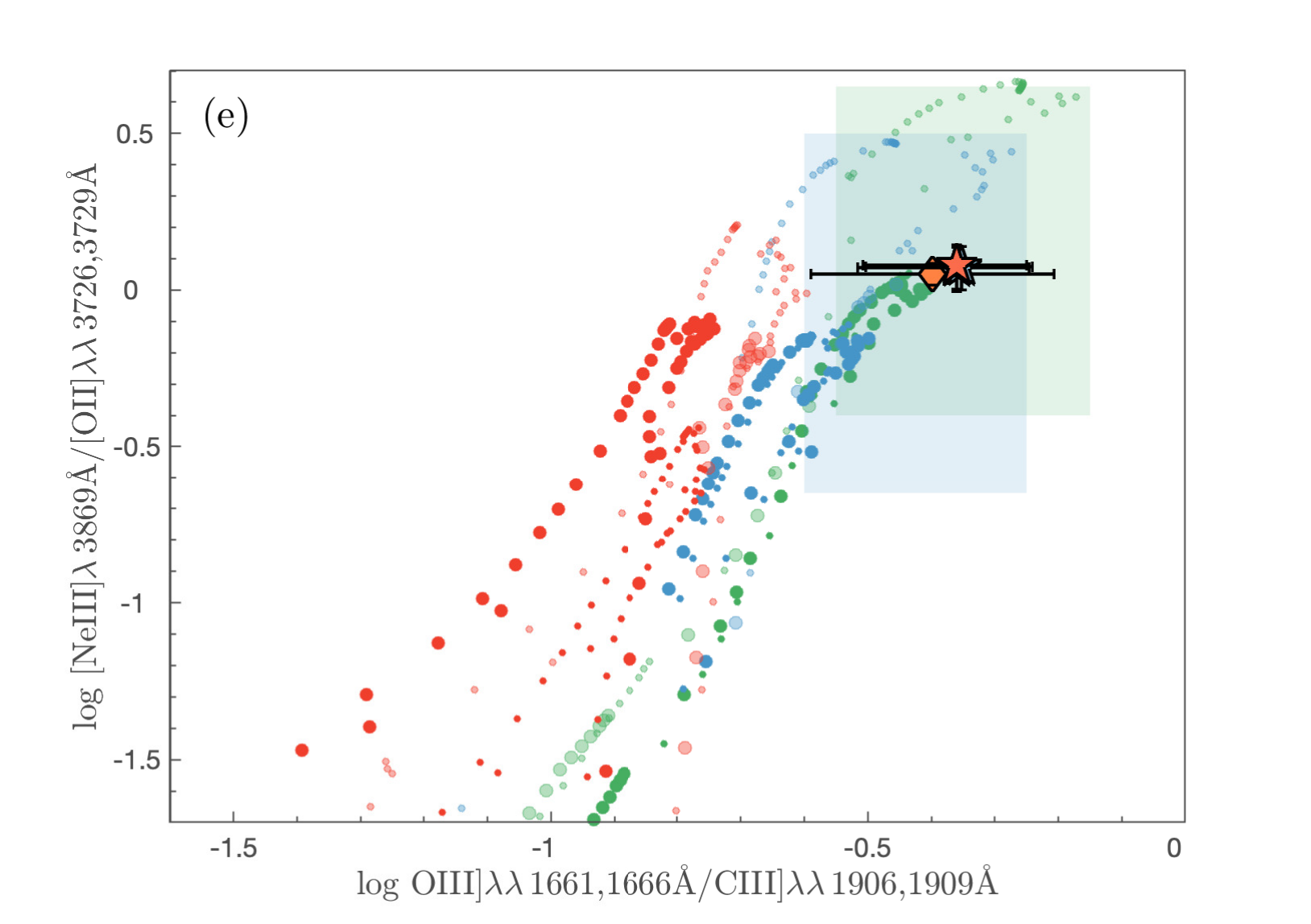}
\includegraphics[width=0.495\textwidth, trim={1.3cm .3cm 2.3cm 1.2cm},clip]{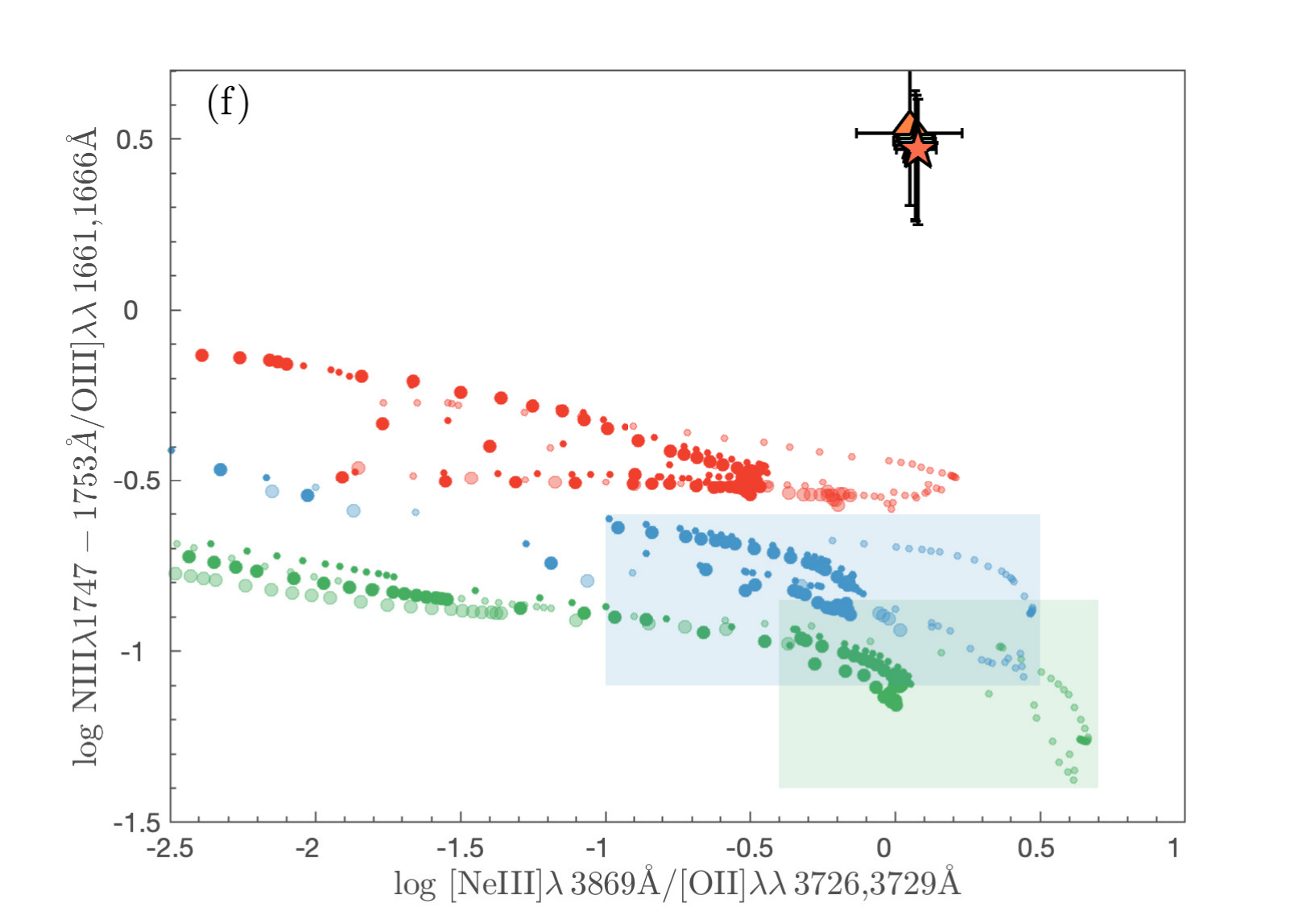}
\caption{The UV diagnostics of prominent emission line luminosities observed in GN-z11.   (a) \ion{C}{iii}]\,$\rm{\lambda\lambda}1907,1909$\AA/\ion{C}{iv}\,$\rm{\lambda\lambda}1548,1550$\AA\,against \ion{C}{iii}]\,$\rm{\lambda\lambda}1907,1909$\AA\,to \ion{He}{ii}\,$\rm{\lambda}1640$\AA\,(b)  [\ion{Ne}{iii}]~$\rm{\lambda}$3869\AA/[\ion{O}{ii}]~$\rm{\lambda\lambda}$3726,3729\AA\,against ([\ion{Ne}{iii}]~$\rm{\lambda}$3869\AA + [\ion{O}{ii}]~$\rm{\lambda\lambda}$3726,3729\AA) / H$\delta$ (c) \ion{C}{iv}\,$\rm{\lambda\lambda}1548,1550$\AA/\ion{C}{iii}]\,$\rm{\lambda\lambda}1907,1909$\AA\,against \ion{O}{iii}]\,$\lambda\lambda$1661,1666\AA/\ion{He}{ii}\,$\rm{\lambda}1640$\AA\,(d) \ion{C}{iv}\,$\rm{\lambda\lambda}1548,1550$\AA/\ion{He}{ii}\,$\rm{\lambda}1640$\AA\,against \ion{O}{iii}]\,$\lambda\lambda$1661,1666\AA/\ion{He}{ii}\,$\rm{\lambda}1640$\AA\,(e) [\ion{Ne}{iii}]~$\rm{\lambda}$3869\AA/[\ion{O}{ii}]~$\rm{\lambda\lambda}$3726,3729\AA\,against \ion{O}{iii}]\,$\lambda\lambda$1661,1666\AA/\ion{C}{iii}]\,$\rm{\lambda\lambda}1907,1909$\AA\, (f) \ion{N}{iii}\,$\lambda1747-1753$\AA/\ion{O}{iii}]\,$\lambda\lambda$1661,1666\AA\,against [\ion{Ne}{iii}]~$\rm{\lambda}$3869\AA/[\ion{O}{ii}]~$\rm{\lambda\lambda}$3726,3729\AA.  As also described in Figure\,\ref{fig:new_agn_discriminator}, the models correspond to starbursts of strengths 10\,000 and 100\,000 M$_{\odot}$ (filled in circles with decreasing symbol sizes) with different colours representing SMC-like, LMC-like and Solar metallicities. The lighter and darker shadings of the same colour denote $n_H$ of 100 and 1000 [cm$^{-3}$], respectively. The shaded regions in each panel approximately denote the parameter space sampled by SMC- and LMC-like metallicity models during the stellar population ages dominated by WR stars. Also shown are the GN-z11 measurements from \citet{Bunker2023} (orange diamond) and the measurements based on this study (red, blue and green stars) for different metallicities.   }
\label{fig:various_uv_diagnostics}
\end{center}
\end{figure*}

In Figure\,\ref{fig:various_uv_diagnostics}, we present several UV emission line ratio combinations across the wavelength coverage of the GN-z11 spectrum to investigate the stellar populations and ionisation conditions of GN-z11.

While certain emission lines, such as \ion{O}{iii}]~$\lambda\lambda$1661,1666\AA, \ion{C}{iii}]~$\rm{\lambda\lambda}1907,1909$\AA, and [\ion{Ne}{iii}]~$\rm{\lambda}$3869\AA\,can be readily analysed by fitting Gaussian profiles to measure their fluxes, the nebular \ion{He}{ii}~$\rm{\lambda}1640$\AA\,and \ion{C}{iv}~$\rm{\lambda\lambda}1548,1550$\AA\,lines present a greater challenge due to their complex line profiles and potential contributions from multiple ionisation sources.

The \ion{He}{ii}~$\rm{\lambda}1640$\AA\, line often comprises two components: a narrow nebular emission component and a broad stellar emission component associated with WR stars \citep{Schaerer1998}. At higher metallicities, the broad WR component becomes increasingly significant\footnote{This is because the minimum initial mass required to form a WR star decreases with increasing metallicity, allowing more stars to enter the WR phase.}, while the number of photons capable of producing nebular \ion{He}{ii}\,$\rm{\lambda}1640$\AA\,decreases. Therefore, without accounting for WR star effects, it is difficult to disentangle the nebular and stellar contributions.

The \ion{C}{iv}~$\rm{\lambda\lambda}1548,1550$\AA\,line doublet, on the other hand, can have strong P-Cygni profiles arising from O-star winds, significantly affecting the line profile \citep{Walborn1984}. To account for this, we simultaneously fit the emission and absorption components of the P-Cygni profile. The WR contamination in \ion{He}{ii}\,$\rm{\lambda}1640$\AA\, by contrast, is at least partly mitigated in our models since they incorporate WR star effects.

The resulting line flux ratios are shown as stars in each panel of Figure\,\ref{fig:various_uv_diagnostics}, while the ratios reported by \cite{Bunker2023} are plotted as diamonds. The light blue and green shaded regions represent the parameter space for line ratio combinations during stellar ages dominated by WR contributions for 0.5Z$_{\odot}$ and  0.25Z$_{\odot}$ metallicities, respectively. Note also that all of these diagnostic diagrams use emission lines that are relatively close in wavelength space to minimise the effects of reddening.

In Figure\,\ref{fig:various_uv_diagnostics} (a)--(c), we present diagnostics involving \ion{C}{iii}]\,$\rm{\lambda\lambda}1907,1909$\AA\,to \ion{C}{iv}\,$\rm{\lambda\lambda}1548,1550$\AA\,collisionally excited line doublets and the \ion{He}{ii}\,$\rm{\lambda}1640$\AA\,Balmer recombination line, which are commonly observed in UV galaxy spectra. The ratios  \ion{C}{iii}]\,$\rm{\lambda\lambda}1907,1909$\AA\,to \ion{He}{ii}\,$\rm{\lambda}1640$\AA\,and \ion{C}{iv}\,$\rm{\lambda\lambda}1548,1550$\AA\,to \ion{He}{ii}\,$\rm{\lambda}1640$\AA\,exhibit sensitivity to metallicity, while the combination of species from different ionisation states, e.g.\,\ion{C}{iii}]\,$\rm{\lambda\lambda}1907,1909$\AA\,to \ion{He}{ii}\,$\rm{\lambda}1640$\AA\,and \ion{C}{iv}\,$\rm{\lambda\lambda}1548,1550$\AA\,to \ion{He}{ii}\,$\rm{\lambda}1640$\AA, primarily trace the ionisation state \citep{Nagao2006, Feltre2016}.

Due to increasing ionisation potentials from \ion{C}{iii}, to \ion{He}{ii}, to \ion{O}{iii}, to \ion{C}{iv} (see \S\,\ref{subsec:uv_line_props}), at fixed other physical parameters, the \ion{C}{iii}]\,$\rm{\lambda\lambda}1907,1909$\AA\,/ \ion{C}{iv}\,$\rm{\lambda\lambda}1548,1550$\AA\,and \ion{C}{iii}]\,$\rm{\lambda\lambda}1907,1909$\AA\,/ \ion{He}{ii}\,$\rm{\lambda}1640$\AA\,ratios are expected to decrease, while the \ion{C}{iv}\,$\rm{\lambda\lambda}1548,1550$\AA\,/ \ion{He}{ii}\,$\rm{\lambda}1640$\AA, and \ion{O}{iii}]\,$\lambda\lambda$1661,1666\AA\,/  \ion{He}{ii}\,$\rm{\lambda}1640$\AA\, ratios rise. This behaviour is clearly evident in panels (a)–(c) for stellar ages not dominated by WR stars.

During the WR phase, however, the sudden influx of ionising photons from WR stars temporarily reverses these trends due to the increased ionisation parameter. After this initial phase, the evolutionary behaviours of the line ratios generally return to the expected patterns, although fluctuations can occur due to the sequential appearance of different WR subtypes.

As noted in \S\,\ref{sec:nebular_models}, we adopt a solar C/O ratio of 0.44 and $\xi_d$ of 0.36 across all metallicities considered. Since our metallicity range is limited to values that permit WR star formation, the overall impact of these assumptions on the results shown in Figure\,\ref{fig:various_uv_diagnostics}  is expected to be minimal. However, the potential effects of varying C/O and $\xi_d$ our results are discussed below.

Variations in the C/O ratio can have a significant and direct impact on emission line ratios involving \ion{C}{iii}]\,$\rm{\lambda\lambda}1907,1909$\AA, \ion{C}{iv}\,$\rm{\lambda\lambda}1548,1550$\AA\,and \ion{O}{iii}]\,$\lambda\lambda$1661,1666\AA. In fact, the prominence of these ultraviolet emission lines makes the line ratios based on them effective tracers of the C/O ratio in star-forming regions \citep{Gutkin2016}. For instance, \cite{Gutkin2016} demonstrate that increasing the C/O ratio from $0.1$(C/O)$_{\odot}$ to 1.4(C/O)$_{\odot}$ leads to a substantial enhancement--by approximately an order of magnitude--in \ion{C}{iii}]\,$\rm{\lambda\lambda}1907,1909$\AA\,and \ion{C}{iv}\,$\rm{\lambda\lambda}1548,1550$\AA\,relative to \ion{He}{ii}\,$\rm{\lambda}1640$\AA. 
In Figure \ref{fig:various_uv_diagnostics}(a), a 40\% increase in the C/O ratio, for example, would result in a more pronounced enhancement of \ion{C}{iii}]\,$\rm{\lambda\lambda}1907,1909$\AA/\ion{He}{ii}\,$\rm{\lambda}1640$\AA\,compared to \ion{C}{iii}]\,$\rm{\lambda\lambda}1907,1909$\AA/\ion{C}{iv}\,$\rm{\lambda\lambda}1548,1550$\AA, as while both \ion{C}{iii}] and \ion{C}{iv} lines strengthen with increasing C/O, the enhancement is greater for \ion{C}{iii}] than for \ion{C}{iv}. Increasing C-abundance while keeping all other parameters fixed has a minimal impact on the \ion{O}{iii}]\,$\lambda\lambda$1661,1666\AA/\ion{He}{ii}\,$\rm{\lambda}1640$\AA. On the other hand, a 40\% decrease in the C/O ratio would produce the opposite trend, which is likely the case given that \cite{Cameron2023} report a C/O ratio of $>0.17$ for GN-z11.
Consequently, the primary shifts would be observed in the \ion{C}{iii}]\,$\rm{\lambda\lambda}1907,1909$\AA/\ion{C}{iv}\,$\rm{\lambda\lambda}1548,1550$\AA\,and \ion{C}{iv}\,$\rm{\lambda\lambda}1548,1550$\AA/\ion{He}{ii}\,$\rm{\lambda}1640$\AA\,ratios, shown in panels (b) and (c), respectively.

For $\xi_d$, the adopted value of 0.36 provides a typical balance for the metallicities considered in this analysis. However, an increase in $\xi_d$ has a complex effect on the line ratios involving \ion{C}{iii}]\,$\rm{\lambda\lambda}1907,1909$, \ion{C}{iv}\,$\rm{\lambda\lambda}1548,1550$\AA. The depletion of coolants from the gas phase reduces cooling efficiency through infrared fine-structure transitions, leading to higher electron temperatures and enhanced cooling via optical and ultraviolet transitions \citep{Feltre2016, Gutkin2016}.

 At low Z$_{ISM}$, a rise in $\xi_d$ decreases the \ion{C}{iii}]\,$\rm{\lambda\lambda}1907,1909$\AA/\ion{He}{ii}\,$\rm{\lambda}1640$\AA\,and \ion{C}{iv}\,$\rm{\lambda\lambda}1548,1550$\AA/\ion{He}{ii}\,$\rm{\lambda}1640$\AA\ ratios due to the depletion of carbon from the gas phase. Conversely, at high Z$_{ISM}$, increasing $\xi_d$ enhances these ratios as the depletion of the coolants from the gas phase leads to higher electron temperatures, thus a higher excitation rate. This effect is, however, tempered to an extent by the increase in dust optical depth associated with higher $\xi_d$, which can lower electron temperatures through enhanced absorption of energetic photons. This interplay between coolant depletion, electron temperature, and dust absorption highlights the complex impact of $\xi_d$ on C-species, in particular. 

On the other hand, non-refractory elements, such as Nitrogen and Sulfur, generally show a mild increase in emission line strengths with rising electron temperatures, as they are less affected by dust depletion and primarily respond to thermal conditions in the gas phase.

The diagnostic combinations involving Neon are presented Figure\,\ref{fig:various_uv_diagnostics} (d) and (e). As discussed in detail in \S\,\ref{subsec:BunkerFig_interp}, the [\ion{Ne}{iii}]~$\rm{\lambda}$3869\AA/[\ion{O}{ii}]~$\rm{\lambda\lambda}$3726,3729\AA\,versus ([\ion{Ne}{iii}]~$\rm{\lambda}$3869\AA + [\ion{O}{ii}]~$\rm{\lambda\lambda}$3726,3729\AA) / H$\delta$ and \ion{O}{iii}]\,$\lambda\lambda$1661,1666\AA/\ion{C}{iii}]\,$\rm{\lambda\lambda}1907,1909$\AA\,act to diagnose the ionisation conditions, which for GN-z11, are within $-2.0\lesssim\log U\lesssim-1.0$.  

Overall, the various ultraviolet line luminosity predictions accounting for the effects of the WR phase in massive stars, shown in Figure\,\ref{fig:various_uv_diagnostics}a-e, exhibit strong agreement with the observed line luminosity ratios reported in this study \citep[largely consistent with the line flux measurements reported in][]{Bunker2023}. Additionally, \cite{Bunker2023} present GN-z11 observations of \ion{C}{iii}],$\rm{\lambda\lambda}1907,1909$\AA to \ion{C}{iv},$\rm{\lambda\lambda}1548,1550$\AA,against \ion{C}{iii}],$\rm{\lambda\lambda}1907,1909$\AA,to \ion{He}{ii},$\rm{\lambda}1640$\AA, alongside the models of \cite{Gutkin2016}, highlighting a lack of models that adequately sample the parameter space hosting GN-z11 \citep[see Figures 4 and 8 of][]{Bunker2023}. We find that this discrepancy can be resolved by incorporating the evolutionary effects of WR stars.

While panels (a)--(e) of Figure\,\ref{fig:various_uv_diagnostics} show good agreement between the model predictions and the measured flux ratios of GN-z11, the \ion{N}{iii}~$\lambda1747-1753$\AA\,/ \ion{O}{iii}]~$\lambda\lambda$1661,1666\AA\,ratio shown in the panel (f) exhibits a significant offset. Specifically, the models under predict the \ion{N}{iii}/\ion{O}{iii}] ratio by more than an order of magnitude, where, our current models, incorporating WR evolution, appear insufficient to reproduce the high \ion{N}{iii}/\ion{O}{iii}] ratio observed in GN-z11.

It is, in fact, the clear presence of \ion{N}{iv}]\,$\rm{\lambda\lambda}1483,1487$\AA\,and \ion{N}{iii}]\,$\lambda1747-1753$\AA\,\citep{Bunker2023}, which are very rarely observed in the local Universe \citep{Mingozzi2022, Mingozzi2024}, that has opened an intense debate on the physical conditions leading significantly elevated Nitrogen enrichment in GN-z11  at only 440 Myrs after the Big Bang. 

\begin{figure}
\begin{center}
\includegraphics[width=0.48\textwidth, trim={.0cm 0.0cm 0.0cm 0.0cm},clip]{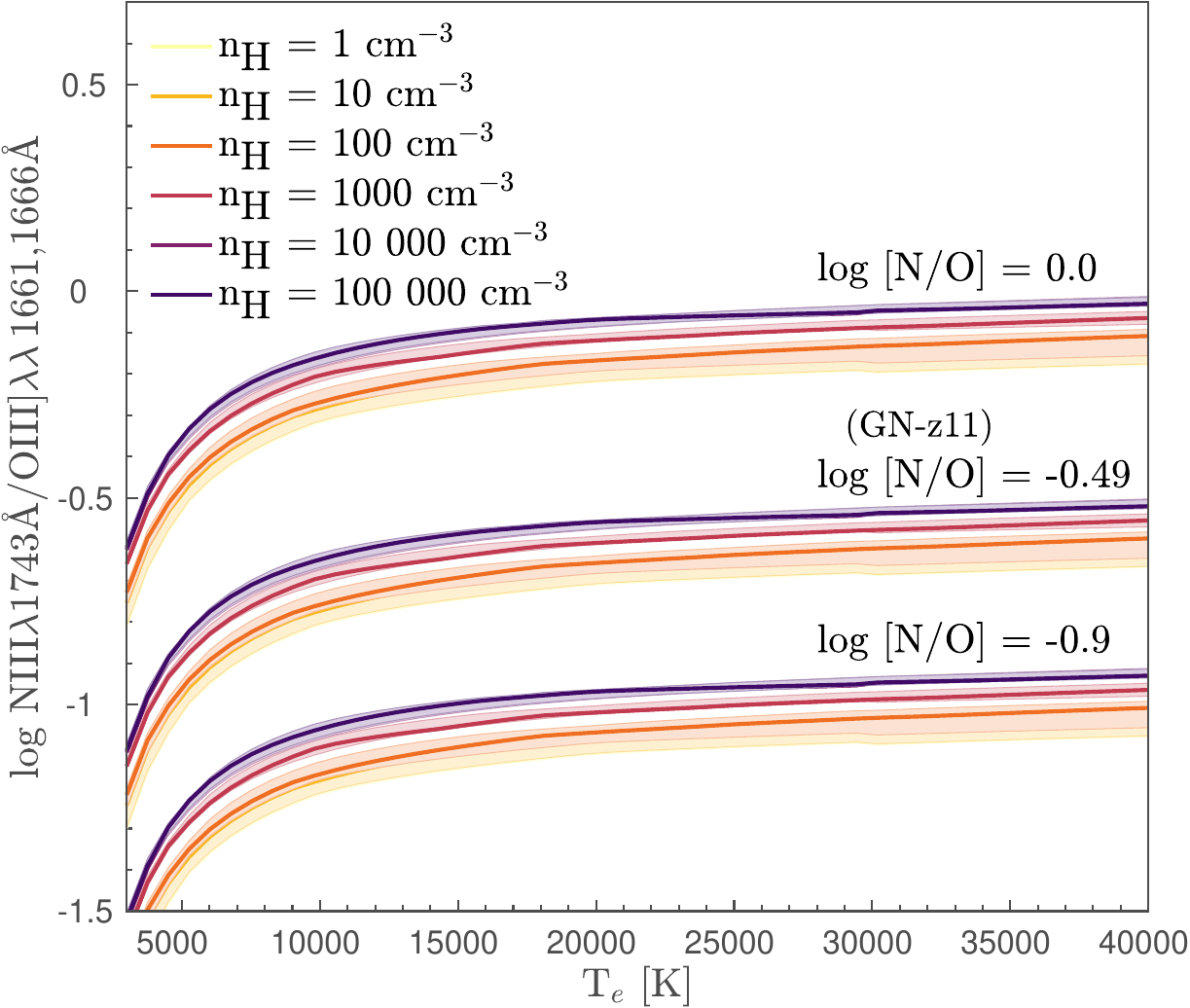}
\caption{\textsc{PyNeb} analysis on the variation of \ion{N}{iii}]\,$\lambda1750$\AA\,multiplet to \ion{O}{iii}]\,$\lambda\lambda$1661,66\AA\,doublet ratio covering all combinations of atomic parameters (i.e.\,energy levels, statistical weights, transition probabilities, effective collision strengths) in a grid of density and temperature for log (N/O) values of -0.9, -0.49 \citep[reported for GN-z11 by][]{Cameron2023}, and 0.0. The colour coding denote different densities, with each colour indicating the variation in the median \ion{N}{iii}]\,$\lambda1750$\AA\,multiplet to \ion{O}{iii}]\,$\lambda\lambda$1661,66\AA\,ratio with increasing electron temperature. The shading of the same colour denote the 16th and 84th percentiles. }
\label{fig:pyneb_no_test}
\end{center}
\end{figure}
Therefore, to examine this discrepancy further, using \textsc{PyNeb} \citep{Luridiana2013}, we calculate the \ion{N}{iii} and \ion{O}{iii} emissivity over all combination of atomic parameters to explore the spread in \ion{N}{iii}]\,$\lambda1747-1753$\AA\,to \ion{O}{iii}]\,$\lambda\lambda$1661,1666\AA\,ratio as a function of the electron temperature in the 5000K to 40\,000K range and densities in the 1 to 10$^5$ [cm$^{-3}$] range, using the Nitrogen to Oxygen abundance from \cite{Groves2008}. The results are shown in Figure\,\ref{fig:pyneb_no_test}. 
As can be seen, for a given Nitrogen to Oxygen scaling, the spread in line flux ratios is minimal with respect to changes in electron temperature and density, and lies far below the observed flux ratio of GN-z11. A significant scaling on N/O is in fact required to bring the \textsc{PyNeb} models in agreement with the observations. 

However, the unusual line ratios shown in the final panel of Figure\,\ref{fig:various_uv_diagnostics} suggest an elevated N/O abundance. While the inclusion of WR stars helps to narrow the gap between the observed line luminosity ratios and model predictions, the observed level of N/O enhancement cannot be fully explained by WR stars alone. Several explanations have been proposed in the literature. For example, \cite{Rizzuti2024} use chemical evolutionary models with varied star formation histories to show that galaxies with extreme star formation rates and differential galactic winds—where the products of core-collapse supernovae are preferentially expelled—can reproduce super-solar N/O abundances.

\section{Discussion}

In this paper, we investigate the unusual emission line luminosity ratios observed in the JADES NIRSpec spectroscopy of GN-z11. The JADES data reveal high line luminosities and a significant detection of the rarely observed \ion{N}{iii}]\,$\lambda1748-1753$\AA\,multiplet, suggesting an unusually high N/O abundance.

Our study is motivated by the analysis of the JADES spectrum of GN-z11 presented in \cite{Bunker2023}. Their findings indicate that, based on the observed emission line ratios and photoionisation models for AGN \citep{Feltre2016} and star formation \citep{Gutkin2016}, neither AGN activity nor star formation alone can explain GN-z11's position in the \ion{C}{iii}]/\ion{He}{ii} versus \ion{C}{iii}]/\ion{C}{iv} diagnostic plane \citep[see Figure,4 of][]{Bunker2023}. Furthermore, using the ([\ion{Ne}{iii}] + [\ion{O}{ii}])/H$\delta$ versus [\ion{Ne}{iii}]/[\ion{O}{ii}] diagnostic ratios, \cite{Bunker2023} report a metallicity range of $0.07 \lesssim Z/Z_{\odot} \lesssim 0.15$ and an ionisation parameter, $\log,U \sim -2$, for GN-z11.

Building on the findings of \cite{Bunker2023}, we develop a suite of stellar and nebular models to further investigate the nature of the stellar populations in GN-z11, utilising the \textsc{Starburst99} stellar population synthesis and \textsc{Cloudy} photoionisation codes. The intense emission line luminosities observed in the spectrum of GN-z11 suggest that the galaxy is undergoing an extreme starburst event. Therefore, our objective, in particular, is to explore whether incorporating the Wolf-Rayet (WR) evolutionary phase of massive stars can resolve the discrepancies reported in \cite{Bunker2023}.

To achieve this, we modify and update \textsc{Starburst99} to incorporate high-resolution stellar spectral libraries, with a particular focus on accurately modelling the WR evolutionary phases of massive stars, as detailed in \S\,\ref{sec:ssp}. The resulting stellar population models are then used as inputs for \textsc{Cloudy}, enabling the self-consistent generation of nebular models across metallicities that support WR star formation (\S\,\ref{sec:nebular_models}).

The equivalent of Figures 4 and 8 from \cite{Bunker2023}, reconstructed using the models developed in this study, are presented in Figures\,\ref{fig:reproduce_buncker_figs} and \ref{fig:reproduce_buncker_figs8}, respectively. As shown in Figure\,\ref{fig:reproduce_buncker_figs}, our starburst models, incorporating improvements to the WR evolutionary phase of massive stars, successfully sample the parameter space occupied by GN-z11. Importantly, we find that the emergence of WR stars is essential to produce a sufficient fraction of ionising photons to populate the region of the \ion{C}{iii}]/\ion{He}{ii} versus \ion{C}{iii}]/\ion{C}{iv} diagnostic plane where GN-z11 resides - a region notably lacking in models in Figure 4 of \cite{Bunker2023}.

Using the metallicity-sensitive ([\ion{Ne}{iii}] + [\ion{O}{ii}])/H$\delta$ versus [\ion{Ne}{iii}]/[\ion{O}{ii}] diagnostic, \cite{Bunker2023} estimate a metallicity range of $0.07 \lesssim Z/Z_{\odot} \lesssim 0.15$ for GN-z11 (see their Figure 8). This metallicity estimate and their assessment of $\log\,U \approx -2$ are in agreement with our model-based derivations.

In the remainder of this paper, we build on the analysis of \cite{Bunker2023} to further investigate the properties of the underlying stellar populations of GN-z11. Figure\,\ref{fig:various_uv_diagnostics} presents different ultraviolet emission line diagnostics considered in this study. The observed line flux ratios shown represent the emission line luminosities derived as a part of this study from subtracting the best-fitting continuum model of each stellar metallicity capable of supporting WR star formation (e.g.\,see Figure\,\ref{fig:best-fitting_gnz11} for the best-fitting continuum model for $Z_s=Z_{\odot}$) alongside the \cite{Bunker2023} measurements.

Overall, the model diagnostics presented in Figure,\ref{fig:various_uv_diagnostics}a-e align well with the observed line luminosities measured in this study, which are consistent with those reported by \cite{Bunker2023} when accounting for the effects of the WR phase in massive stars.

In addition to the \textsc{Parsec} models, we employed the \textsc{Geneva} stellar evolutionary tracks for comparison with the  \textsc{Parsec} models. Overall, massive stars tend to enter the WR phase earlier in the \textsc{Geneva} models, yielding slightly younger best-fitting ages than those derived using \textsc{Parsec}. This discrepancy is likely driven by the higher mass-loss rates assumed for massive stars in the \textsc{Geneva} framework, which accelerates their evolutionary progression. Minor differences are also apparent in the predicted emission line ratios, with adjustments in n$_{\rm{H}}$, stellar metallicity, and burst strengths required to achieve similar sampling. These variations likely arise primarily from differences in the treatment of mass loss, along with other model-dependent assumptions. Despite these discrepancies, the \textsc{Geneva} predictions are broadly consistent with those of \textsc{Parsec}.

Part of the puzzle of stellar populations in GN-z11, however, remains unsolved as the \textsc{Parsec} or \textsc{Geneva} models cannot reproduce the notable underestimate of \ion{N}{iii}/\ion{O}{iii}] shown in Figure\,\ref{fig:various_uv_diagnostics} f. The unusual line ratios suggest an elevated N/O abundance. While the presence of WR stars partially bridges the gap between the observed line luminosity ratios and the model predictions, the observed magnitude of the N/O enhancement cannot be fully explained by WR stars alone. 

To investigate this further, we examined the emissivities of \ion{N}{iii}] and \ion{O}{iii}]  over all combination of atomic parameters to explore the spread in \ion{N}{iii}]\,$\lambda1747-1753$\AA/\ion{O}{iii}]\,$\lambda\lambda$1661,1666\AA\,ratio in electron temperature and density space (Figure\,\ref{fig:pyneb_no_test}). For reference, the $\log$(N/O) ratio of $-0.49$ reported by \cite{Cameron2023} for GN-z11, which is approximately twice the solar value of $\log$(N/O) $=-0.86$ \citep{Asplund2009}, is also indicated in Figure\,\ref{fig:pyneb_no_test}. These results suggest that a significant scaling of N/O is required to bring the observed line ratios in agreement with the model predictions. 

As discussed earlier, several explanations have been proposed in the literature to account for elevated N/O abundances. For example, \citet{Rizzuti2024} use chemical evolution models incorporating a range of star formation histories to show that galaxies with intense star formation and differential galactic winds—where core-collapse supernova ejecta are preferentially removed—can achieve super-solar N/O ratios. \citet{Cameron2023} explore multiple scenarios, including enrichment from Wolf-Rayet (WR) stars, runaway stellar collisions in dense clusters forming very massive stars, and tidal disruption events, as possible mechanisms for nitrogen enhancement. A similar pathway involving massive stars is proposed by \citet{Vink2024}, while \citet{Watanabe2024} suggest that nitrogen-rich winds from rotating WR stars that undergo direct collapse could also play a role (see \S,\ref{sec:intro} for further discussion).

Therefore, it is plausible that the enrichment observed in GN-z11 arises from a combination of these processes. For instance, incorporating higher wind velocities for stars entering the WR phase in our models could help reconcile the predicted and observed line ratios associated with nitrogen.

\section{Summary}
In this study, we analyse the JADES NIRSpec spectrum of GN-z11, which reveals high emission line luminosities and a significant detection of the rarely observed \ion{N}{iii}],$\lambda1748-1753$\AA\,multiplet, suggesting an unusually high N/O abundance.

Our stellar and nebular models, developed using \textsc{Starburst99} and \textsc{Cloudy}, demonstrate that the WR evolutionary phase of massive stars plays a crucial role in explaining the strong ionising radiation and emission line ratios. The inclusion of WR stars allows our models to reproduce GN-z11's position in the \ion{C}{iii}]/\ion{He}{ii} vs. \ion{C}{iii}]/\ion{C}{iv} diagnostic plane, which was not fully explained by previous models.

Moreover, our model-based metallicity and ionisation parameter estimates ($0.07 \lesssim Z/Z_{\odot} \lesssim 0.15$, $\log,U \approx -2$) are in agreement with those derived in \cite{Bunker2023}. Likewise, the stellar ages inferred from our model fitting are in good agreement with the SED-based age of $\sim19$ Myr presented in their study.

We find that while WR stars contribute significantly to the observed line ratios and are able to explain the position of GN-z11 in various diagnostic diagrams, they alone cannot fully explain the unusually high \ion{N}{iii}/\ion{O}{iii}] ratio. This suggests that an additional mechanism is required to enhance the N/O abundance in GN-z11.

\section*{Acknowledgements}
We thank Caroline Foster for the initial discussions on analysis and the presentation of the paper.  This research
was supported by the Australian Research Council Centre of Excellence for All Sky Astrophysics in 3 Dimensions (ASTRO 3D, CE170100013), and the ARC grant DP190102714. JB acknowledges support from Funda\c{c}\~{a}o para a Ci\^{e}ncia e a Tecnologia (FCT) through the research grants UIDB/04434/2020 and UIDP/04434/2020. DOI: 10.54499/UIDB/04434/2020 and DOI: 10.54499/UIDP/04434/2020. 2020.03379.CEECIND/CP1631/CT0003 and DOI: 10.54499/2020.03379.CEECIND/CP1631/CT0003. AJB acknowledges funding from the “FirstGalaxies” Advanced Grant from the European Research Council (ERC) under the European Union’s Horizon 2020 research and innovation program (Grant agreement No. 789056). SO acknowledges support from the NRF grant funded by the Korea government (MSIT) (No. RS-2023-00214057 and No. RS-2025-00514475).
This study uses the \href{https://www.python.org}{Python} programming language. We acknowledge the use of
\textsc{Numpy} \citep{harris2020array}, \textsc{Scipy} \citep{2020SciPy-NMeth}, \textsc{Matplotlib} \citep{Hunter:2007}, \textsc{Astropy} \citep{2022ApJ...935..167A, 2018AJ....156..123A}, and \textsc{Topcat} \citep{Taylor2005}.

\section*{DATA AVAILABILITY}
This analysis is conducted using the data and emission line fluxes presented in \cite{Bunker2023}.

\section*{AUTHOR CONTRIBUTION STATEMENT}
MLPG devised the project, carried out the analysis and drafted the paper. MLPG and JB contributed to the data analysis. MLPG, JB, SC and AJB contributed to the interpretation of the results. All authors discussed the results and commented on the manuscript.

{
\bibliographystyle{apsrmp}
\bibliography{references}
}
\bsp




\label{lastpage}

\end{document}